\documentclass[longauth]{aa} 

\usepackage{graphicx}
\usepackage{txfonts}

\newcommand{\hii}{H\,{\scriptsize II}}
\newcommand{\uchii}{UC H\,{\scriptsize II}}

\newcommand{\kms}{km s$^{-1}$}
\newcommand{\vlsr}{v$_{lsr}$}

\newcommand{\at}{{ATLASGAL}}

\newcommand{\msol}{$M_{\odot}$}
\newcommand{\lsol}{$L_{\odot}$}

\newcommand{\mysou}{G328.2551-0.5321}
\newcommand{\methanol}{CH$_3$OH}

\begin{document}

\title{The search for high-mass protostars with ALMA revealed up to kilo-parsec scales (SPARKS)} 

   \subtitle{I. Indication for a centrifugal barrier in the environment of a single high-mass {envelope}}

   \author{T. Csengeri 
          \inst{1}
          \and
          S. Bontemps
          \inst{2}
           \and
          F. Wyrowski
           \inst{1}
          \and
           A. Belloche
          \inst{1}
          \and
         K. M. Menten
          \inst{1}
          \and
 S. Leurini
         \inst{3} 
         \and
         H. Beuther
         \inst{4}
         \and 
         L. Bronfman
          \inst{5}
        \and {B. Commer\c con}          
        \inst{6}
         \and {E. Chapillon}          \inst{2,7}
           \and{S. Longmore}   \inst{8}  
       \and {A. Palau}    \inst{9}               
         \and{J.\,C. Tan}  \inst{10,11}
       \and{J. S. Urquhart} \inst{12}
             }

   \institute{
        Max-Planck-Institut f\"ur Radioastronomie,
              Auf dem H\"ugel 69, 53121 Bonn, Germany
              \email{csengeri@mpifr-bonn.mpg.de}
                \and
          OASU/LAB-UMR5804, CNRS, Universit\'e Bordeaux, all\'ee Geoffroy Saint-Hilaire, 33615 Pessac, France 
          \and
          INAF - Osservatorio Astronomico di Cagliari, Via della Scienza 5, I-09047 Selargius (CA), Italy
            \and
          Max Planck Institute for Astronomy, K\"onigstuhl 17, 69117 Heidelberg, Germany
          \and
            Departamento de Astronom\'{i}a, Universidad de Chile, Casilla 36-D, Santiago, Chile
          \and
          Univ. Lyon, ENS de Lyon, Univ Lyon1, CNRS, Centre de Recherche Astrophysique de Lyon UMR5574, F-69007, Lyon, France
           \and{IRAM, 300 rue de la piscine, 38406, Saint-Martin-d'H\'eres, France}
            \and
          Astrophysics Research Institute, Liverpool John Moores      
               \and
               {Instituto de Radioastronom\'ia y Astrof\'isica, Universidad Nacional Aut\'onoma de M\'exico, P.O. Box 3-72, 58090, Morelia, Michoac\'an, M\'exico}
 \and {Dept. of Space, Earth \& Environment, Chalmers University of Technology, Gothenburg, Sweden}
\and {Dept. of Astronomy, University of Virginia, Charlottesville, VA, USA}
 \and
School of Physical Sciences, University of Kent, 
              Ingram Building, Canterbury, Kent CT2 7NH, UK
}

   \date{Received , 2017; accepted , 2017}

 \abstract{
 The conditions leading to the formation of the most massive O-type stars, are still an enigma in modern astrophysics.
 To assess the physical conditions of high-mass protostars in their main accretion phase, here we present a case study of a young massive clump selected from the ATLASGAL survey, \mysou. The source exhibits a bolometric luminosity of $1.3\times10^4$\,\lsol, which allows us to estimate its current protostellar mass to be between $\sim$11 and 16\,\msol.
   We show high angular-resolution observations with ALMA reaching
    a physical scale of $\sim$400\,au. To reveal the structure of this high-mass protostellar envelope in detail at a $\sim$0.17\arcsec\ resolution, we use the thermal dust continuum emission and spectroscopic information, amongst others from the CO ($J$=3--2) line, which is sensitive to the high velocity molecular outflow, the SiO ($J$=8--7), and SO$_2$ ($J$=$8_{2, 6}- 7_{1, 7}$) lines tracing shocks along the outflow, as well as several CH$_3$OH and HC$_3$N lines that probe the gas of the inner envelope in the closest vicinity of the protostar. 
The dust continuum emission reveals a single high-mass protostellar envelope, down to our resolution limit. We find evidence for a compact, marginally resolved continuum source, which is surrounded by azimuthal elongations that could be consistent with a spiral pattern. We also report on the detection of 
a rotational line of CH$_3$OH within its $\varv_{\rm t}=1$ torsionally excited state.
{This shows two bright peaks of emission} spatially offset from the dust continuum peak, {and exhibiting a distinct velocity component $\pm$4.5\,\kms\ offset compared to the source \vlsr.}  Rotational diagram analysis and models based on local thermodynamic equilibrium (LTE) assumption require high \methanol\ column densities reaching $N$(\methanol)={$1.2-2\times10^{19}$}\,cm$^{-2}$, and kinetic temperatures of the order of {160-200}\,K at the position of these peaks. A comparison of their morphology and kinematics with those of the outflow component of the CO 
line, and the SO$_2$ 
line suggests that the high excitation \methanol\ spots are associated with the innermost regions of the envelope. While the HC$_3$N $\varv_{\rm 7}=0$ ($J$=37--36) line is also detected in the outflow, the HC$_3$N $\varv_{\rm 7}=1e$ ($J$=38--37) rotational transition within the molecule's vibrationally excited state shows a compact morphology.  
 We find that the velocity {shifts} at the position of the observed high excitation \methanol\ spots correspond well to the expected Keplerian velocity around a central object with {15\,\msol\ consistent with the mass estimate based on the source's bolometric luminosity}.
 We propose a picture where the CH$_3$OH emission peaks trace the accretion shocks around the centrifugal barrier, pinpointing the interaction region between the collapsing envelope and an accretion disk. The physical properties of the accretion disk inferred from these observations suggest a specific angular momentum several times larger than typically observed towards low-mass protostars. This is consistent with a  scenario of global collapse setting on at larger scales that could carry a more significant amount of kinetic energy compared to the core collapse models of low-mass star formation.  Furthermore, our results suggest that vibrationally exited HC$_3$N emission could be a new tracer for compact accretion disks around high-mass protostars. 
} 


   \maketitle
%

\section{Introduction}

Whether high-mass star formation proceeds as a scaled-up version of low-mass star formation
is an open question in today's astrophysics. 
Signatures of infall and accretion processes associated with the formation of high-mass stars
are frequently observed: 
ejection of material \citep{Beuther2002, Zhang2005, Beltran2011, Duarte2013} with
powerful jets \citep{Guzman2010, Moscadelli2016, Purser2016} 
and the existence of (massive) rotating structures, such as toroids and 
disks has been reported towards massive young stellar objects (MYSOs) \citep{Beltran2005, Sanna2015, Cesaroni2017}. 
Most of these studies focus, however, on sources with high luminosities ($L_{\rm bol}$$>$$3\times10^4$\,$L_\odot$), and are frequently  associated with at least one embedded {\uchii} region (see also \citealp{Mottram2011b}). 
 Some of them harbour already formed O-type YSOs typically accompanied by radio emission
 and surrounded by hot molecular, as well as ionised gas. Some examples are 
  G23.01-00.41 \citep{Sanna2015}, 
  G35.20-0.74N \citep{Sanchez-Monge2013}, and
 G345.4938+01.4677 (also known as IRAS 16562--3959, \citealp{Guzman2014}) 
 that have been studied in detail
  at high angular-resolution. 

High-mass star formation activity is typically accompanied by
the emergence of radio continuum emission \citep{Rosero2016}.
The earlier evolutionary stage of high-mass star formation, 
that precedes the emergence of strong ionising radiation from an UC-H{\scriptsize{II}} region,
is characterised by
typically lower bolometric luminosity (e.g.\,\citealp{Molinari2000,Sridharan2002,M07}). 
This stage can be considered as an analog of the
 Class 0 stage of low-mass protostars \citep{Duarte2013}, which could be 
 the main accretion phase, dominated by the cold and dusty envelope, and is accompanied by powerful ejection of material \citep{Bontemps1996,Andre2000}. 
Due to the high column densities, extinction is very high towards these objects,
and therefore high-mass protostars in this early stage are elusive (e.g.\,\citealp{Bontemps2010, Motte2017}).
Rare examples of them are found to be single down to $\sim$500\,au scales.
They typically only probe a limited mass range: one of the best studied sources is the protostar CygX-N63 with a current envelope mass of $\sim$55\,$M_\odot$ potentially forming a star with a mass of $\lesssim$20\,$M_\odot$ \citep{Bontemps2010, Duarte2013}.
  
We report here on the discovery of a \emph{single high-mass protostellar envelope} with the largest  
 mass observed so far, 
 reaching the $100$\,$M_\odot$ mass range on 0.06\,pc scale, and potentially forming 
 a star with a mass of $\sim$50\,$M_\odot$,
corresponding to an O5-O4 type star.
We resolve 
its immediate surroundings using high angular resolution observations reaching $\sim$400\,au scale
 with the Atacama Large Millimeter/submillimeter Array (ALMA),
  which show evidence for a flattened, rotating envelope, and 
 for accretion shocks implying the presence of a disk at a few hundred au scales.

   \begin{figure*}
   \centering
   \includegraphics[height=8.30cm]{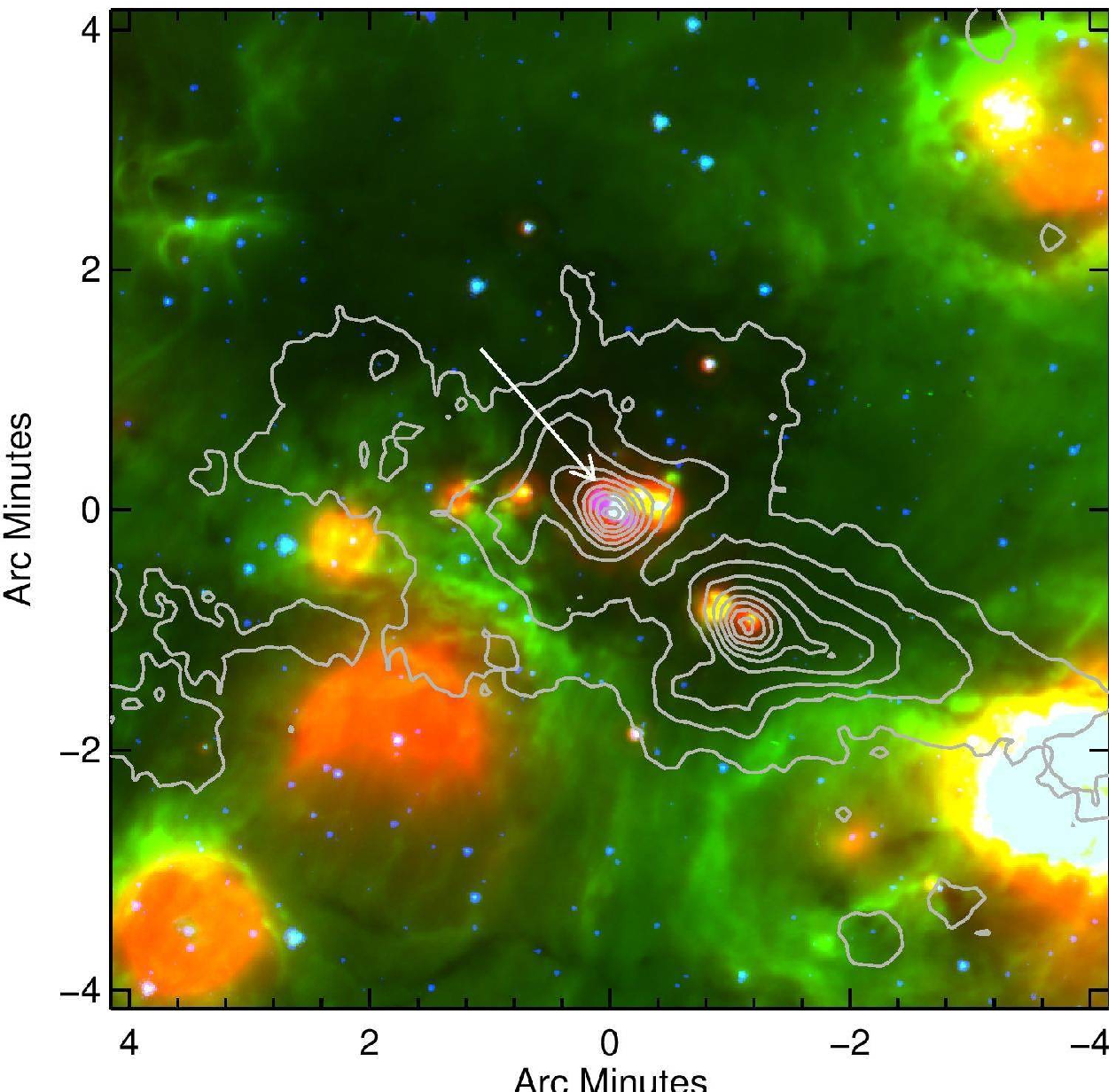}
   \includegraphics[height=8.3cm]{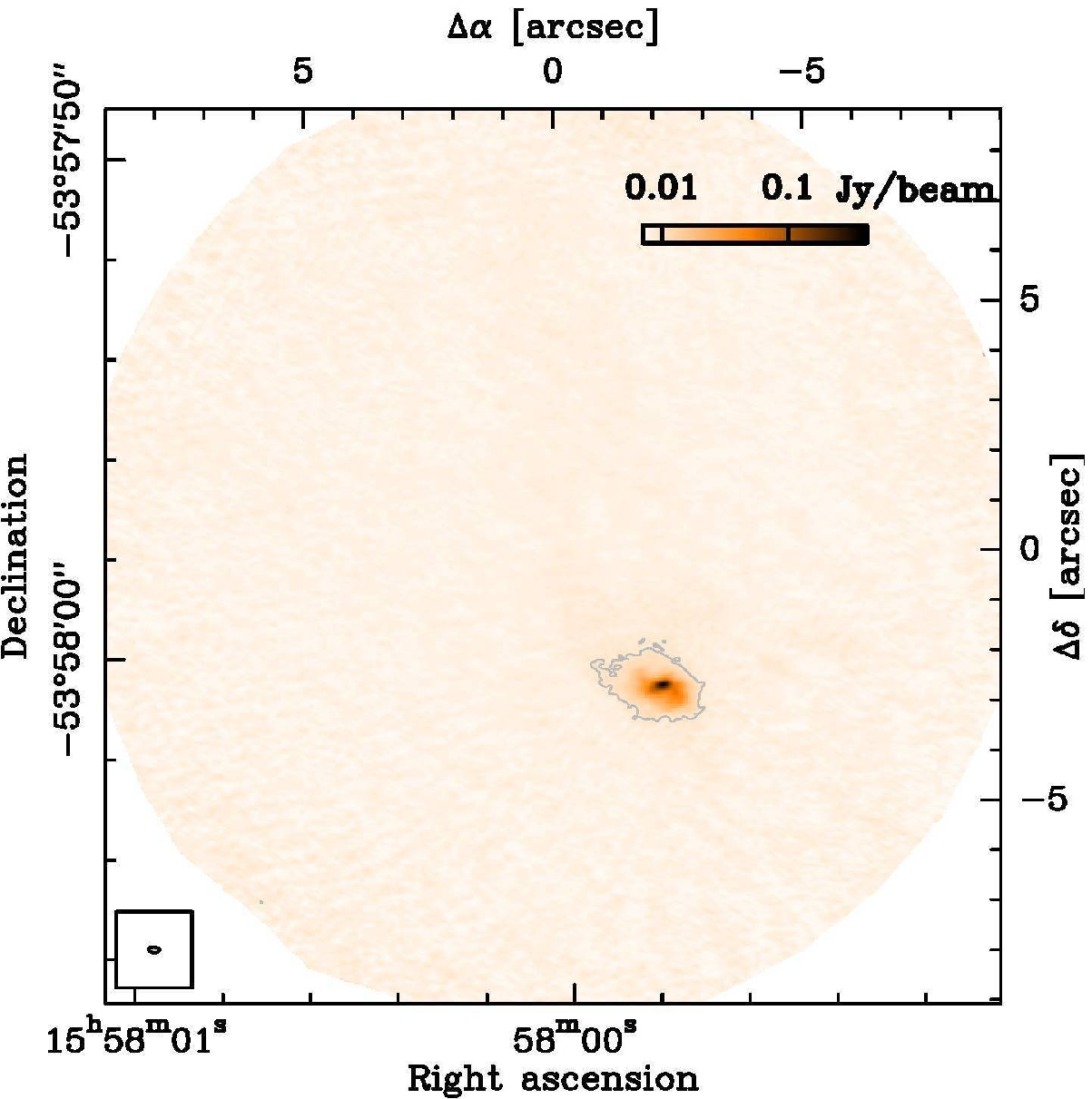}
      \caption{Overview of the region centred {on $\ell$=$+328.2551$, $b$=$-0.5321$ in Galactic coordinates}. {\sl Left:} The three-color composite image is from the \emph{Spitzer}/GLIMPSE \citep{Benjamin2003} and MIPSGAL \citep{Carey2009} surveys (blue: 4.5\,$\mu$m, green: 8\,$\mu$m, red: 24\,$\mu$m) and is shown in Galactic coordinates. The contours show the 870\,$\mu$m {continuum} emission from the ATLASGAL survey \citep{schuller2009, Csengeri2014}. The arrow marks the dust continuum peak of the targeted clump.  {\sl Right:} Line-free continuum emission imaged at 345\,GHz with the ALMA 12m array. The color scale is linear from $-3\sigma$ to $120\sigma$. The contour shows the 7$\sigma$ level.
       The FWHM size of the synthesised beam is shown in the lower left corner. 
      }
              \label{fig:overview}%
    \end{figure*}

\section{Observations and data reduction}

{We study at high angular-resolution} the mid-infrared quiet massive 
 clump\footnote{{Mid-infrared quiet massive clumps are defined by weak or no emission in the 21$-$24\,$\mu$m wavelength range. We follow here the definition of \citet{Csengeri2017a} which is based on \citet{M07}. As discussed there, our mid-infrared flux limit corresponds to that of an embedded star with $10^4$\,\lsol.}}, \mysou,
 selected from the complete sample of such sources
 identified from the APEX Telescope Large Area Survey of the Galaxy (ATLASGAL) \citep{schuller2009, Csengeri2014, Csengeri2017a}. 
 {The SPARKS project 
(Search for High-mass Protostars with ALMA up to kilo-parsec scales, Csengeri et al., in prep, \textsl{a})
targets 35 of these sources
corresponding to the early evolutionary phase of high-mass star formation \citep{Csengeri2017b}.
  Located at a distance of 2.5$^{+1.7}_{-0.5}$\,kpc ,
 our target is embedded in the 
 MSXDC G328.25-00.51 dark cloud \citep{Csengeri2017a}. We show an overview of the region in Fig.\,\ref{fig:overview}, left panel. }

{\mysou\ has been observed with ALMA
in Cycle\,2, and the phase
center  was $(\alpha,\delta)_{\rm J2000}=(15^{\rm h}58^{\rm m}00.05^{\rm s}$, $-53^\circ57'57\rlap{.}{''}8)$.}      
We used 11 of the 7\,m  antennas on the 2014 July 8 and 16, as well as 34
and 35 of the 12\,m antennas on 2015 May 3, and 2015 September 1, respectively. 
The 7\,m array observations are discussed in detail in \citet{Csengeri2017b}. 
Here we also present the 12\,m array observations, for which
the baseline range is 15\,m (17\,k$\lambda$) to 1574\,m (1809\,k$\lambda$). 
The total time on source was 7.4 minutes, and the {system temperature} ($T_{\rm sys}$) varies between 120 and 200\,K. 

The spectral setups used for the 7\,m and 12\,m array observations are identical, and the
signal was correlated in low-resolution wide-band mode
in Band 7, yielding $4\times1.75$\,GHz
effective bandwidth with a spectral resolution of 0.977\,MHz which corresponds to $\sim$0.9\,\kms\
velocity resolution.
The four basebands were centred on
347.331, 345.796, 337.061, and 333.900\,GHz, respectively.  

The data 
have been calibrated in CASA 4.3.1 with the pipeline (version 34044). 
For the imaging, we used Briggs weighting with a robust parameter of $-2$, corresponding to uniform weighting, favouring a smaller beam size, 
and used the CLEAN algorithm for deconvolution. 
We created line-free continuum maps by excluding channels with line emission above 3$\sigma_{\rm rms}$ per channel  
 determined on the brightest continuum source {on the cleaned datacubes} in an iterative process.
 The synthesised beam is 0.22\arcsec$\times0.11$\arcsec\ with 86$^\circ$ position angle measured from north to east, which is the convention we follow from here on.
The geometric mean of the major and minor axes corresponds to a beam size of $0.16$\arcsec ($\sim$400\,au at the adopted distance for the source).

To create cubes of molecular line emission we subtracted the continuum determined in emission free channels 
 around the selected line. To favour sensitivity, we used here a robust parameter of $0.5$ for the imaging, which gives a 
synthesised beam of 0.34\arcsec$\times0.20$\arcsec, corresponding, on average, to a $0.26$\arcsec\ resolution ($\sim$650\,au). 
For the SiO (8--7) datacube we lowered the weight on the longest baselines with a tapering function to gain in signal to noise. The resulting synthesised beam is 0.69\arcsec$\times$0.59\arcsec (corresponding to $\sim$1600\,au resolution).

We mainly focus here on molecular emission that originates  
from scales typically smaller than the largest angular scales of 
 $\sim$\,7\arcsec, where the sensitivity of our 12\,m array observations drops. 
 The interferometric filtering has, however, a significant impact on both the continuum and the CO (3--2) emission,
 which are considerably more extended than {the largest angular scales probed by the ALMA configuration used for these observations}.
 For these two datasets, we therefore {combine} the 12\,m array data  
 with the data taken in the same setup with the 7\,m array \citep{Csengeri2017b}.
 For this purpose we used the standard procedures in CASA for a joint deconvolution, 
 and imaged the data with a robust parameter of $-2$ providing the highest angular resolution. 
The resulting beam is $0.23\times0.12$\arcsec with 86$^\circ$ position angle, with a geometric mean of
$0.17$\arcsec\ for the continuum maps, and $0.31\times0.18$\arcsec with 88$^\circ$ position angle, with a geometric mean of $0.24$\arcsec\ for the CO (3--2) line.
We measure an $rms$ noise {level} in the final continuum image of {$\sim1.3$\,mJy/beam} in an emission free region close to the centre on the {image corrected for primary beam attenuation}. The primary beam of the 12m array at this frequency is 18\rlap{.}{\arcsec}5. {We measure a} 3$\sigma_{\rm rms}$ column density sensitivity  
of 
{$2-8\times10^{23}$\,cm$^{-2}$ 
for {a physically motivated range of plausible} temperatures {corresponding to} $T_{\rm d}=30-100$\,K}, respectively, and within a beam size of 0.16\arcsec\footnote{
We use $N$(H$_2$) = $\frac{F_\nu\,R}{B_\nu(T_d)\,\Omega\,\kappa_{\nu}\,\mu_{\rm H_2}\,m_{\rm H}}
\rm{[cm^{-2}]}$, where $F_\nu$ is the $3\sigma_{\rm rms}$ flux density, $B_{\rm \nu} (T)$ is the Planck-function, 
$\Omega$ is the solid angle
of the beam calculated by $\Omega = 1.13\times \Theta^2$, where 
$\Theta$ is the geometric mean of the beam major and minor axes; 
$\kappa_{\nu}=0.0185$\,g\,cm$^{-2}$ including the gas-to-dust ratio, R, of 100;
 $\mu_{\rm H_2}$ is 
 the mean molecular weight per 
hydrogen molecule and is equal to 2.8; and
 $m_{\rm H}$ is the mass of a hydrogen atom. 
}, and a 5$\sigma$ mass sensitivity of 0.03--0.13\,\msol\ for the same temperature range.

   \begin{figure*}
   \centering
   \includegraphics[height=3.5cm]{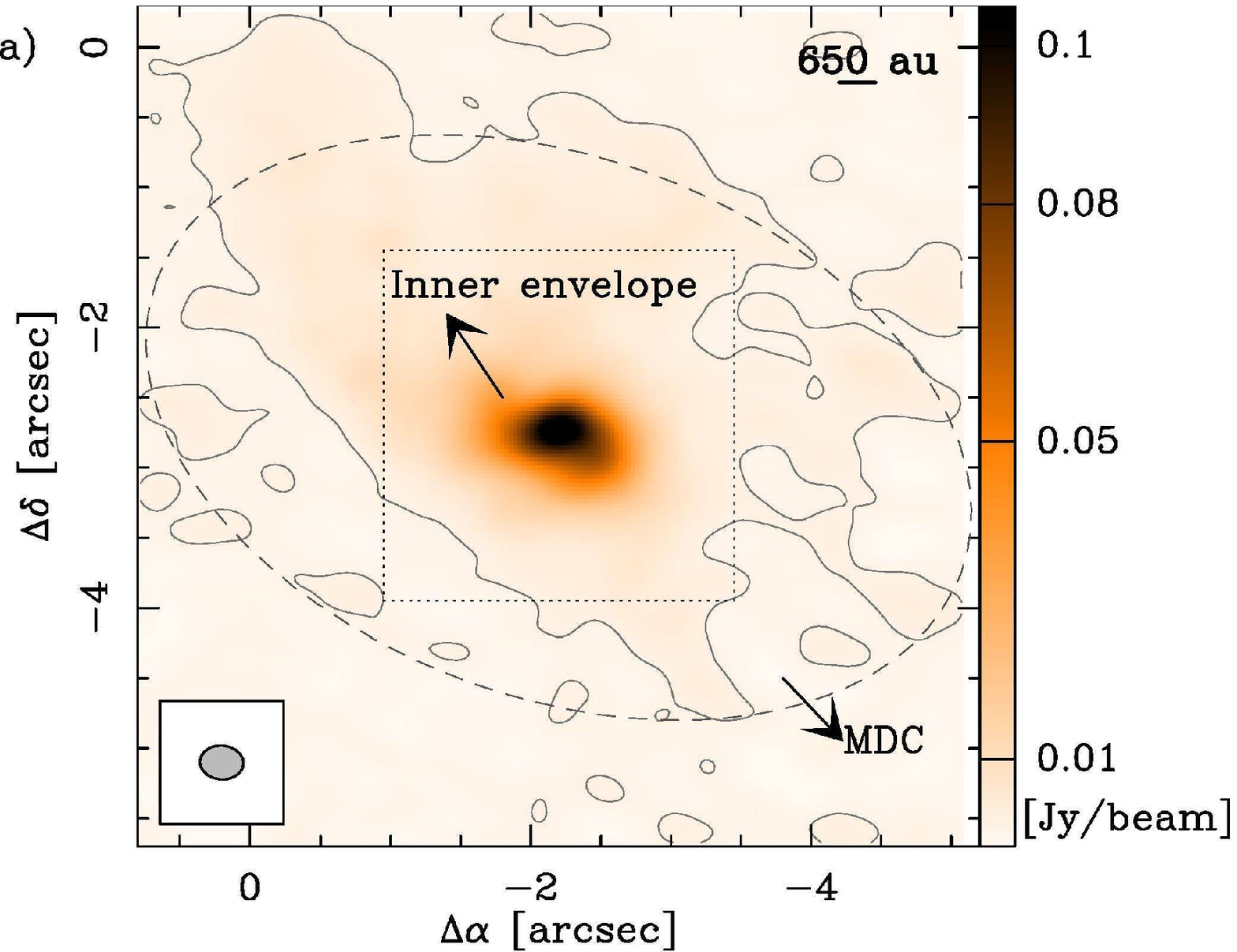}
  \includegraphics[height=3.5cm]{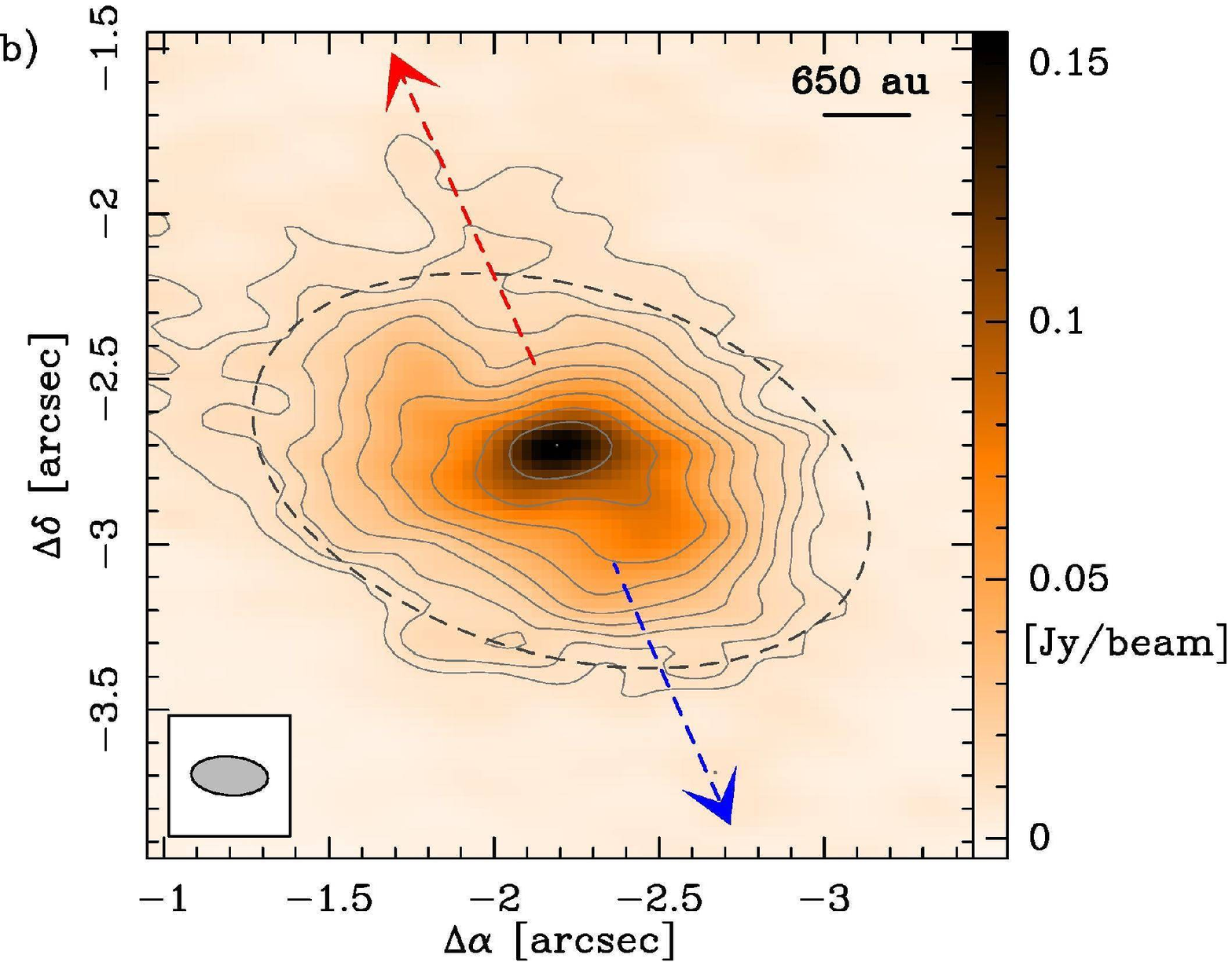}
   \includegraphics[height=3.5cm]{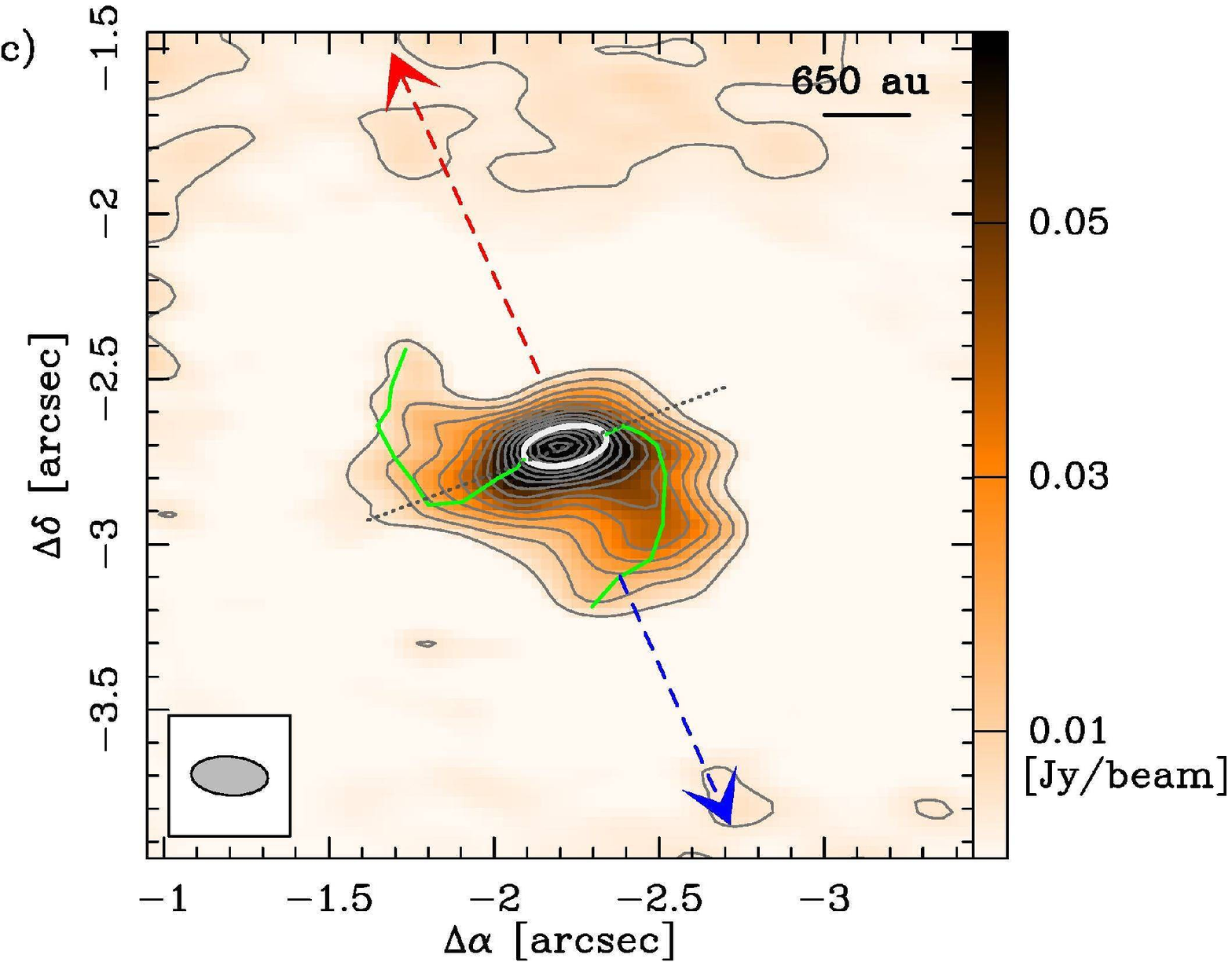}
   \includegraphics[height=3.5cm]{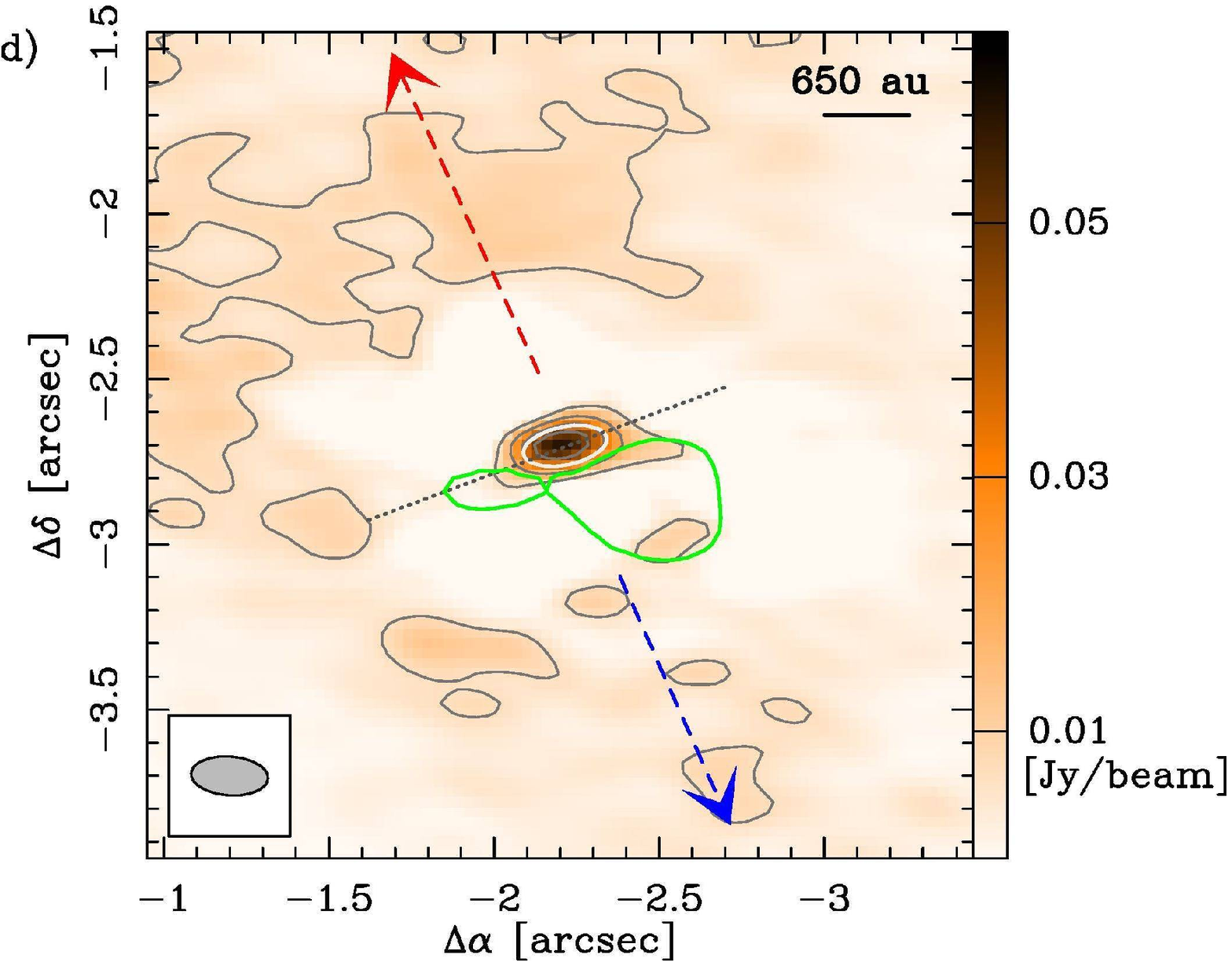}
       \caption{A zoom on Fig.\,\ref{fig:overview} of the protostar {centred on  $\Delta\alpha=-2.17\arcsec$, and $\Delta\delta=-2.76\arcsec$ offset from the phase
center.} 
      {\sl a)} {Continuum emission smoothed to a resolution of 0.27\arcsec\ showing the extended emission at the scale of the MDC.
      The color scale is linear between $-1$ and 105\,mJy/beam, and the contour displays the 1.4\,mJy/beam level corresponding to $\sim$5\,$\sigma$. Dashed ellipse shows the $FWHM$ of the  MDC from Table\,\ref{tab:cont} adopted from \citet{Csengeri2017b}. 
       The region of the inner envelope shown in panels \textsl{b} to \textsl{c} is outlined by a dashed box.    {The $FWHM$ beam is shown in the lower left corner of all panels}.
            {\sl b)} Line-free continuum emission in the original, unsmoothed 12m and 7m array combined map where the beam has a geometric mean of 0.17\arcsec. The color scale is linear between $-3\sigma$ and $120\sigma,$
                      contours start at $7\sigma$ and increase on a logarithmic scale up to $120\sigma$ by a factor of 1.37.}
  The red and blue dashed lines show the direction of the CO outflow (see Fig.\,\ref{fig:outflow} and Sect.\,\ref{sec:outflow}).  {The dark gray {dashed} ellipse shows  $2\times$ the major and minor axes of the fitted 2D Gaussian.}  {\sl c)} Residual continuum emission after removing a 2D Gaussian with a fixed lower peak intensity from the envelope component in order to enhance the contrast of the inner envelope. The color scale is linear from $0$ to $50\sigma$. The contours start at $3\sigma$ and increase by  $6\sigma$. White ellipse shows the $FWHM$ of the fitted 2D Gaussian to the residual from panel \textsl{b}, {and the green line outlines the azimuthal elongations.} The black dotted line marks the direction perpendicular to the outflow. 
      {\sl d)} Residual continuum emission after removing the Gaussian fit to the envelope component (see the text for details). The color scale is the same as in panel \textsl{c}. Contours start at $5\sigma$ and increase by  $10\sigma$.     {Green contours show 80\% of the peak of the velocity integrated emission of the 334.436\,GHz $\varv_{\rm t}=1$ CH$_3$OH line shown in Fig.\,\ref{fig:methanol}}.   
             }
              \label{fig:dust}%
    \end{figure*}

\section{Results}

We discuss here the dust continuum image obtained with ALMA, which 
reveals a {protostellar envelope} that stays single down to our resolution limit of
400\,au in physical scale (Sect.\,\ref{sec:cont}).
We then report the detection of a high-energy rotational line of CH$_3$OH within its $\varv_{\rm t}=1$ torsionally excited state, and compare its observed properties with those of rotational CH$_3$OH lines within its $\varv_{\rm t}=0$ state (Sect.\,\ref{sec:mol}).  
We use then {the local thermodynamic equilibrium (LTE)
approximation to estimate the physical
conditions of the \methanol\ emitting gas (Sect.\,\ref{sec:phys}).
Finally, to constrain the origin of the \methanol\ emission, we compare it to other molecules, such as CO, SO$_2$ and HC$_3$N
tracing outflowing gas associated with the  
protostellar activity (Sect.\,\ref{sec:outflow}). 
The full view of the molecular complexity of this source will be discussed in a forthcoming paper (Csengeri et al.\,in prep, \textsl{b}).

\subsection{Dust continuum}\label{sec:cont}
\subsubsection{A  flattened envelope and a compact dust continuum source}\label{sec:shape}

 We show the line-free continuum emission of \mysou\ in Fig.\,\ref{fig:overview}, right panel, which
 reveals a single compact object down to 400\,au physical scales. The source drives a prominent bipolar outflow (Sect.\,\ref{sec:outflow}) suggesting that it hosts a protostar undergoing its main accretion phase.
 We resolve well the structure of the envelope, and show a zoom on the brightest region in Fig.\,\ref{fig:dust}{\sl a}. 
 
To extract the properties of the bulk emission of the dust, we use here a 2D Gaussian fit in the image plane as a first approach. This reveals the position of the continuum peak at 
       $(\alpha,\delta)_{\rm J2000}=(15^{\rm h}57^{\rm m}59.802^{\rm s}$, $-53^\circ58'00\rlap{.}{''}51$), which is $-2.17$\arcsec,$-2.76$\arcsec\ offset from the phase centre, 
and gives a 
full-width at half-maximum ($FWHM$) of 0.96\arcsec$\times$0.56\arcsec\ with a position angle of $74$$^\circ$. We calculate
the envelope radius as $R_{\rm 90\%}^{\rm env}=1.95\times\Theta/\sqrt{8\,\rm{ln} 2}$,
where $\Theta$ is the beam deconvolved geometric mean of the major and minor axes $FWHM$. This gives a radius, $R_{\rm 90\%}^{\rm env}$, of 0.59\arcsec\ corresponding to 1500\,au.  {In the following we refer to this component, thus the structure within 1500\,au, as the inner envelope (Fig.\,\ref{fig:dust}{\sl b}).}

{In order to investigate the structure of the inner envelope,
we remove a low-intensity 2D Gaussian fit, i.e. allowing only positive residuals for the brightest, central structure (Fig.\,\ref{fig:dust}\textsl{c}).
This reveals azimuthal elongations 
exhibiting a warped S-shape connecting to the continuum peak.}
 
   \begin{figure}[!h]
   \centering
     \includegraphics[width=0.9\linewidth]{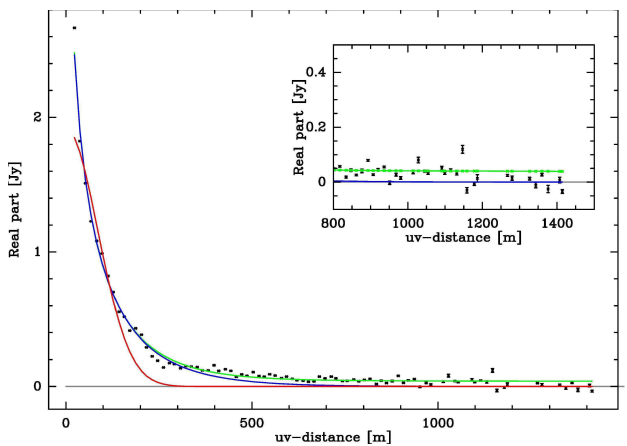}
\caption{{Real part of the visibility measurements versus $uv$-distance shown for the ALMA 12\,m array data. The data points show an average of line-free channels. The red and blue lines show fits to the envelope, an elliptic Gaussian, and a single component power-law fit, respectively. The green line shows a two component fit with a power-law and a compact disk source (see the text for more details).}}  \label{fig:uvfit}%
    \end{figure}

\subsubsection{A compact source within the envelope}
 
When removing the bulk of the dust emission, i.e. our first 2D Gaussian fit to the envelope {shown in Fig.\,\ref{fig:dust}\textsl{b}}, we find a compact residual source with a peak intensity larger than 10$\sigma$ in Fig.\,\ref{fig:dust}\textsl{d}. 
We determine its properties with another 2D Gaussian fit in the image plane data and identify a compact source that is resolved only along its major axis. The beam deconvolved  
 $R_{\rm 90\%}^{\rm disk}$ radius is $\sim$0.1\arcsec, corresponding to an extent of 250\,au. 
The properties and the corresponding flux densities of the inner envelope and the compact component based on our analysis in the image plane are summarised in Table\,\ref{tab:cont}. 

{These results suggest} that the envelope itself is not well described by a Gaussian flux density distribution, and
to test whether the residual source is better represented by a different flux density profile, we also perform a fitting procedure in the $uv$-domain. 
We fit the visibilities averaged over line-free channels as a function of $uv$-distance for the ALMA 12m array data (Fig.\,\ref{fig:uvfit}). {We find that a single power-law component provides a relatively good fit to the data up to $\sim$300\,m long baselines.} The visibility points show,{however}, a significant residual on the longer baselines further suggesting the presence of a compact component (see inset of Fig.\,\ref{fig:uvfit}).

{We used various geometries to fit the compact component, however, we can only constrain that it is unresolved along its minor axis. }
{Our models show that the compact source is consistent with a disk} component with a flux density of 43 mJy, and a 0.2\arcsec\ major axis $FWHM$. 
As a comparison, our analysis in the image plane attributes a somewhat larger flux density to this compact component, but finds a similar spatial extent.  
Later we argue that
this component, that is {significantly more compact} than the 
{in the inner envelope}, likely corresponds to a compact accretion disk 
around the central protostar (Sect.\,\ref{sec:d3}). 

\begin{table*}[!h]
\centering
\caption{Dust continuum measurements.}\label{tab:cont}
\begin{tabular}{l c c c c c c c c c}
\hline\hline
 Component & Peak Intensity & Integrated flux d. & $FWHM$ & p.a.$^{a}$ & Physical size & Used data \\ & 		         [Jy/beam]      &  [Jy]                 & [\arcsec$\times$\arcsec] &  &\\ \hline
Clump &  8.32  & $14.95$ & 28.3\arcsec$\times$23.4\arcsec & $-$31$^\circ$ & $0.32$\,pc & APEX/LABOCA$^{b}$ \\
MDC & 2.10  & 4.0 & 6.1\arcsec$\times$3.85\arcsec &   70$^\circ$ & $0.06$\,pc & ALMA 7m array$^{c}$\\
Inner envelope$^{d}$  &  0.1  & 2.1 & 0.96\arcsec$\times$0.56\arcsec & $74^\circ$ & 1500\,au$^{e}$ & ALMA 7m+12m array\\
Residual (disk) $^{d}$   &   0.063  & 0.068 & 0.26\arcsec$\times$0.12\arcsec & $101^\circ$ & 250\,au$^{e,f}$ & ALMA 7m+12m array
\\ 
\hline
\end{tabular}
 \tablefoot{\\
 \tablefoottext{a}{The position angle of the fitted Gaussian is measured from north to east.}\\ 
 \tablefoottext{b}{The corresponding parameters are extracted  from the catalog of \citet{Csengeri2014} based on the ATLASGAL data.}\\
 \tablefoottext{c}{The listed parameters are from the ALMA 7\,m array from \citet{Csengeri2017b}.}\\
 \tablefoottext{d}{Parameters obtained with a 2D Gaussian fit in the image plane in this work.}\\
 \tablefoottext{e}{Corresponds to the beam deconvolved $R_{90\%}$, see the text for details.}
  \tablefoottext{f}{This estimate is based on the resolved major axis.}}
\end{table*}

\subsubsection{The mass reservoir for accretion}\label{sec:temp}

To understand the distribution of dust emission on various scales, we compare the recovered emission from the 12\,m array observations, the 12\,m and 7\,m array combined data, and the total flux density from the single dish observations from \at . 
We recover a total flux density of 2.9\,Jy in the field with the 12\,m array observations only, which is a significant fraction (73\%) of the 
$\sim$4.0\,Jy measured in the 7\,m and 12\,m array combined data. The
 870\,$\mu$m single-dish peak intensity measured on the \at\ emission map is 10.26\,Jy at this position, from which 8.32\,Jy has been assigned to the clump in the catalog of \citet{Csengeri2014}. This means that the 12m and 7m array combined observations recover $\sim$50\% of the total dust emission from the clump. Clearly, there is a large concentration of emission on the smallest scales which agrees with our previous results comparing the clump and the core scale properties in \citet{Csengeri2017a}.

Based on this information, we describe the structure of the source in the following.
We attribute the large scale emission to the clump, whose parameters are obtained from the \at\ data at $0.32$\,pc scales (Fig.\,\ref{fig:overview}\textsl{a}). The smaller scale structure is attributed to a Massive Dense Core (MDC) forming a single protostellar envelope which has been first identified based on the ALMA 7m array observations in \citet{Csengeri2017b} at $\sim$0.06\,pc scales. In Fig\,\ref{fig:dust}\textsl{a}, in order to show the lower column density material at $N$(H$_2$)$\sim2.5-10\times10^{22}$\,cm$^{-2}$ for $T=30-100$\,K, we smoothed the data to illustrate the extent of the MDC. 
The original, not smoothed ALMA 12\,m and 7\,m array combined data reveals the brightest emission with a 1500\,au radius corresponding to the inner regions of the envelope showing the highest column densities. 

Here, we attempt to provide a more robust mass estimate on the available mass reservoir for accretion based on the MDC properties {within a radius of $\sim$0.06\,pc.}
To do this, we constrain the dust temperature ($T_{\rm d}$) 
from a two component (cold and warm) 
greybody fit to the far-infrared spectral energy distribution (SED)
(see App.\,\ref{app:sed}), where in particular the 
wavelengths shorter than 70\,$\mu$m are a sensitive probe to the amount of heated dust in the vicinity of the protostar. From the fit to the SED we obtain a cold component at 
22\,K\footnote{In \citet{Csengeri2017a} we used  $T_{\rm d}=$25\,K for all cores.}, 
and a warm component at $\sim$48\,K that we assign to the inner envelope. These models show, that the amount of warm gas is only a small fraction ($<5$\%) of the mass reservoir of the MDC. 
Adopting therefore, $T_{\rm d}=$22\,K for the MDC we obtain 
$M_{\rm MDC}= 120$\,\msol\footnote{
 We used Eq.\,2 from \citet{Csengeri2017a} with the same parameters for the dust (dust emissivity, $\kappa_\nu = 0.0185$\,cm$^{2}$\,g$^{-1}$ from \citealp{OH1994} accounting for a gas-to-dust ratio, $R$, of 100.) }. Using a different dust emissivity (e.g.\, \citealp{Peretto2013}), and/or a gas-to-dust ratio of 150 would increase this value by 50-100\%. The uncertainty in the distance estimation of this source would either decrease this value by 36\%, or increase it by roughly a factor of three. On the measured physical sizes the effect of distance uncertainty is less dramatic, it would either decrease the inner envelope radius by 20\% or increase it by 70\%, if the source is located at the farthest likely distance.

Since the MDC is gravitationally bound, its mass should be available for accretion onto the central protostar. Assuming an efficiency of 20-40\% \citep{Tanaka2017}, we can  expect that an additional $24-48$\,$M_{\odot}$ could still be accreted on the protostar. Therefore, this makes our target one of the most massive protostellar envelopes known to date, which is likely in the process of forming an O-type star. 
  
\begin{table*}[!ht]
\centering
\caption{Summary of the   molecular transitions studied in this work.}\label{tab:lines}
\begin{tabular}{rcccccrrrrrrrr}
\hline\hline
Molecule & Quantum number & Frequency &  Log$_{10}$ Aij & $E_{\rm up}/k$ & $n_{\rm cr}^{a}$ & Database \\
& & [GHz] &  [s$^{-1}$]& [K] & [cm$^{-3}$] & \\
\hline
CH$_3$OH$-A$ $\varv_{\rm t}=0$    & $2_{-2}-3_{-1}$  & 335.13369 & $-4.57$ & 45 & $1.1\times10^7$ & CDMS \\
CH$_3$OH$-A$ $\varv_{\rm t}=0$  & $7_1-6_1$  & 335.58200  & $-3.29$ & 79& $3.6\times10^7$ & CDMS\\
CH$_3$OH$-A$ $\varv_{\rm t}=0$	 & $14_7-15_6$  & 336.43822 & $-4.25$	& 488 & 	$6.7\times10^5$ & CDMS \\
CH$_3$OH$-A$ $\varv_{\rm t}=0$	  & $12_{-1}-12_{0}$  & 336.86511  & $-2.84$ & 197 & $1.0\times10^8$ &	CDMS \\
CH$_3$OH$-E$ $\varv_{\rm t}=0$	  & $3_3-4_2$ & 337.13587   & $-4.61$ & 62 & $6.9\times10^7$ 	& CDMS \\
CH$_3$OH$-E$ $\varv_{\rm t}=1$ & $3_{0}$--$2_{1}$ & 334.42656  & $-4.26$ & 315 & & CDMS \\ 
CH$_3$OH$-A$ $\varv_{\rm t}=2$    & $7_1-6_1$  & 336.60589 & $-3.79$ &  747 &  & JPL  \\
$^{13}$CH$_3$OH$-A$ $\varv_{\rm t}=0$$^c$ & $12_{-1}$--$12_{0}$ & 335.56021  & $-3.40$	& 193 & $1.0\times10^8$	& CDMS \\
$^{13}$CH$_3$OH$-A$ $\varv_{\rm t}=0$$^c$ & $14_{-1}$--$14_{0}$  & 347.18828 &  $-3.39$ & 254 & $6.9\times10^6$	& CDMS\\
HC$_3$N $\varv_{\rm 7}=0$$^b$	 &		37--36	&  	336.52008  &$-2.52$	& 307	& & CDMS\\
HC$_3$N $\varv_{\rm 7}=1e$ 	&38--37	&	346.45573 & 		$-2.48$	 &	645 &	& CDMS \\ 
SO$_2$ $v = 0$	 &	$8_{2, 6}- 7_{1, 7}$ &	334.67335 & 	$-3.26$	& 43 & $7.1\times10^7$& JPL\\
CO & $3-2$ & 345.79599 &  $-$3.61 & 33 & $4.00\times10^4$ & CDMS \\
\hline
\end{tabular}\\
\tablefoot{\\
\tablefoottext{a} {Calculated at $T=100$\,K using collisional rate coefficients from the LAMDA database \citep{Schoier2005} where available.} \\
\tablefoottext{b} {For HC$_3$N the molecular datafile lists cross sections up to $J_{\rm up}=21$, and $T=80$\,K.}\\
\tablefoottext{c} {Calculated from the collisional rate coefficients of the main isotopologue.}}
\end{table*}

\begin{table}[!h]
\centering
\caption{Observational parameters for the CH$_3$OH
$\varv_{\rm t}=1$ lines, and results of the LTE modelling for \methanol. }\label{tab:spec}
\begin{tabular}{l c c | r c c c}
\hline\hline
\multicolumn{3}{c}{Observed parameters} &  \multicolumn{3}{c}{LTE fit parameters}\\
 & $v_{\rm lsr}$$^{a}$  &  $\Delta v$&    N & size & $T_{\rm ex}$ \\
  &[km\,s$^{-1}$]  & [km\,s$^{-1}$]       &  [cm$^{-2}$] & [\arcsec]& [K] \\
\hline 
A   &    $-48.1$$\pm$0.1 ($-4.6$)  & 4.5$\pm$0.2 & $1.6\times10^{19}$ &  0.4 & 160   \\
B   &    $-39.9$$\pm$0.1 ($+3.6$) & 5.6$\pm$0.1 & $2\times10^{19}$ & 0.4 & 170                 
    \\
\hline
\end{tabular}
{
 \tablefoot{
 \tablefoottext{a}{The number given in parenthesis corresponds to the difference between the line velocity and the 
$-43.5$\,\kms\ $v_{\rm lsr}$ of the source. }}
}\end{table}

   \begin{figure*}[!ht]
   \centering
     \includegraphics[height=4.0cm]{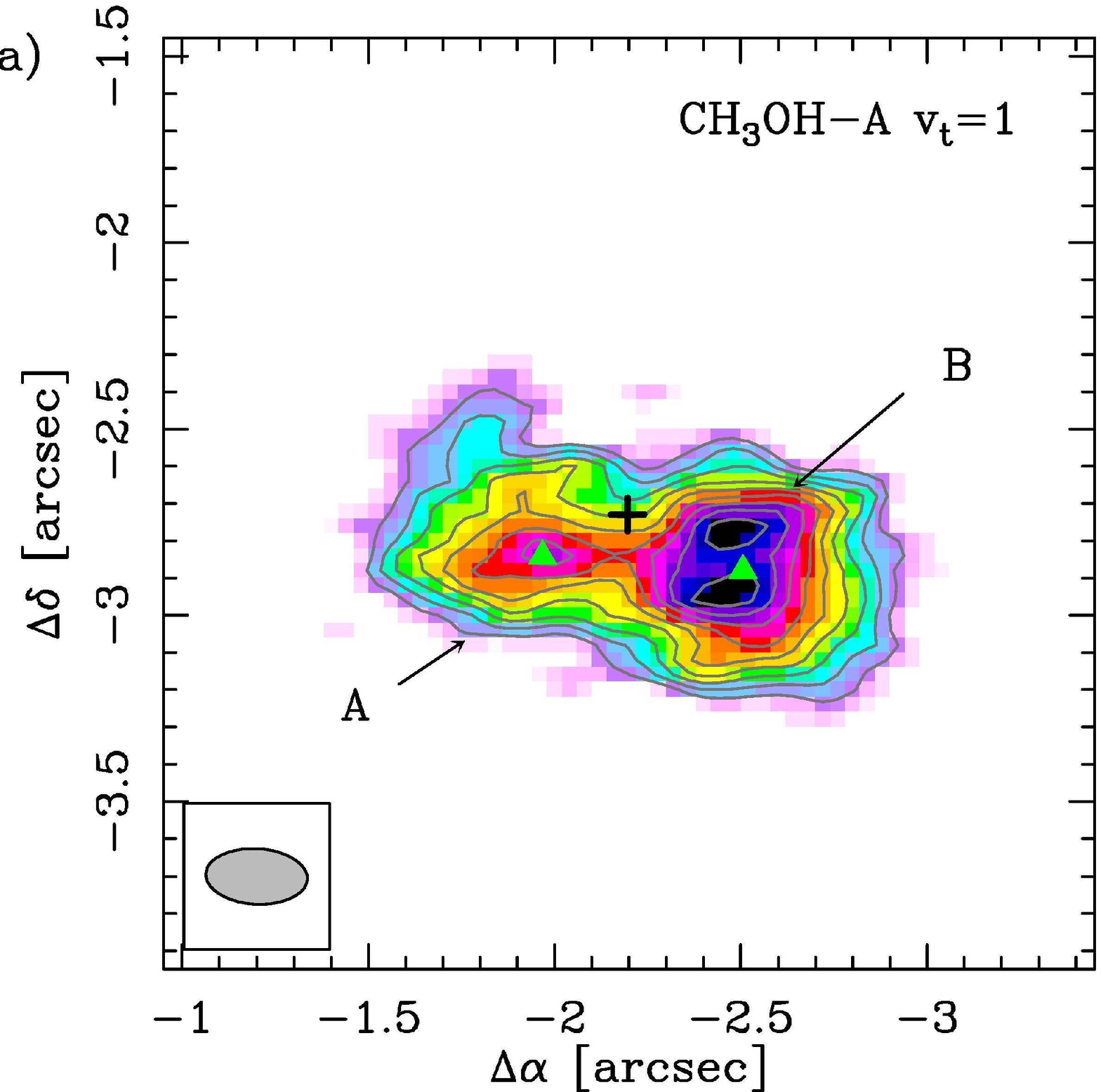}
     \includegraphics[height=4.0cm]{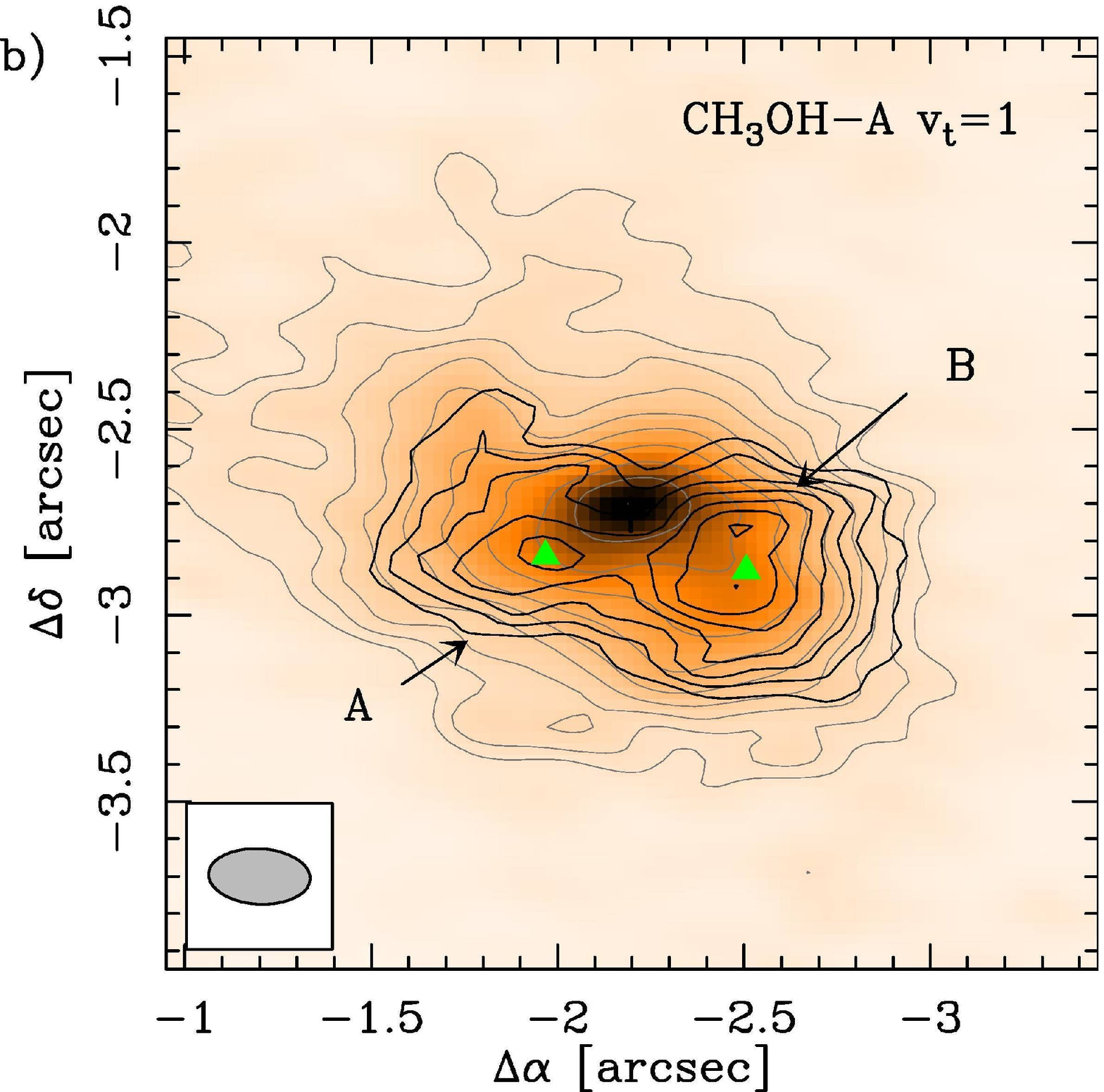}
     \includegraphics[height=4.0cm]{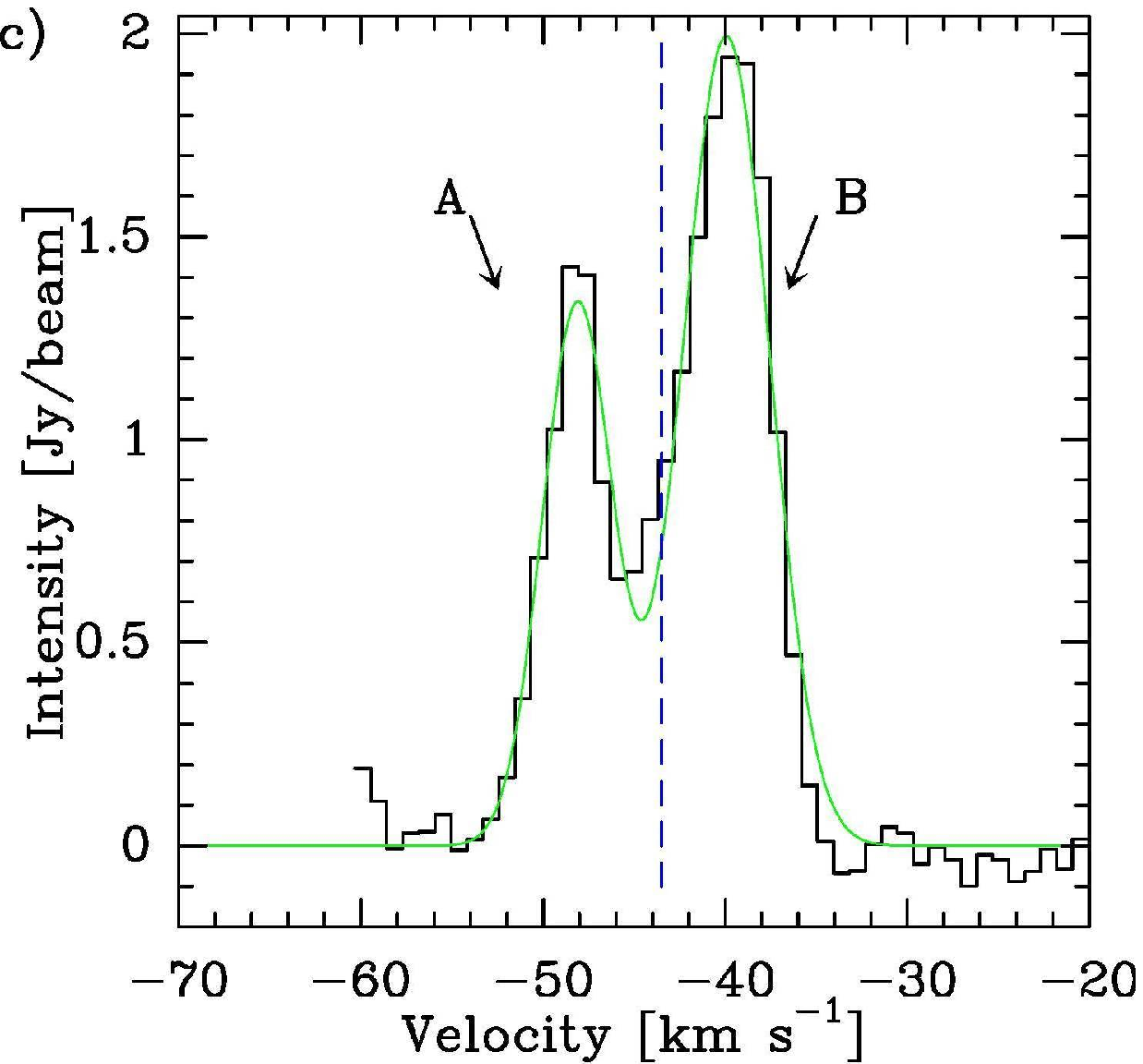}
     \includegraphics[height=4.0cm]{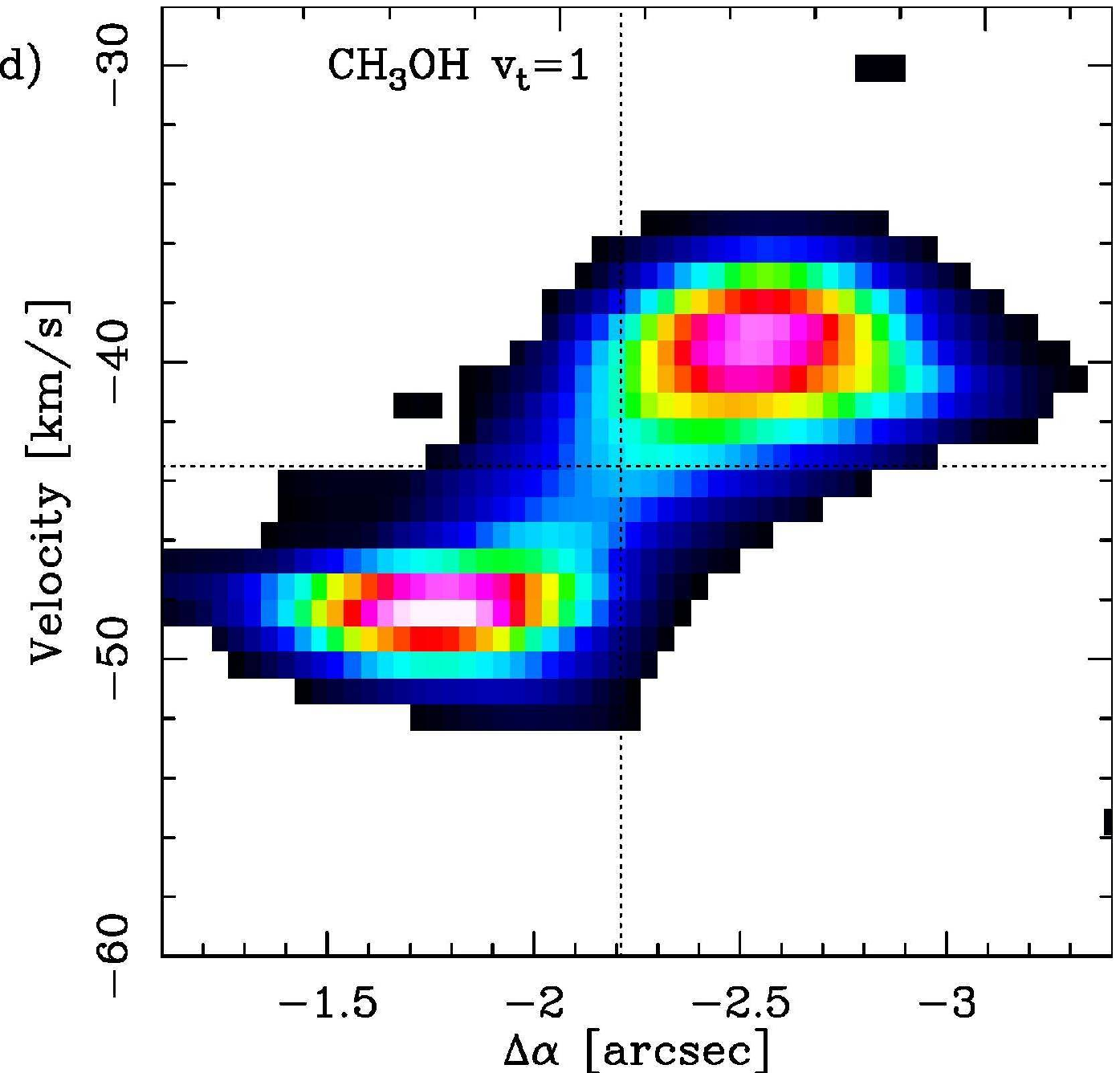}
\caption{{\sl a)}  Color scale shows the integrated intensity {map} of the $\varv_{\rm t}=1$ \methanol\ line at 334.4\,GHz. The green triangles indicate the positions where the spectrum has been extracted for the rotational diagram analysis on the \methanol\ spots, and are labeled as \emph{A} and \emph{B} components. The black cross marks the position of the dust continuum peak. The beam is shown in the lower left corner.
      {\sl b)} The color scale shows the continuum emission from Fig.\,\ref{fig:dust}{\sl b}, contours and markers are the same as on panel \textsl{a}.
      {\sl c)} Integrated spectrum of the torsionally excited 
      \methanol\ transition at 334.4\,GHz over the area shown in panel \textsl{a}. 
      The green lines show the two component Gaussian
      fit to the spectrum. The blue dashed line shows the \vlsr\ of the source. {\sl d)} {Position-velocity diagram along the $\Delta\alpha$ axis and averaged over the shown extent of the cube corresponding to $\sim$2.5\arcsec}. The dotted lines mark the position of the dust peak and the \vlsr\ of the source.}  \label{fig:methanol}%
    \end{figure*}

   \begin{figure*}[!ht]
   \centering
   \includegraphics[width=0.95\linewidth]{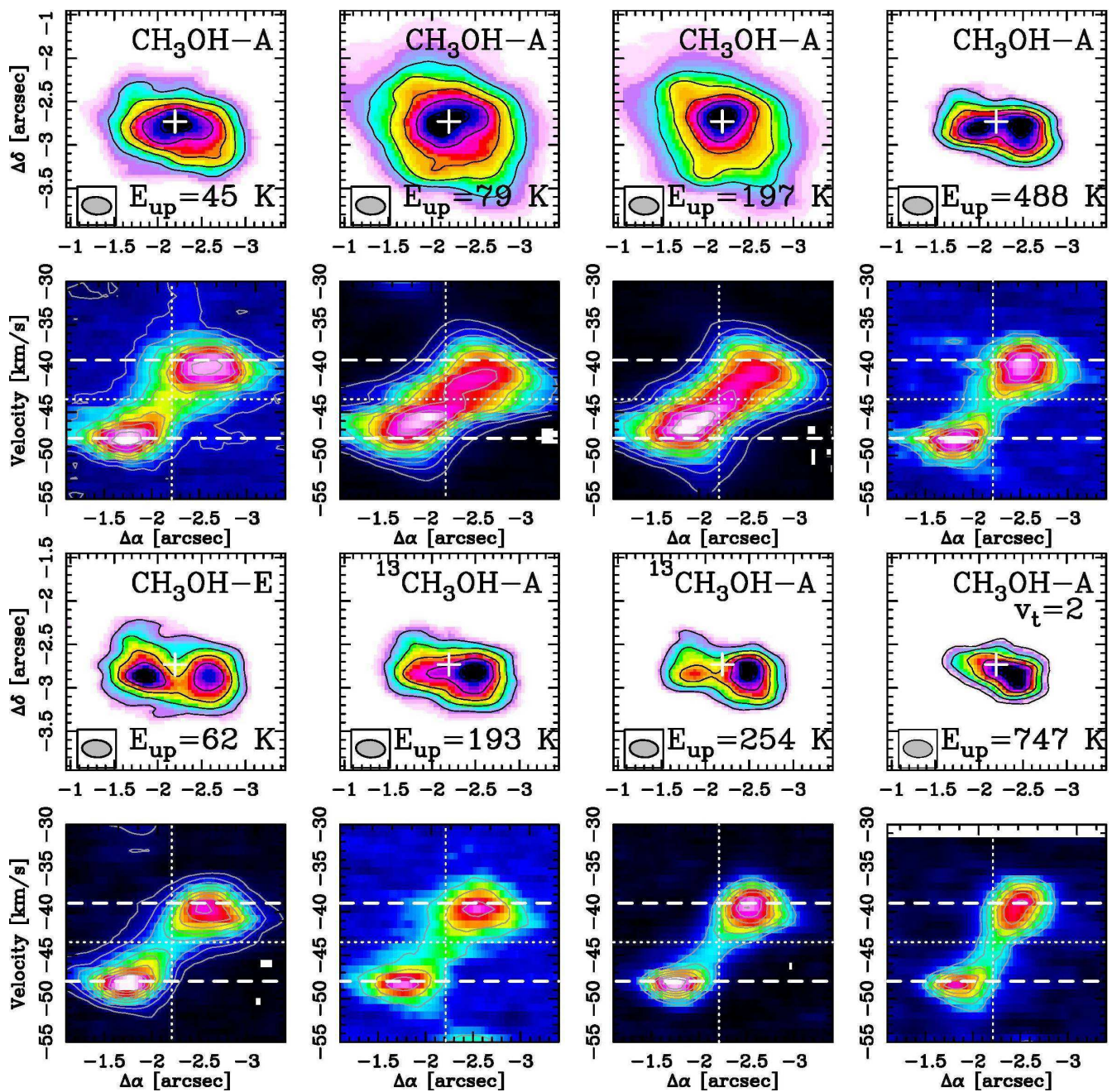}
      \caption{CH$_3$OH 0th moment map and position velocity diagrams for the transitions listed in Table\,\ref{tab:lines}. The top row shows the 0th moment map calculated over a velocity range of $[-55;-30]$\,km\,s$^{-1}$, contours start at 30\% of the peak and increase by 15\%. White cross marks the position of the continuum peak. The beam is shown in the lower left corner. The symmetry of the methanol molecule is labeled on each panel, as well as the upper level energy of the transition. The subsequent row shows the position velocity map {along the $\Delta\alpha$ axis and averaged over the shown extent of the cube corresponding to $\sim$2.5\arcsec}. The {right-ascension} offset of the continuum peak and the $v_{\rm lsr}$ velocity of the source are marked as white dotted line. Dashed lines show the  $v_{\rm lsr}\pm4.5$\,km\,s$^{-1}$ corresponding to the peak velocity of the CH$_3$OH spots.}
              \label{fig:methanol_all}%
    \end{figure*}

\subsection{Methanol emission}\label{sec:mol}

The total observed bandwidth of 7.5\,GHz reveals emission from several molecular species.
{In this study we focus} on a selected list of molecules 
summarised in  Table\,\ref{tab:lines}. {We first discuss the torsionally excited \methanol\ emission in Sect.\,\ref{sec:methanol1}, and to better constrain its physical origin in Sect.\,\ref{sec:methanol2} we discuss all unblended torsional ground state transitions of \methanol.}

\subsubsection{Torsionally excited methanol shows two bright spots offset from the protostar}\label{sec:methanol1}

Our continuous frequency coverage between 333.2 and 337.2\,GHz, as well as between 345.2 and 349.2\,GHz, 
includes several 
transitions of CH$_3$OH and its $^{13}$C isotopologue.
Most interestingly,  towards the inner envelope,  we detect and spatially resolve emission  from a rotational transition of \methanol, from its first torsionally excited, $\varv_t=1$, state at 334.42\,GHz with an upper energy level of 315\,K
 (Fig.\,\ref{fig:methanol}\textsl{a}).
Its spatial morphology shows two prominent emission peaks (marked as A and B in Fig.\,\ref{fig:methanol}\textsl{a}), which
spatially coincide with the {azimuthal elongations within} the envelope. 
The emission drops, however, significantly towards the continuum peak, i.e.\,the protostar  (Fig.\,\ref{fig:methanol}, \textsl{b}).  
We find that the observed morphology is dominated by two velocity components, 
which show an offset, on average, of $-4.6$, and $+3.6$\,km\,s$^{-1}$ compared to the $v_{\rm lsr}$ of the source\footnote{We adopt the $v_{\rm lsr}$ of the dense gas seen on the clump/core scale used in \citet{Csengeri2017a}.}
(Table\,\ref{tab:spec}). {As discussed in Sect.\,\ref{sec:methanol2}, optical depth effects are unlikely to be at the origin of the observed velocity pattern.}
We  spatially resolve the emission from these spots, and
 estimate the peak of its distribution to fall between a projected distance of 300 and 800\,au symmetric from the protostar.

The $pv$-velocity diagram along {the $\Delta\alpha$ axis and averaged perpendicularly to this axis over the shown extent of $\sim2.5$\arcsec} 
reveals that the emission 
is dominated by the two velocity components, and shows a pattern
 consistent with rotational motions (Fig.\,\ref{fig:methanol}\textsl{d}). We compare these observations to simple models of the gas
 kinematics in more detail in Sect.\,\ref{sec:d3}.
  
\subsubsection{Pure rotational lines of methanol}\label{sec:methanol2}


{To further investigate the origin of the 334.426\,GHz $\varv_{\rm t}=1$ methanol emission, 
in Fig.\,\ref{fig:methanol_all} we show maps of all the detected transitions of CH$_3$OH
 and its $^{13}$C isotopologue in the torsional ground state, $\varv_t=0$. They}
probe a range of upper energy levels between 45\,K and 488\,K, and based on our
LTE modelling (Sect.\,\ref{sec:weeds}) they are unlikely to be blended
with emission from other species.
{We use these lines, in particular, to test} whether a high optical depth toward the protostar could mimic the observed velocity pattern and morphology {of the torsionally excited state line. }

{These maps reveal three transitions of the CH$_3$OH$-A$ symmetry state 
with upper level energies of $E_{\rm up}<200$\,K that peak on the continuum source, 
while its higher energy transitions show the two
prominent peaks like the $\varv_{\rm t}=1$ CH$_3$OH line\footnote{The CH$_3$OH$-E$ transitions in the band have an order of magnitude smaller Einstein coefficients which can explain 
why despite their lower energy level, all the CH$_3$OH$-E$ lines {show two peaks of emission}. 
}. 
Our LTE modelling in Sect.\,\ref{sec:weeds}, indeed shows that
the three lowest energy transitions of CH$_3$OH$-A$  
have high optical depths. }

{The other transitions are, however, optically thin and they show two peaks of emission offset from the protostar similarly to the $\varv_{\rm t}=1$ CH$_3$OH line, while the emission drops towards the position of the protostar. In addition, their $pv$-diagrams are also similar to the $\varv_{\rm t}=1$ CH$_3$OH line revealing 
the two velocity components. 
Among these lines, we have the two $^{13}$CH$_3$OH transitions
detected with a high signal-to-noise ratio which are the least affected by optical depth effects.
This leads us to conclude that the two prominent spots traced by the $\varv_{\rm t}=1$ state \methanol\ line cannot be a result of a large optical depth of \methanol\ towards the continuum peak. } 

{
We calculate the critical densities for these \methanol\ transitions in Table\,\ref{tab:lines}, and find that they all trace high density gas (if thermalised), strictly above $10^5$\,cm$^{-3}$, but typically on the order of $10^7$\,cm$^{-3}$. We notice that the different transitions have a varying contribution as a function of upper energy level from the central source, which suggests that they may trace two physical components, one associated with the inner envelope showing the bulk emission of the gas likely at lower temperatures, and another, warmer and denser component associated with the two peaks of the CH$_3$OH $\varv_{\rm t}=1$ line. }

\subsection{Physical conditions of the methanol spots}\label{sec:phys}

We use here two  
methods to measure the physical conditions towards the methanol
spots, and the position of the protostar as well. For this we extracted
the spectrum covering the entire observed 7.5\,GHz. To convert it from Jy/beam to K scales we used a factor of 198 K/Jy calculated for the 0.23\arcsec\ averaged beam size at 335.2\,GHz, and 347.2\,GHz. Taking a mean conversion factor for the entire bandwidth adds less than 10\% inaccuracy in the measurement of the brightness temperatures.

\subsubsection{Rotational diagram analysis of CH$_3$OH transitions}\label{sec:rot_dia}

We perform a rotational diagram analysis  (\citealp{garay2010, Gomez2011}) to estimate 
the rotational temperature of the CH$_3$OH emission, and its column density, $N$(CH$_3$OH) at the 
CH$_3$OH peak positions indicated in Fig.\,\ref{fig:methanol}, and towards the position of the protostar.
We extracted the integrated intensities using a Gaussian fit to the 
CH$_3$OH transitions listed in Table\,\ref{tab:lines}.  
We include the $^{13}$CH$_3$OH lines with an isotopic ratio of 60 \citep{Langer1990, Milam2005}, and 
exclude the 335.582\,GHz and 336.865\,GHz transitions from the fit. 
{This is because as a combination of their large optical depths and interferometric filtering due to their spatial extent (see in Fig.\,\ref{fig:methanol_all}) we observe considerably lower fluxes than expected if the \methanol\ emission is thermalised.} 
For the rest of the transitions we assume optically thin emission.
We show the rotational diagram in Fig.\,\ref{fig:rot_dia}, where the error bars show the measured error on the Gaussian fit to the spectral lines, and are on the order of 10\%. We obtain similar values for the two methanol peaks of $T_{\rm rot}=160-175$\,K, and $N$(CH$_3$OH)=$\sim8\times10^{18}$\,cm$^{-2}$. 
{The individual measurements have relatively large uncertainties, 
however, our} results suggest that the two methanol spots have on the order of magnitude similar temperatures and column densities. Ignoring the effect of potentially more severe blending, we performed the same measurement on a spectrum extracted towards the continuum peak, which suggests similarly low rotational temperature as towards the brightest \methanol\ spot, and shows a somewhat lower column density of $N$(CH$_3$OH)=$\sim7\times10^{18}$\,cm$^{-2}$. 
{While Fig.\,\ref{fig:methanol_all} suggests systematic differences in the \methanol\ emission between the high excitation \methanol\ spots and the continuum peak, 
the population diagram analysis shows
that the three positions have similar column densities and rotational temperatures of methanol.
}
This is particularly interesting since the radiation field, and hence the temperature is expected to be the strongest at the position of the protostar.

   \begin{figure}[!h]
   \centering
   \includegraphics[width=0.85\linewidth]{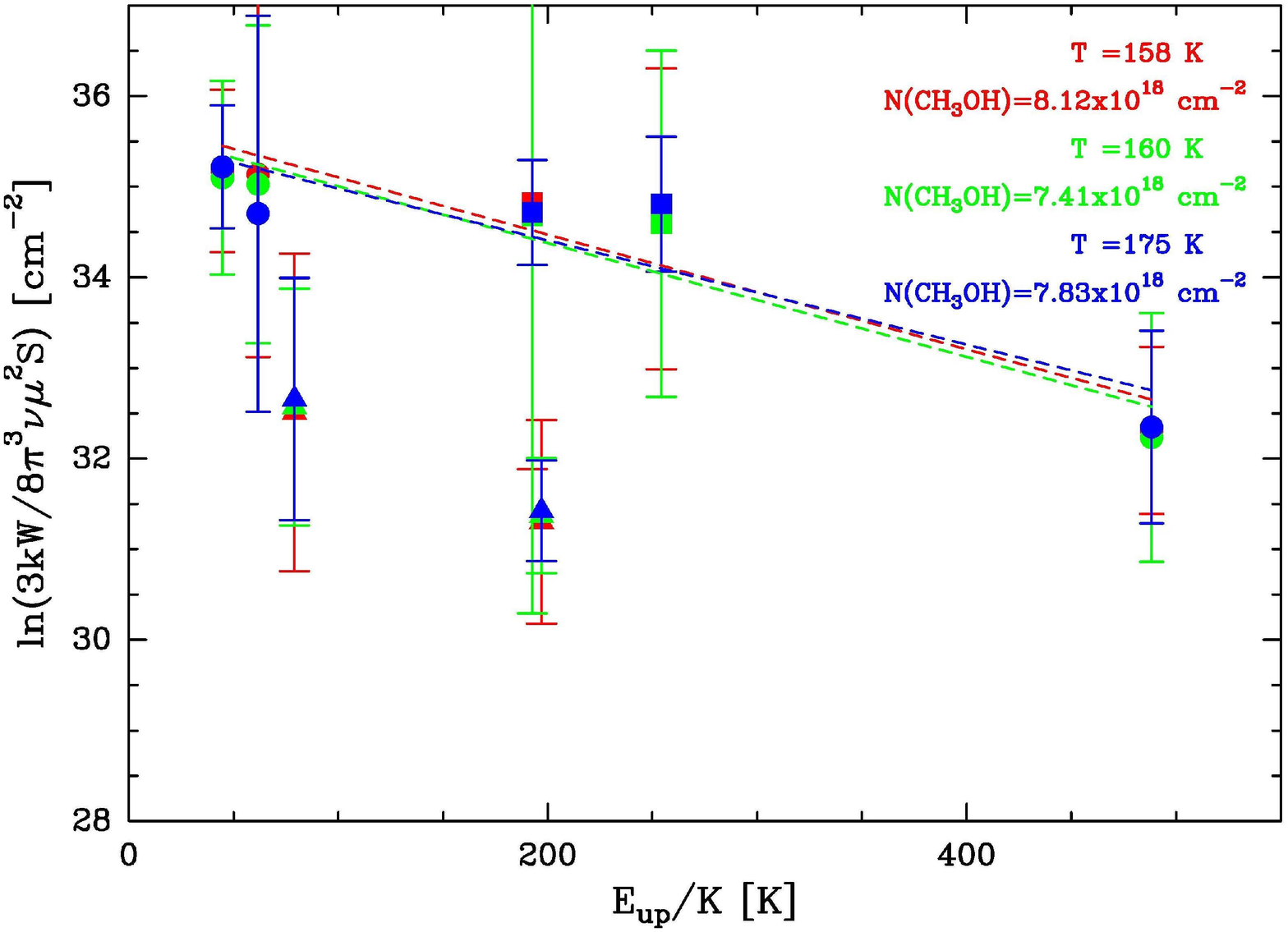}
      \caption{Rotational diagram of the CH$_3$OH (and isotopologue) transitions from Table\,\ref{tab:lines}. {The colours correspond to the measurements on different positions; green shows the position marked as A, red shows the position marked as B, and blue corresponds to the central position marked by a cross in Fig.\,\ref{fig:methanol}.} The filled circles show the CH$_3$OH lines, squares the $^{13}$CH$_3$OH lines, and triangles the optically thick lines that are not used for the fit. The error bars show the linearly propagated errors from the Gaussian fit to the {integrated intensities}. }
              \label{fig:rot_dia}%
    \end{figure}

\subsubsection{LTE modelling with WEEDs}\label{sec:weeds}

{Our observations cover a 7.5\,GHz bandwidth, and the spectra extracted towards the \methanol\ peak positions show line forests of other molecular species, typically COMs.
Therefore, to analyse the CH$_3$OH emission, we modelled the entire spectrum using the  WEEDs package \citep{Maret2011} assuming LTE conditions, which are likely to apply due to the high volume densities. The molecular composition of the gas towards the \methanol\ peaks will be described in a forthcoming paper, together with the detailed analysis.
}

{In short, we} performed the modelling in an iterative process, and started  first with the
CH$_3$OH lines. The input parameters are the molecular column density, kinetic temperature, source size, $v_{\rm lsr}$, and line-width. From these parameters we fix the source size to 0.4\,\arcsec, 
which means that the emission is resolved, as it is suggested by the data (Fig.\,\ref{fig:methanol_all}). 
The modelled transitions may have
different source sizes, however, as long as they are resolved by our observations, the actual source size
does not influence the result. 
The line-width of the CH$_3$OH line is obtained by the Gaussian fits to the extracted spectra. 
After obtaining a first, reasonably good fit to the listed CH$_3$OH transitions,
we started to subsequently add other molecular
species, mainly COMs that are responsible for the lower intensity lines 
(Csengeri et al in prep,\textsl{b}).

We created a grid of models for the CH$_3$OH column density, $N$(CH$_3$OH),  between $10^{17}$ and $10^{20}$\,cm$^{-2}$, and kinetic temperatures between 50 and 300\,K.
We sampled by 25 linearly spaced values both parameter ranges, and then 
visually assessed the results. These models show that the detection or non detection of certain transitions allows us to put a rather strict upper limit on the kinetic temperature. Above $T_{\rm kin}>200$\,K, our models predict 
that other transitions of the $\varv_{\rm t}=1$ state should be detectable at these column densities, the brightest ones
are at 334.627\,GHz ($J$=$22_{3, 0, 1}-22_{2, 0, 1}$), and 334.680\,GHz ($J$=$25_{-3,0,1}-24_{-2, 0, 1}$).
Although these frequency ranges are affected by blending with COMs, models with $T_{\rm kin}>200$\,K predict these rotational transitions within its $\varv_{\rm t}=1$ state  to be brighter than the observed features in the spectrum at their frequency. 

Our results give column densities on the order of {1.2--2$\times10^{19}$\,cm$^{-2}$}, and a kinetic temperature around {160--200}\,K for the two positions. We find that the peak brightness temperatures for the transitions observed across the band are 
more sensitive to the methanol column density than to the variation in kinetic temperature. 
While this modelling takes into account blending with other transitions, our results are
 consistent with the estimates of column density and rotational temperature 
 obtained from the rotational diagram analysis (Sect.\,\ref{sec:rot_dia}).

   \begin{figure*}[!h]
   \centering
   \includegraphics[width=0.32\linewidth,angle=0]{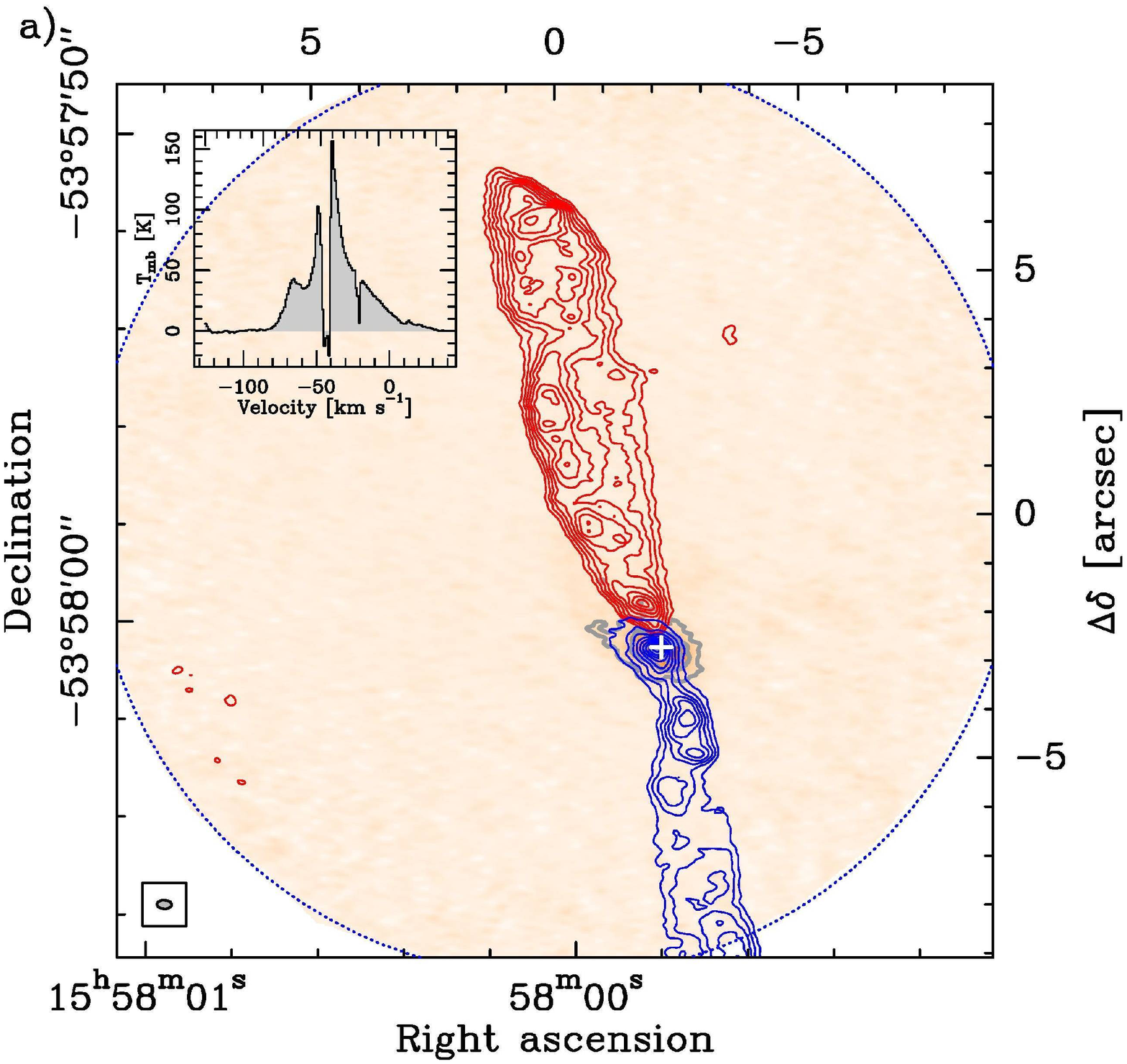}
   \includegraphics[width=0.32\linewidth,angle=0]{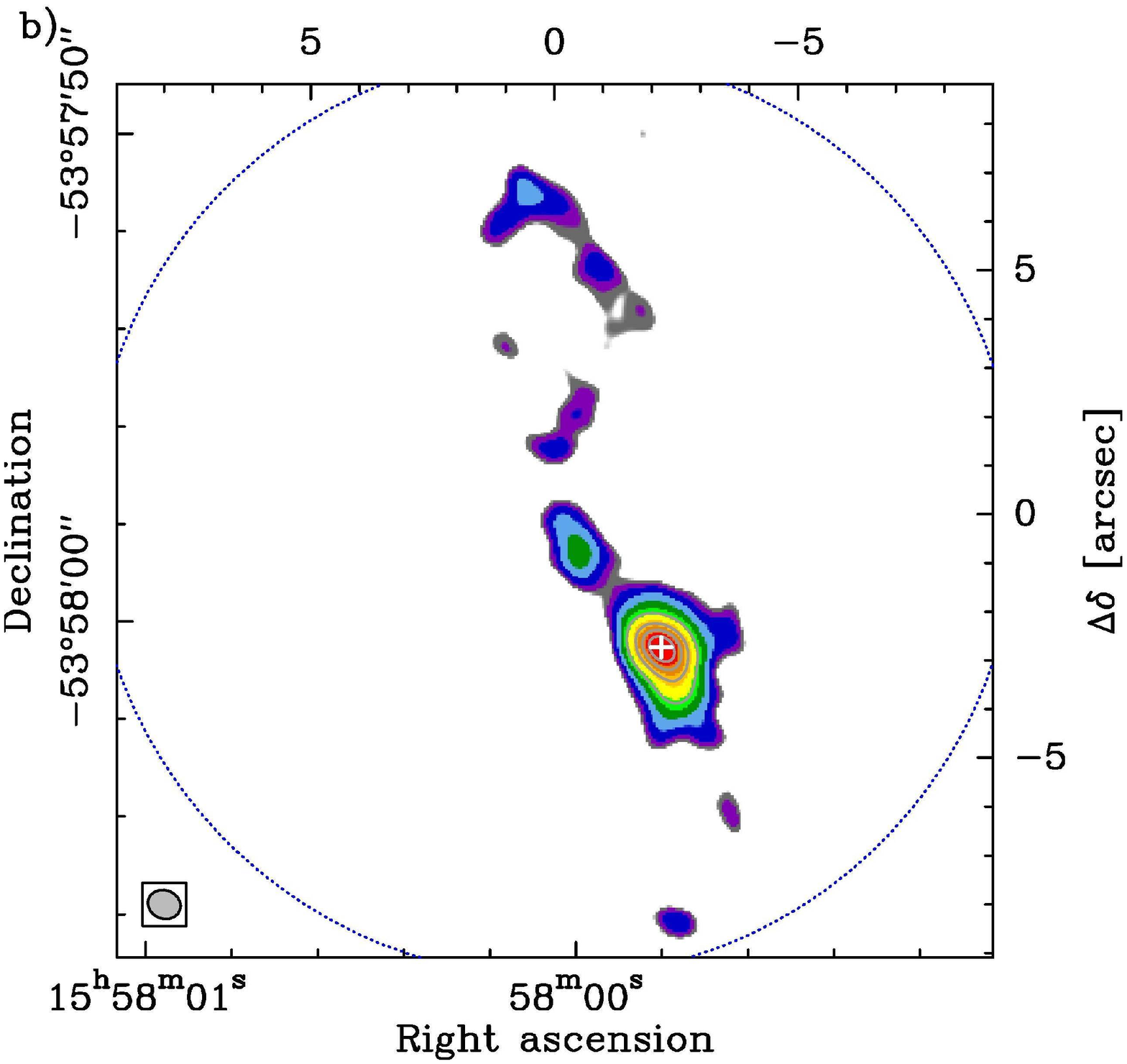}      
   \includegraphics[width=0.32\linewidth,angle=0]{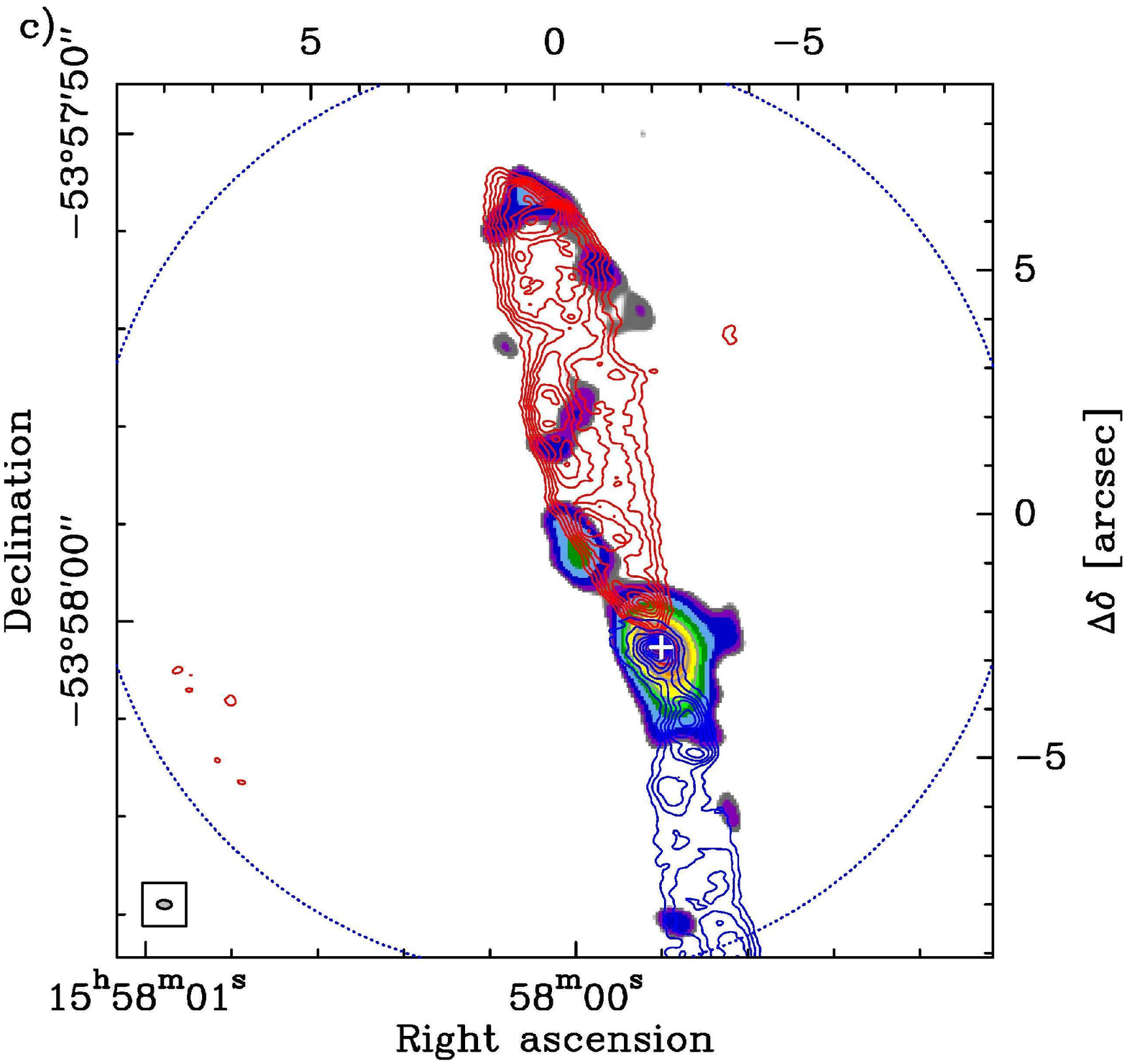}

      \caption{{\sl a)} Color scale shows the continuum emission from Fig.\,\ref{fig:overview} panel \textsl{b}. The contours show the CO (3--2) integrated emission between $-80$ and $-65 \,$\kms\ for the blue, and between $-30$ and $+36\,$\kms\ for the
red, respectively.  White cross marks the position of the dust continuum peak. The inset shows a spectrum of the integrated emission over the area of the lowest contours. {\sl b)}  The contours show the velocity integrated SiO (8--7)  emission starting from 4$\sigma$ (1$\sigma$=0.26 Jy/beam\,\kms) , and increase by 2$\sigma$ levels. {\sl c)} Overlay of the CO (3--2) contours on the  velocity integrated SiO (8--7) emission shown in panel \textsl{b}. {The beam is shown in the lower left corner of each panel. On panel \textsl{c} it corresponds to that of the CO (3--2) map.}
}
              \label{fig:outflow}%
    \end{figure*}

   \begin{figure*}[!ht]
   \centering
   \includegraphics[width=4.5cm,angle=0]{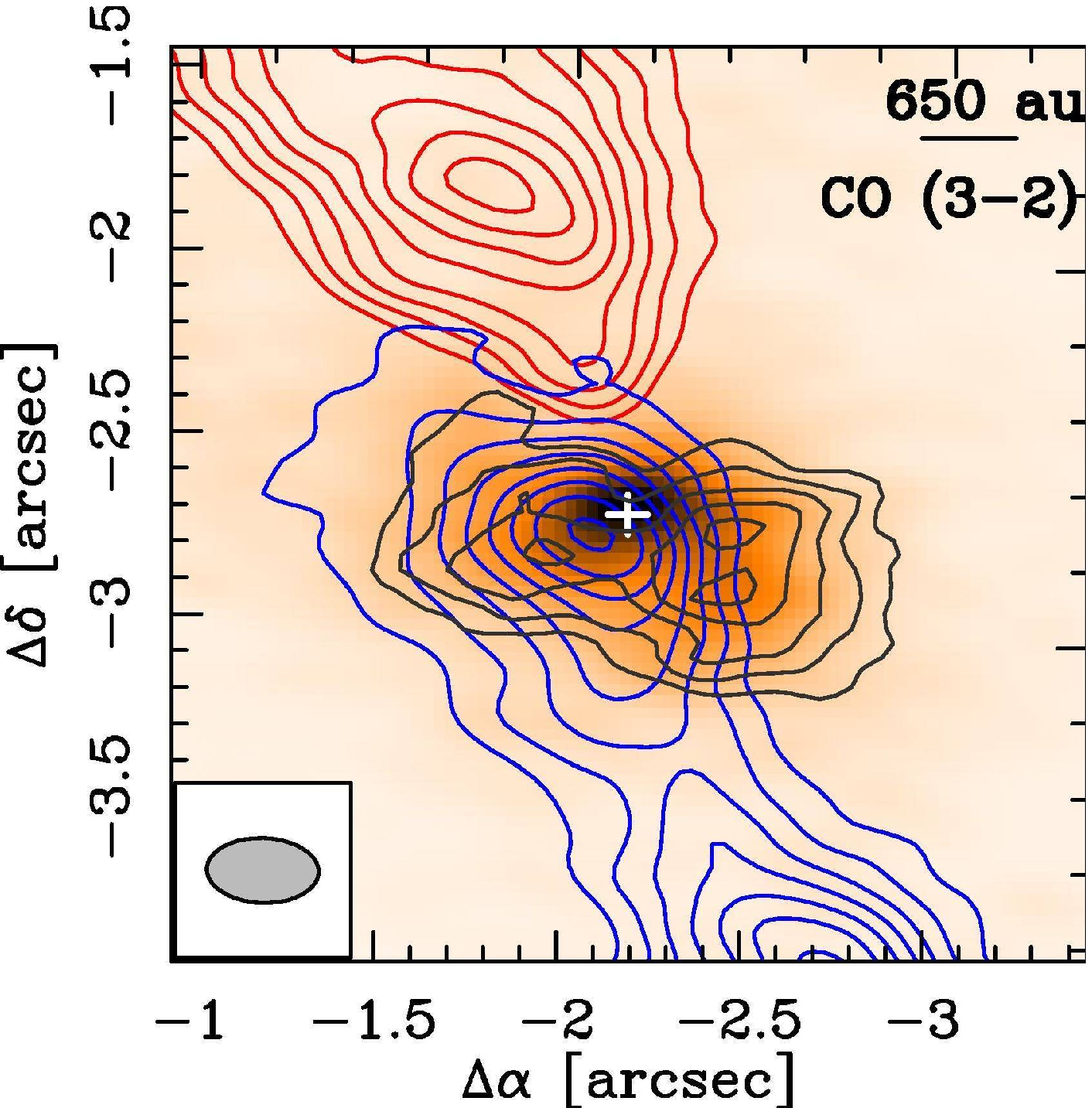}
   \includegraphics[width=4.5cm,angle=0]{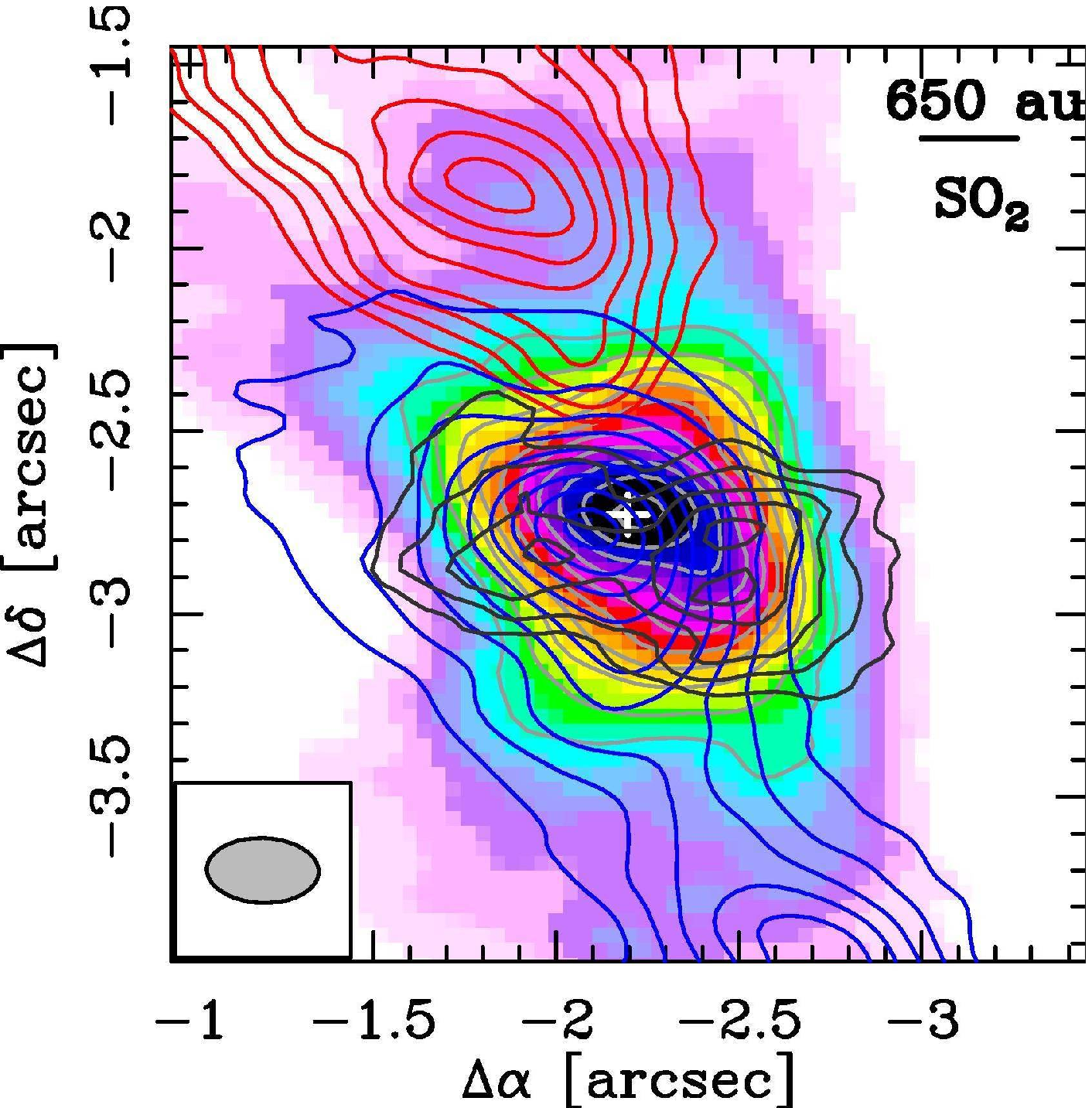}
   \includegraphics[width=4.5cm,angle=0]{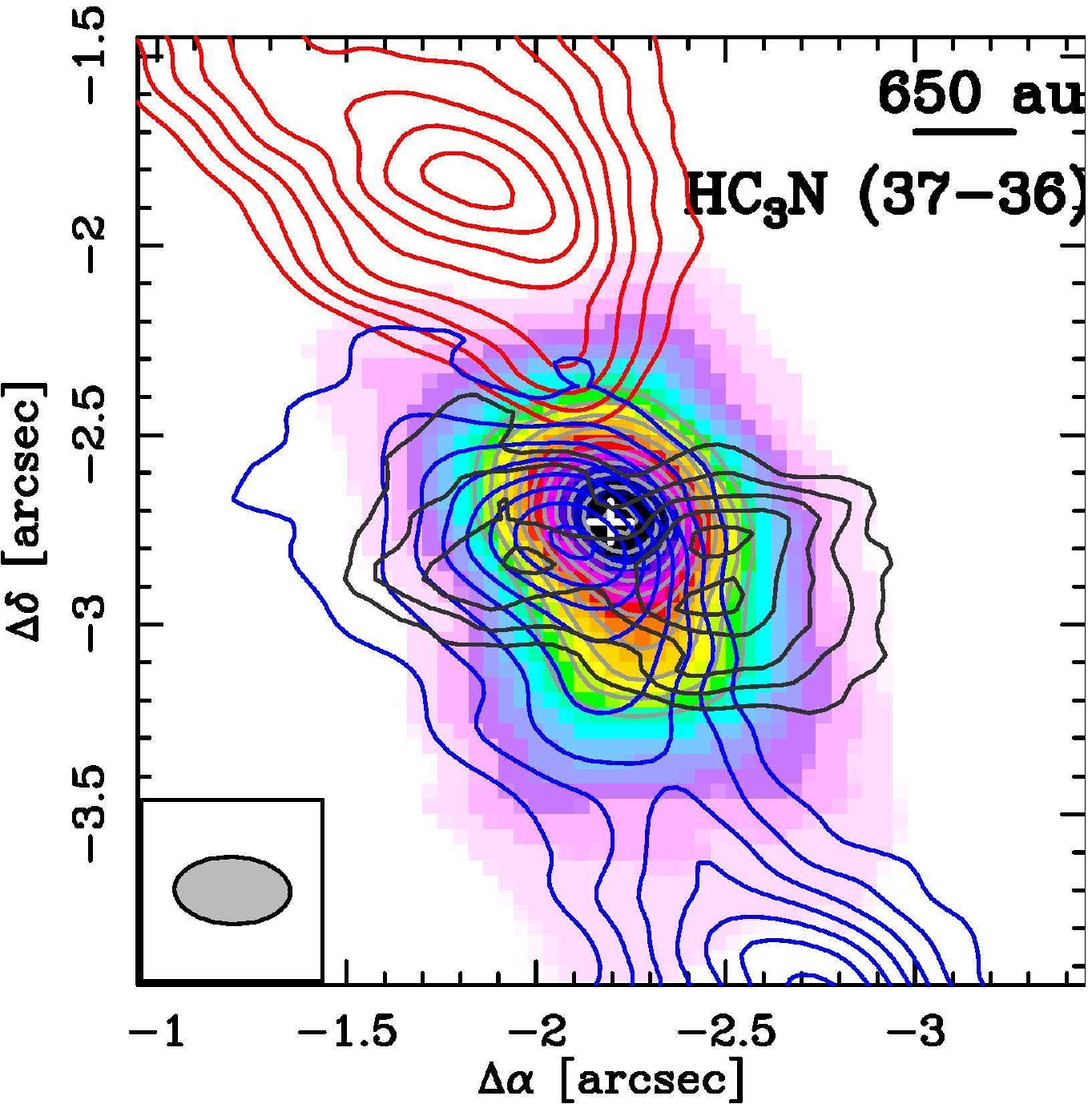}
   \includegraphics[width=4.5cm,angle=0]{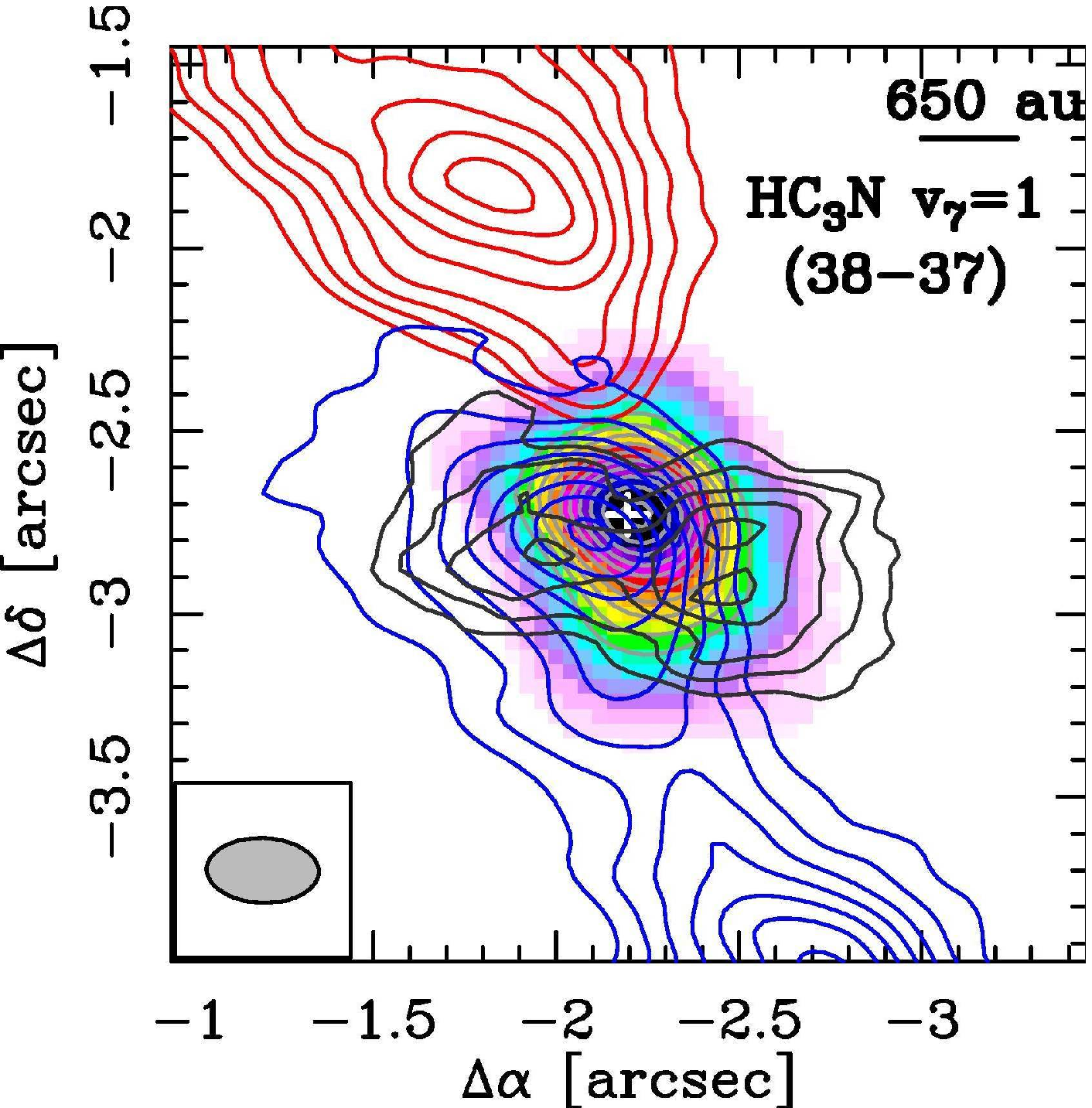}
      \caption{A zoom on the envelope showing  the CO (3--2) outflow lobes indicated in blue and red
    contours, the \methanol\ $\varv_{\rm t}=1$ line in black contours. The background shows the dust continuum, and {moment 0} maps {calculated over a velocity range of $-55$ to $-35$\,\kms} of SO$_2$,  HC$_3$N ($J$=37-36), and HC$_3$N $\varv_7=1$ ($J$=38-37) from left to right, respectively. 
    White cross marks the position of the dust continuum peak. The beam of the CO ($J$=3--2) data is shown in the lower left corner. }
              \label{fig:outflow_tracers}%
    \end{figure*}
    \begin{figure*}[!ht]
   \centering
   \includegraphics[width=4.5cm,angle=0]{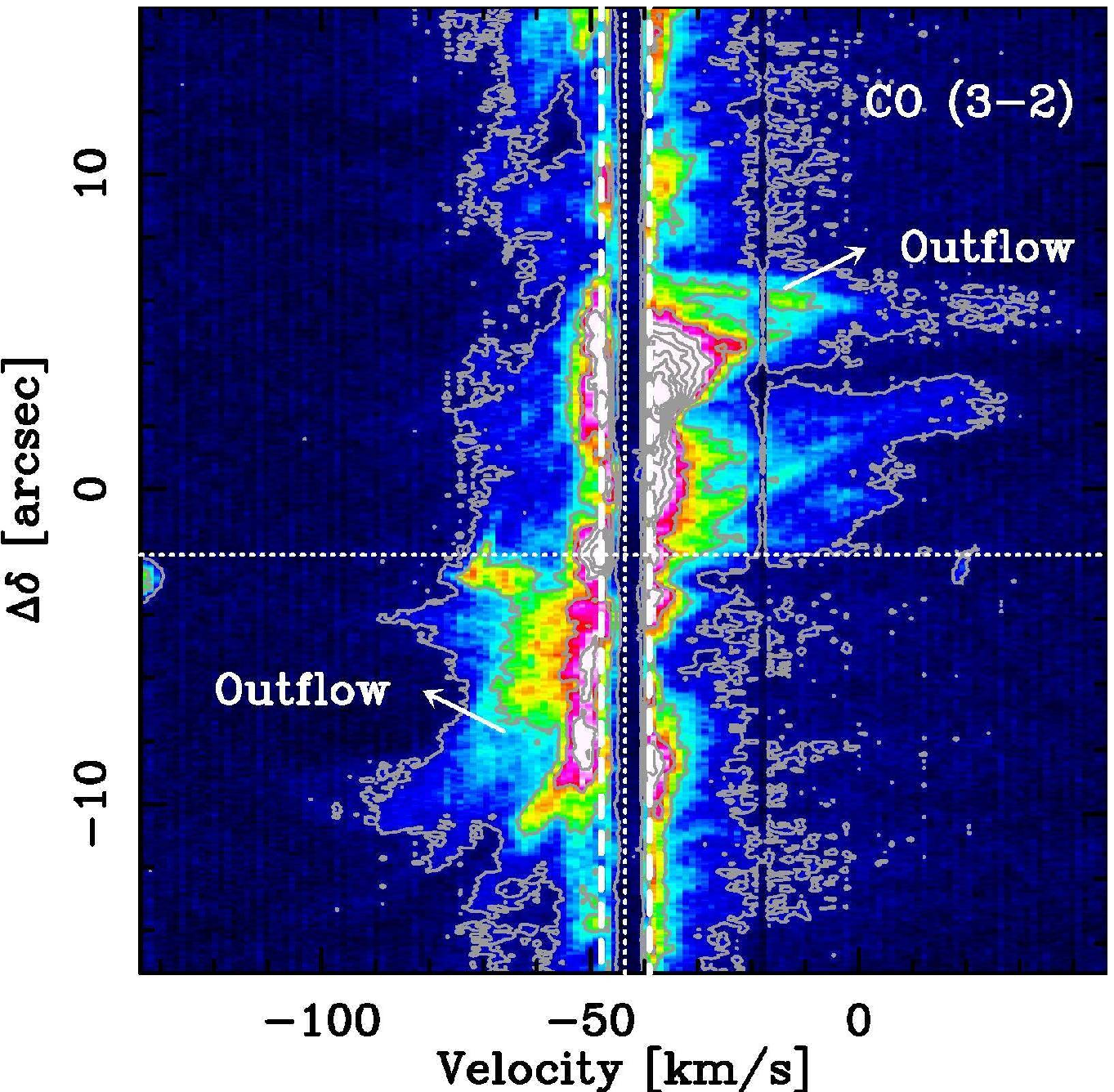}
   \includegraphics[width=4.5cm,angle=0]{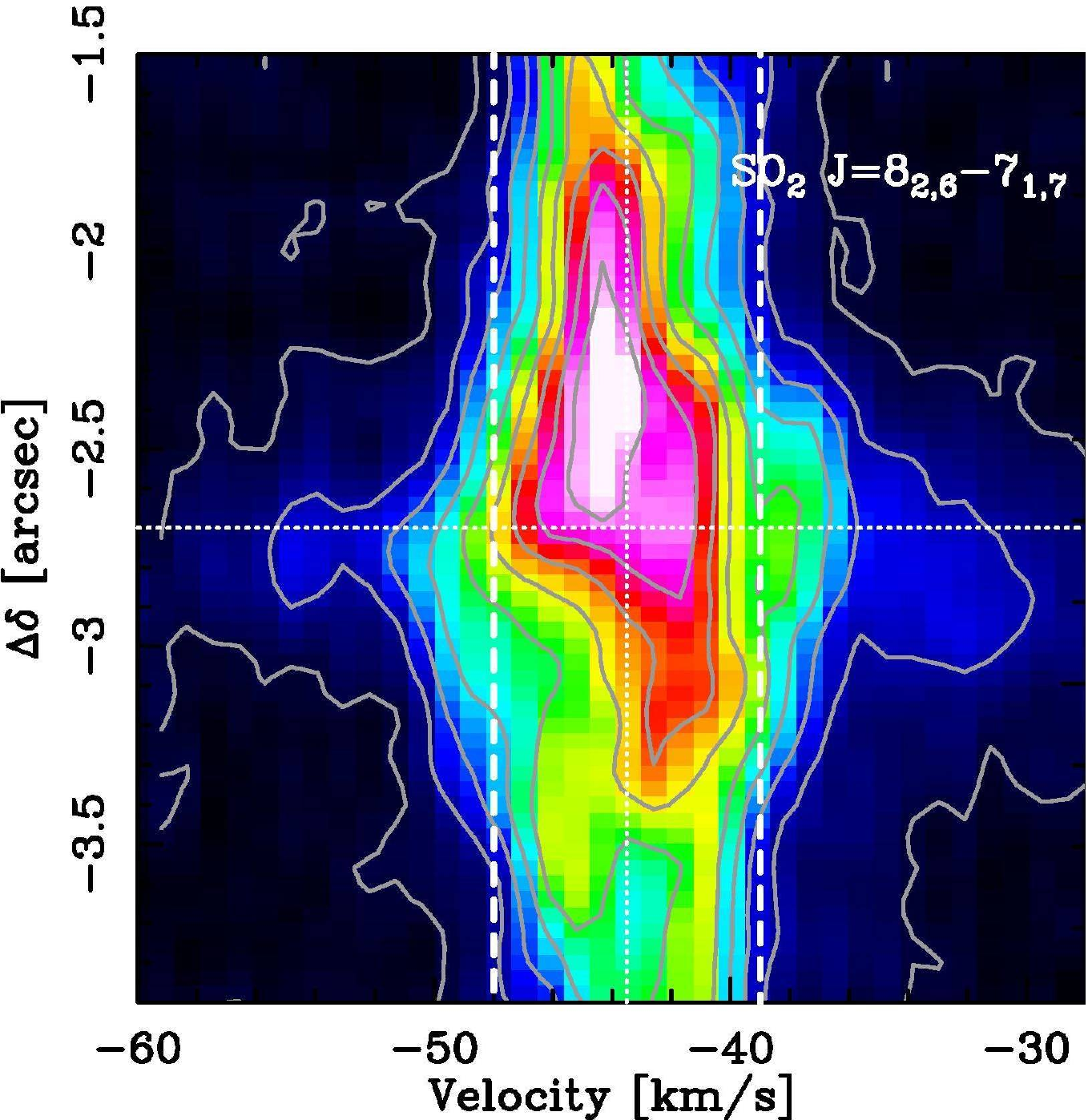}
   \includegraphics[width=4.5cm,angle=0]{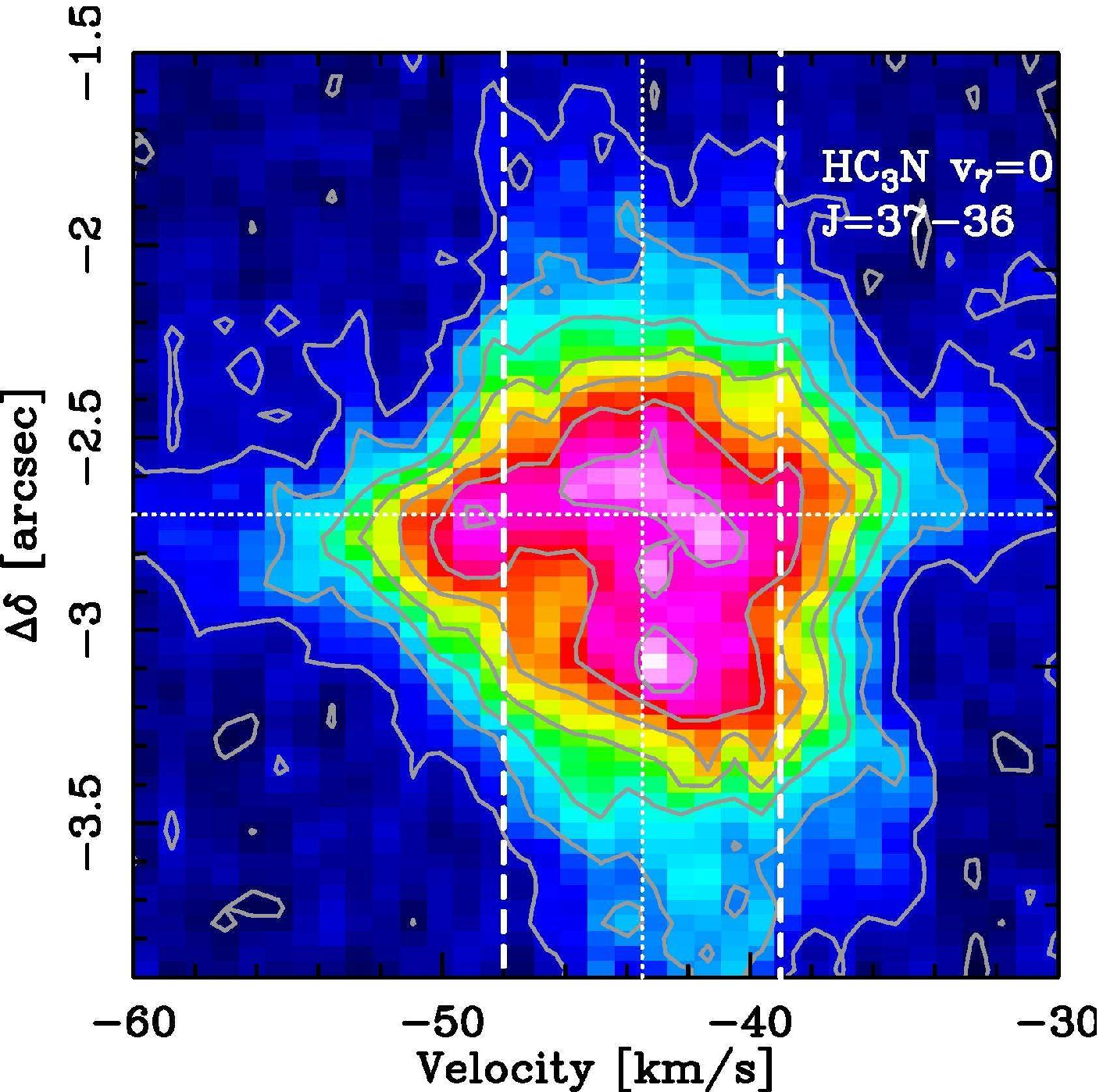}
   \includegraphics[width=4.5cm,angle=0]{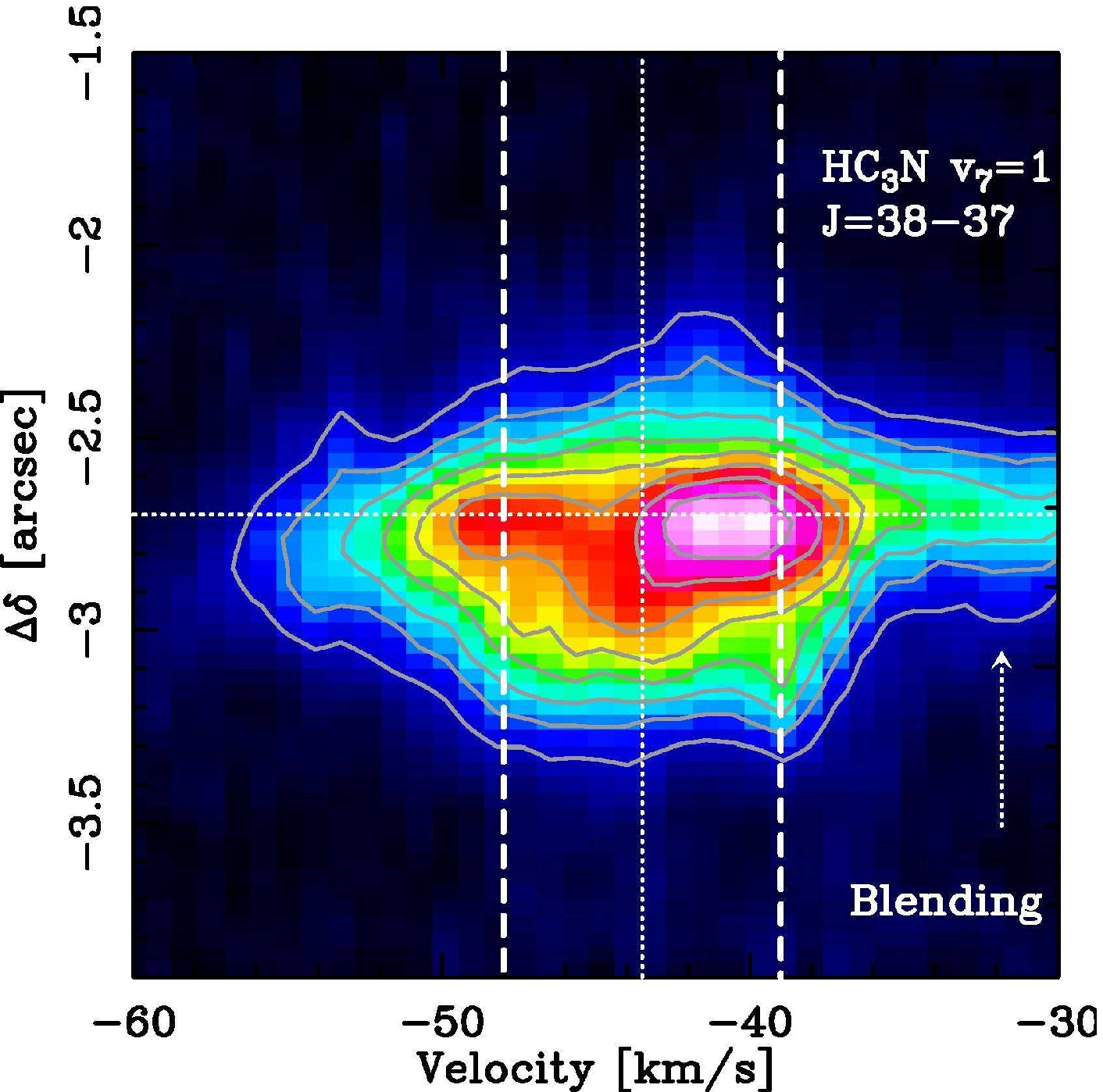}
   \includegraphics[width=4.5cm,angle=0]{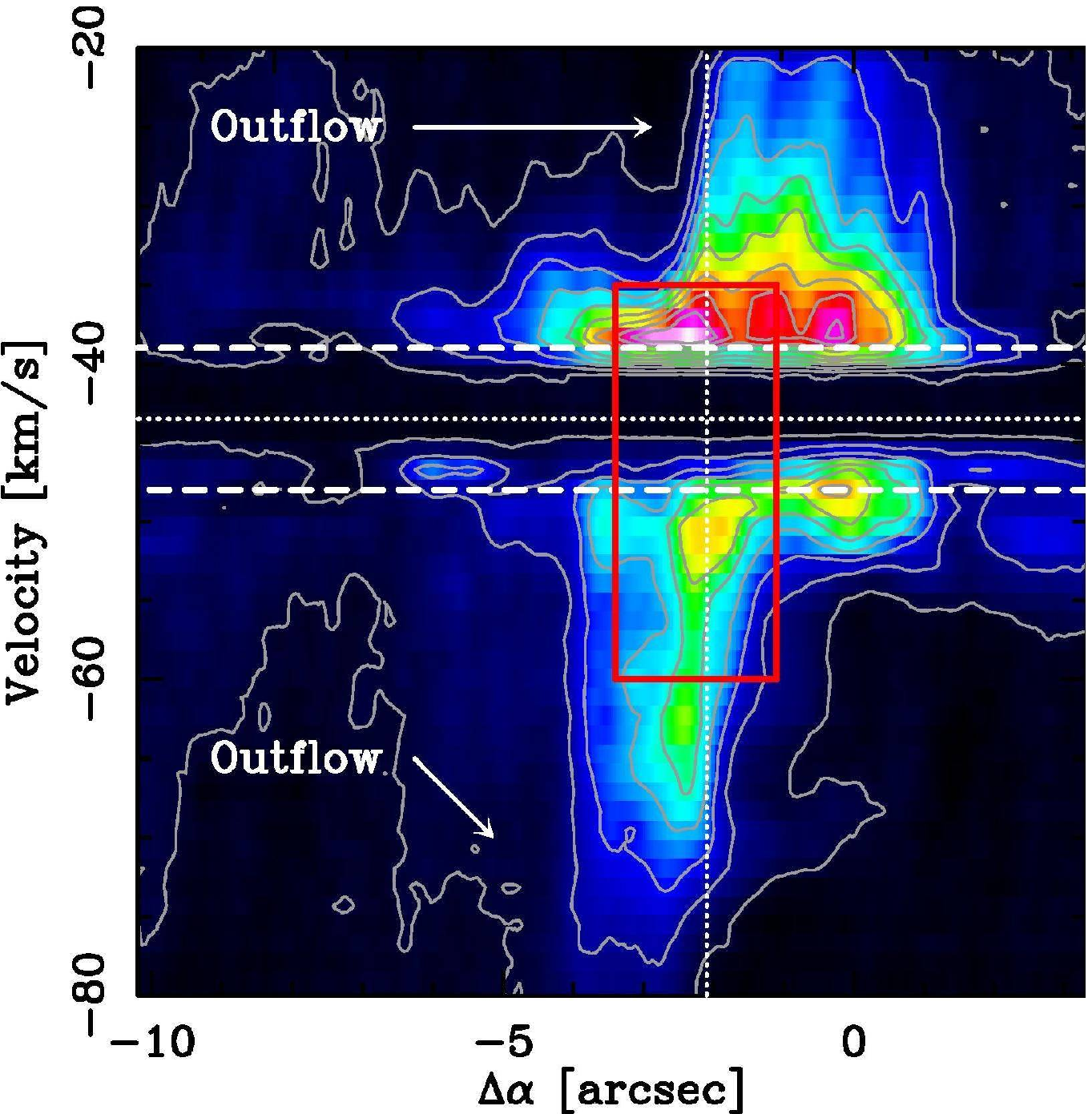}
   \includegraphics[width=4.5cm,angle=0]{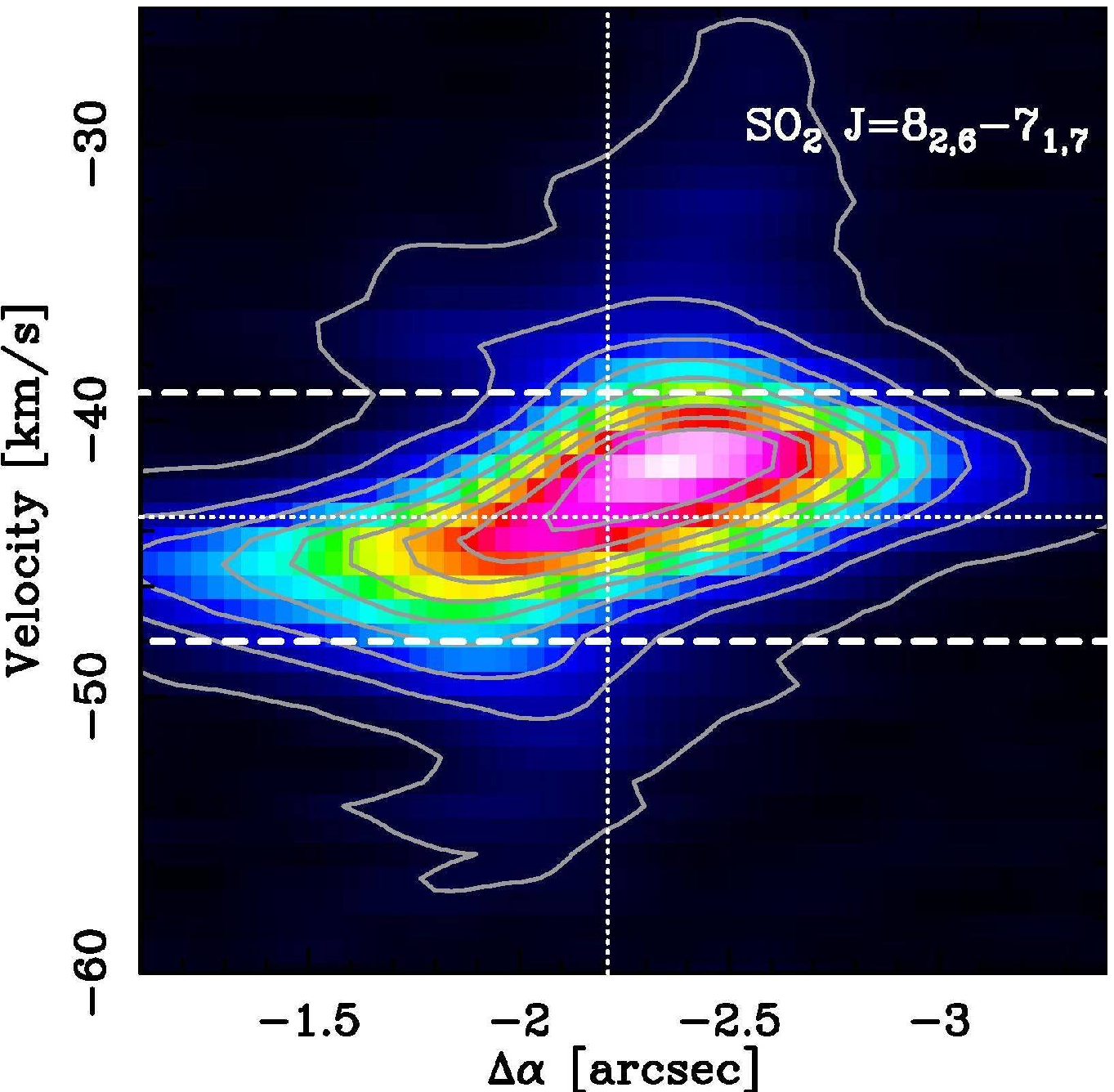}
   \includegraphics[width=4.5cm,angle=0]{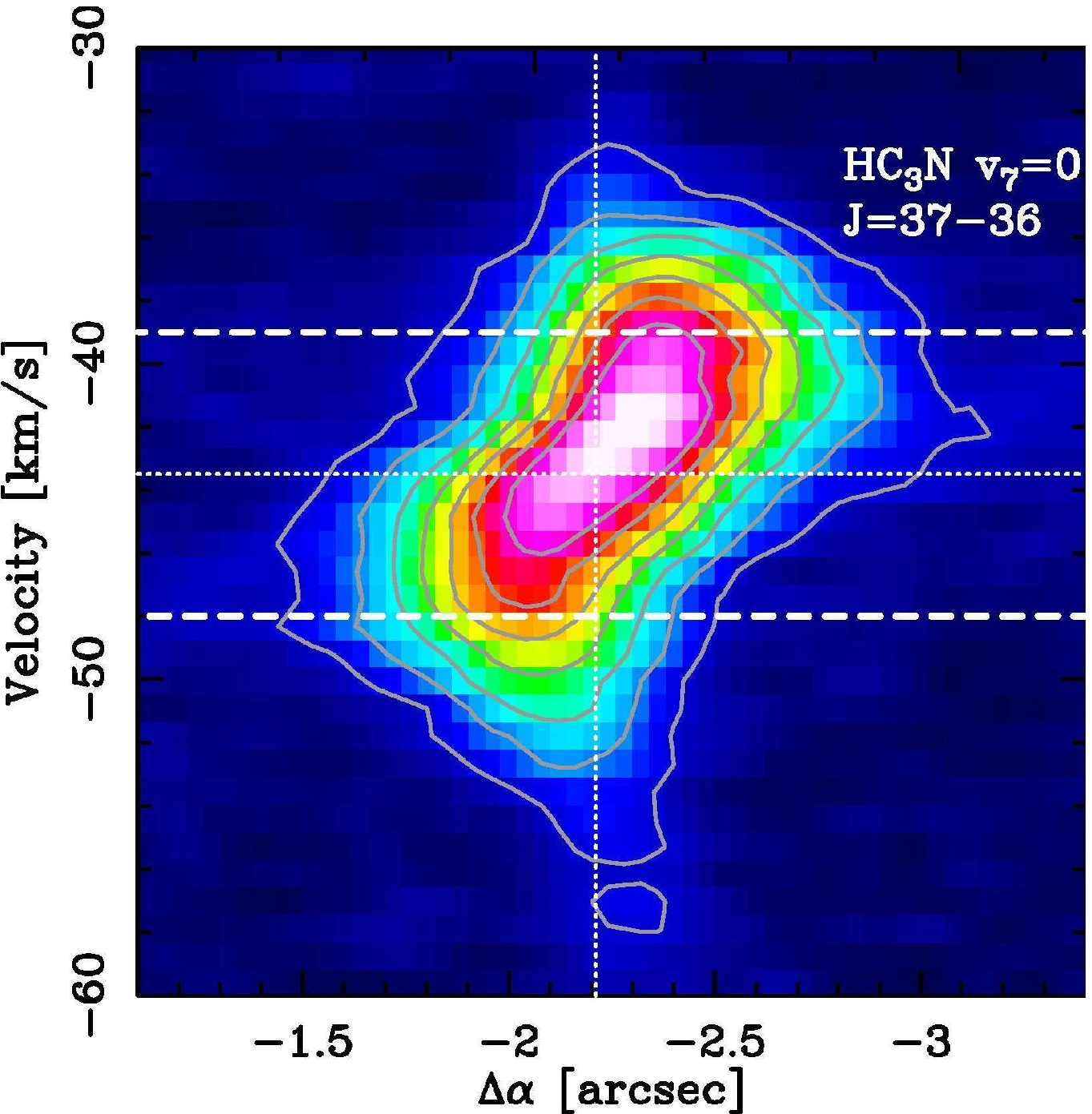}
   \includegraphics[width=4.5cm,angle=0]{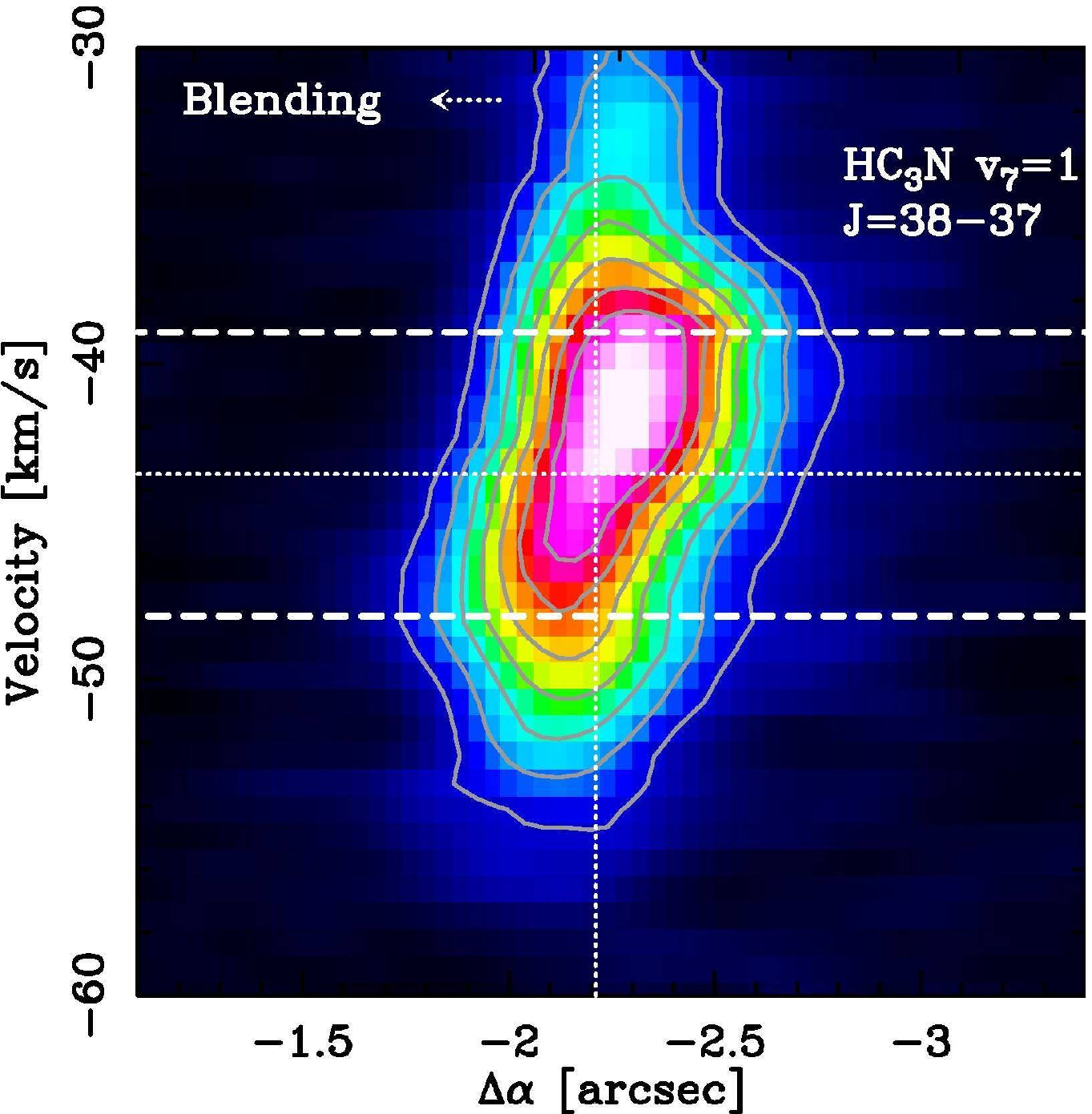}

      \caption{Position-velocity ($pv$) diagrams along the $\Delta\delta$ axis and averaged over the shown extent of the cube (top row) and along the $\Delta\alpha$ axis (bottom row) of the CO (3--2), SO$_2$,  HC$_3$N ($J$=37-36), and HC$_3$N $\varv_7=1$ ($J$=38-37) {transitions} from left to right, respectively. The dotted lines show the \vlsr\ of the source, and the position of the dust continuum peak. The dashed lines indicate velocities of \vlsr$\pm$4.5\,\kms\ roughly corresponding to the velocities of the \methanol\ peaks.  The red rectangle on the lower left panel corresponds to the region shown in the other panels.
      The contours start at 20\% of the peak and increase by 10\%, except for the panel of the CO and SO$_2$ lines, where the lowest contours start at 5\% of the peak.
}\label{fig:pv}
 
          \label{fig:outflow_kinematics}%
    \end{figure*}

\subsection{Protostellar activity}\label{sec:outflow}
\subsubsection{Outflowing gas traced by the CO ($J$=3--2) line}\label{sec:co}
 The protostellar activity of the compact continuum source
 is revealed by a single bipolar molecular outflow shown in Fig.\,\ref{fig:outflow}. 
 The CO\,(3--2) line shows emission over a broad velocity range, $\Delta v$, of $\pm$50\,km\,s$^{-1}$ with respect to the source rest velocity ($v_{\rm lsr}$) (Fig.\,\ref{fig:outflow}\textsl{a}). Imaging the highest velocities of this gas reveals a single and confined bipolar molecular outflow.  The orientation of the high-velocity CO emission clearly outlines the axis of material ejection. The high velocity CO (3--2) emission coincides well with the integrated emission from the shock tracer SiO (8--7) (Fig.\,\ref{fig:outflow}\textsl{b}), which likely traces the bow shocks along the outflow-axis {(Fig.\,\ref{fig:outflow}\textsl{c})}. 
  
 In particular the brightest CO emission of the red-shifted northern lobe appears to be confined (Fig.\,\ref{fig:outflow}\textsl{c}). 
Coinciding with the terminal position of the northern lobe, we
detect emission at the source rest velocity of the shocked gas tracer, SiO (8--7),
that is analogous to the bow-shocks observed in the vicinity of low-mass 
protostars. These indicate the shock front of the outflowing gas impacting the ambient medium.
 
Assuming that the maximum velocity observed in the CO (3--2) line 
corresponds to the speed at which the material has been ejected from the vicinity of the protostar, 
we can estimate when this material has been ejected. 
We measure an angular separation of $\sim$10\arcsec\
between the central object and the bow-shock seen
in the SiO (8--7) line
corresponding to a projected physical distance of 25000\,au. 
Based on the observed line-wings of the CO (3--2) line,
the measured velocity extent of the flow is $\sim$50\,km\,s$^{-1}$.
We estimate an inclination angle, $i$, of $56^\circ$, where $i=0^\circ$ describes 
a face-on geometry, and $i=90^\circ$ corresponds to an edge-on view. This is obtained from the axis 
ratio of the measured envelope size from the dust emission assuming that it has a circular morphology\footnote{The outflow  is rather confined with a small opening angle of $\sim30^\circ$. Simple geometric considerations based on the outflow orientation, opening angle and the fact that there is no significant overlap along the line-of-sight between the blue and red shifted emission, we can exclude an inclination range between $i<15^\circ$, and $i>75^\circ$. An inclination angle range between $15^\circ$ and $75^\circ$ would result in $t_{\rm dyn}=6.3\times10^2-8.8\times10^3$\,yr.}.
After correcting for the projection effects, we obtain
a dynamical age estimate of $t_{\rm dyn}\sim3.5\times10^3$\,yr 
for the protostar. This estimate is, however, 
affected by the uncertainty of the inclination angle, and
that of the jet velocity creating the bow shocks 
compared to the high velocity entrained gas seen by CO.
While the jet velocity could be 
higher than traced by the entrained
CO emitting gas leading to an even shorter time-scale,
the highest velocities seen in CO may not reflect the expansion speeds of the outflow lobes
as material accelerates.
Considering the mass of the central object (Sect.\,\ref{sec:d1}) and 
the typically observed {infall} rates 
of the order of $10^{-3}$\,{\msol}/yr \citep{Wyrowski2012, Wyrowski2016}, the larger values of the 
age estimate, of the order of a few times 10$^3$ years to 10$^4$ years, seem the most plausible. 
This estimate, at the order of magnitude, 
supports the picture of the protostar being very young.

\subsubsection{Sulphur-dioxide and cyanoacetylene}\label{sec:so2-hc3n}

To understand the origin of the CH$_3$OH $\varv_{\rm t}=1$ line, 
we compare its distribution 
to other species, such as SO$_2$ and cyanoacetylene, HC$_3$N, in Fig.\,\ref{fig:outflow_tracers}. Transitions from the latter species are detected both from the vibrational ground and excited states. 
These lines are not affected by blending, and probe 
various excitation conditions (see Table\,\ref{tab:lines}).
It is clear that from the investigated molecules, the methanol emission corresponds the best to the distribution of the dust continuum; both SO$_2$ and HC$_3$N show a different morphology. 

We show the  SO$_2$ ($8_{2, 6}$--$7_{1, 7}$) line at 334.673\,GHz which peaks on the position of the protostar and shows the most extended emission among the species discussed here, with a north-south elongation spatially coinciding with the outflow axis. Similarly, the HC$_3$N lines  peak on the protostar. The $\varv_7$=0  ($J$=37--36) transition shows a north-south elongation following the outflow axis, and is more compact compared to the SO$_2$ line. The higher excitation state 
$\varv_7$=1  ($J$=38--37) transition also shows a north-south elongation along the outflow; it is, however, even more compact than the vibrational ground state line.

In Fig.\,\ref{fig:outflow_kinematics} we show horizontal and vertical averages of the datacubes as a function of velocity ($pv$-diagrams). Due to the relatively simple source geometry, we show the averages along $\Delta\alpha$ and $\Delta\delta$ axes. For the CO emission we use the cube covering the primary beam, while for the other species we only use the region shown in Fig.\,\ref{fig:outflow_tracers}.  The high-velocity outflowing gas is clearly visible in the CO (3--2) line, and the kinematic pattern of both the SO$_2$, and HC$_3$N $\varv_7=0$ lines confirms that they show emission associated with the outflowing molecular gas. The velocity range of the HC$_3$N $\varv_7=1$ emission is clearly broader than that of the \methanol\ lines; it is, however not as broad as the SO$_2$, and HC$_3$N $\varv_7=0$ emission. The emission around $v_{\rm lsr}\sim-33$ \kms\ close to the HC$_3$N $\varv_7=1$ line is {a contamination from COMs with line forests, such as acetone and ethylene glycol}. 

   \begin{figure*}[!ht]
   \centering
   \includegraphics[width=0.99\linewidth,angle=0]{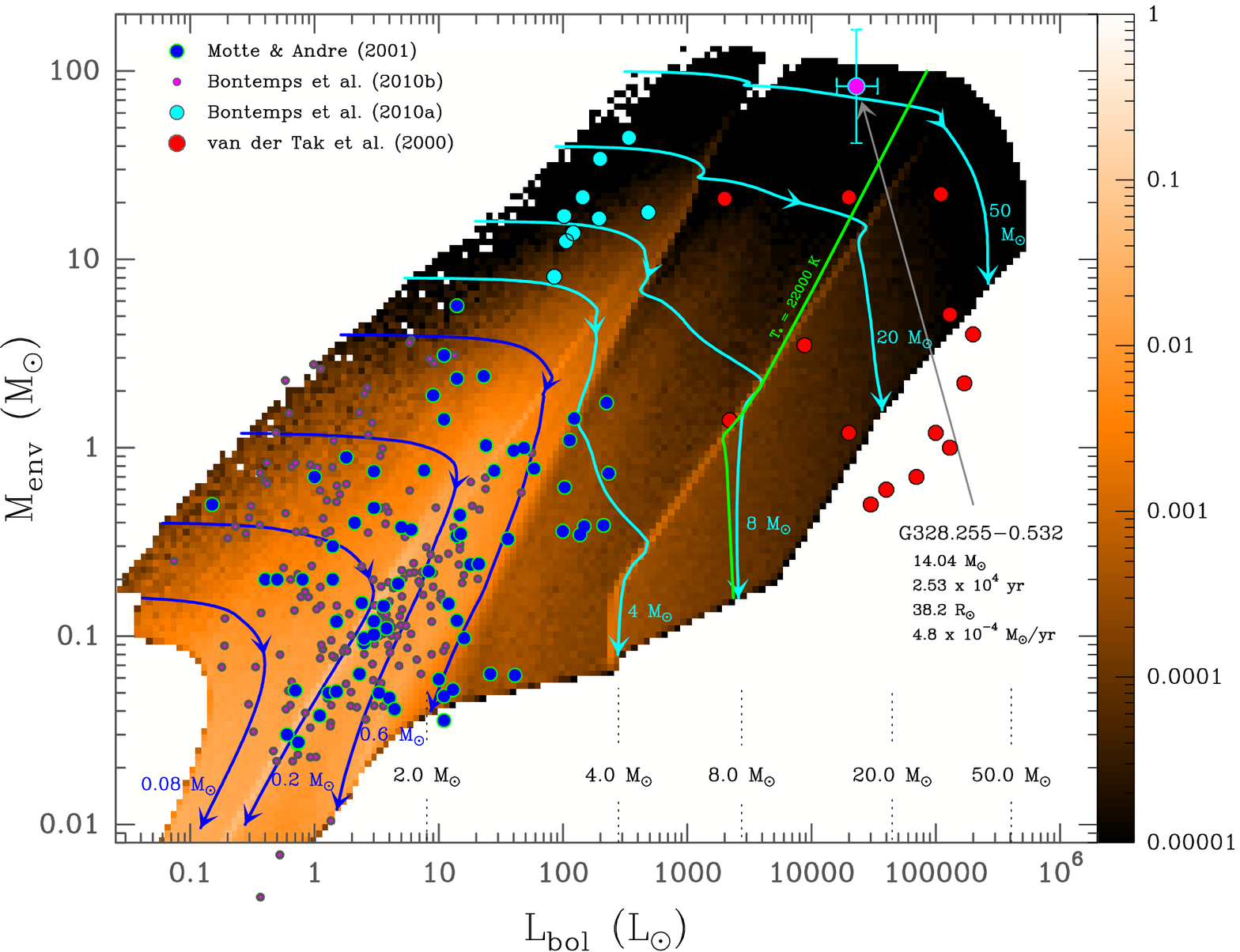}
      \caption{Evolutionary diagram showing  $M_{\rm env}$ versus $L_{\rm bol}$. {
      The color scale shows the predicted distribution of protostars as a fraction relative to 1 accounting for the typical stellar initial mass function (\citealp{Kroupa2003}) and for a star formation and accretion history (constant star formation and decreasing accretion rates) as described in \citet{Duarte2013}. (Figure adapted from \citealp{Motte2017}). 
            The protostar of \mysou\ is shown in magenta filled circle.}}
              \label{fig:evolutionary_diagram}%
    \end{figure*}

\section{Discussion}

We identify a single high-mass protostellar envelope within the $\sim${600}\,\msol\ mid-infrared quiet clump, \mysou. In Sect.\,\ref{sec:d1} we argue that the protostar is still in its main accretion phase, resembling the Class\,0 phase of low-mass protostars (c.f.\,\citealp{Duarte2013}). We investigate the origin of the CH$_3$OH emission in Sect.\,\ref{sec:d2} and propose that in particular the torsionally excited state line traces shocks due to the infall from the envelope to an accretion disk. In Sect.\,\ref{sec:d3} we compare its kinematics with a vibrationally excited state transition of HC$_3$N, that we propose as a potential new tracer for the accretion disk. 

\subsection{The most massive envelope of a young protostar}\label{sec:d1}
{A luminosity of $L_{\rm bol}=1.3\times10^4$\,$L_{\odot}$} (see App.\,\ref{app:sed}) associated with a massive envelope, and a strong outflow point to a still strongly accreting, young massive protostar. 
Due to the high extinction towards the protostar, we use  evolutionary diagrams (Fig.\,\ref{fig:evolutionary_diagram}) based on protostellar evolution models \citep{Hosokawa2009} to estimate the mass of the central object. 

According to models of protostellar evolution the observed luminosity is too high to originate only from accretion \citep{Hosokawa2010}, instead it is rather dominated by the Kelvin-Helmholtz gravitational contraction of the protostellar core. Using the typical accretion rate observed towards high-mass protostars and YSOs, 
i.e. of the order of $10^{-3}$\,$M_{\odot}\,\rm yr^{-1}$ \citep{Wyrowski2016}, 
the Hosokawa tracks indicate that the protostar has to be bloated, and close to the maximum of its radius during its protostellar evolution, 
suggesting that it is at the onset of the Kelvin-Helmholtz contraction phase \citep{Hosokawa2010}. During this regime, the luminosity does not depend strongly on the actual accretion rate and mainly depends on the mass of the protostar. 

  Using these models for the typically expected accretion rate of $\dot M_{\rm acc} = 10^{-3}$ \,$M_{\odot}\,\rm yr^{-1}$, we find a central mass {around} 11 $M_{\odot}$ for $L_{\rm bol}=1.3\times10^4$\,$L_{\odot}$. Evolutionary tracks with accretion rates of $\dot M_{\rm acc} = 1 - 30 \times10^{-4}$ \,$M_{\odot}\,\rm yr^{-1}$ give a similar protostellar mass range, between 11.1 and 15.2  $M_{\odot}$, respectively. The protostellar radius adjusts, roughly in a proportional way, to the average accretion rate with a range of radius from 8 to 260 $R_{\odot}$ for $\dot M_{\rm acc} = 1 - 30 \times10^{-4}$ \,$M_{\odot}\,\rm yr^{-1}$. We can therefore  assume that the current protostellar mass is relatively well determined, and lies in the range between 11 and 16 $M_{\odot}$. Taking the dynamical age estimate from Sect.\,\ref{sec:co} and the current propostellar mass, we can put an upper limit on the accretion rate by $\dot{M}_{\rm acc} = M_{\rm proto}/\tau_{\rm dyn}$, which is between $3.1\times10^{-3}$ and $4.2\times10^{-3}$\,\msol/yr. This estimate is, however considerably affected by the uncertainties in the dynamical age estimate (see Sect.\,\ref{sec:co}).
  
 Several types of maser emission also support the presence of an already high-mass protostellar object; 
 the evolutionary models indicate, however, that due to the bloating of the protostar no 
 strong ionising emission is expected despite its high mass. 
 This is consistent with the lack of radio continuum detection towards this object.
  The field has been covered at 3.6 and 6 cm  by a 
  radio survey of southern Red MSX Sources (RMS) \citep{Urquhart2007, Lumsden2013} that targeted a nearby
  MYSO/UCHII region only showing radio emission 1.1\arcmin\ offset compared to our position.
  Similarly, radio continuum observations at 8.4\,GHz only report a $4\sigma$ upper limit of 0.6\,mJy \citep{Phillips1998}.
  Based on this upper limit and adopting a spherical model of ionised plasma with typical values of $T_{\rm e}$=10$^4$\,K electron temperature, $EM$=$10^8-10^{10}$\,pc\,cm$^{-6}$ emission measure, we find that only a very compact {\hii} region with a radius of 100\,au could have remained undetected in these observations. This is an independent confirmation for the young nature of the protostar.

Based on the estimated current core mass of $120\,M_{\odot}$, 
we may expect a final stellar mass of $\sim$50\,$M_{\odot}$ (Fig.\,\ref{fig:evolutionary_diagram}).
The protostar of \mysou\ is therefore a particularly interesting object, and we suggest that it is 
one of the rare examples of a bloated, high-mass protostar, 
 precursor of a potentially O4-O5 stellar type star. 
 There are only very few candidates of such bloated protostars in the literature (e.\,g.\,\citealp{Palau2013}, and references therein), and most of them correspond to objects detected in the optical, and the source presented here is much more embedded.

Throughout this work we refer to the continuum peak as a single high-mass protostellar envelope. However, we can not exclude that the source would be fragmented at smaller, i.e. $<$400\,au scales which would lead to the formation of a close binary from a single collapsing envelope. While O-type stars have a high multiplicity (e.\,g.\,\citealp{Sana2017}), we do not find any clear evidence at the observed scales that would hint to multiplicity on smaller scales. For example, 
the small outflow opening angle could suggest that either the system is still young, 
or there is a single source driving the outflowing gas. Alternatively, gravitational fragmentation of the massive inner envelope could lead to the formation of companions at a subsequent evolutionary stage.

\subsection{Accretion shocks at the inner envelope}\label{sec:d2}
{
All \methanol\ lines follow the extension of the envelope, and their $pv$-diagrams 
are consistent with rotational motions (Fig.\,\ref{fig:methanol_all}). Fig.\,\ref{fig:renzo} shows the contours of the CH$_3$OH $\varv_{\rm t}=1$ line in different velocity channels, and reveals a clear velocity gradient over the bulk of the envelope. When looking at the most extreme velocity channels, we find that the emission follows well the {azimuthal elongations} of the envelope. This gradient is spatially resolved over the blue lobe, that is moving towards our direction and connects the envelope to the compact dust component.  The receding arm located on the near side of the envelope is more compact. Altogether this is consistent with a picture of spiral streams developing in the collapsing envelope as the material undergoes infall in a flattened geometry.}
The development of such spirals is frequently seen in numerical simulations of accretion to a central, dominant protostar. They are typically associated with a flattened structure (e.\,g.\,\citealp{Krumholz2007, Hennebelle2009, Kuiper2011, Hennebelle2016b}). 
A similar pattern has been observed towards the low-mass sources BHB07-11, in the B59 core \citealp{Alves2017}, and Elias 2-27 \citep{Perez2016};  
and also on somewhat larger scales of infalling envelopes towards more evolved  high-mass star forming regions with an order of magnitude higher luminosity \citep{Liu2015}.

The peak of the torsionally excited state line, together with the $^{13}$C isotopologue lines, pinpoint the location of the highest methanol column densities. {We explain these observations by two physical components for the \methanol\ emission, the more extended emission associated with the bulk of the inner envelope, and the peak of the $\varv_{\rm t}=1$ line observed at the largest velocity shift compared to the source \vlsr. In the following we investigate the physical origin of the $\varv_{\rm t}=1$ emission peaks.} 

Our LTE analysis suggests a high \methanol\ column density up to $2\times10^{19}$\,cm$^{-2}$ {towards these positions}, which is at least three orders of magnitude higher than typically observed in the quiescent gas (e.g.\,\citealp{Bachiller1995, Liechti1997}). {Such high \methanol\ column densities are, however, observed on small scales towards high-mass star forming sites, as} reported by e.g.\,\citet{Palau2017} in the disk component of a high-mass protostar, IRAS20126+4104, and {towards} other high-mass star forming regions \citep{Beltran2014}. Towards the hot-cores in the most extreme star forming region, Sgr B2, even higher values are reported above $10^{19}$\,cm$^{-2}$  \citep{Bonfand2017} .

    \begin{figure}[!h]
   \centering
   \includegraphics[width=0.45\linewidth]{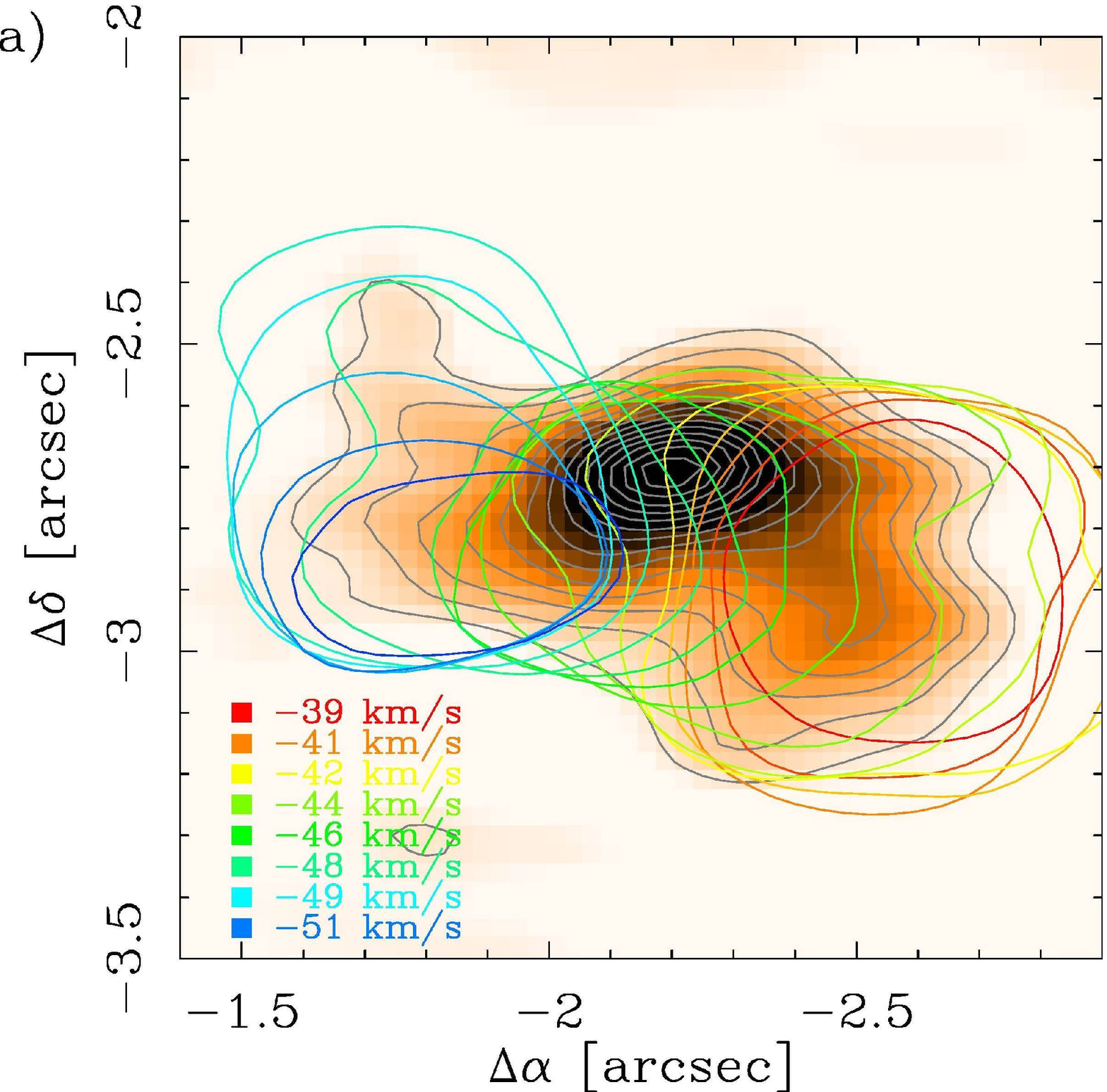}
   \includegraphics[width=0.45\linewidth]{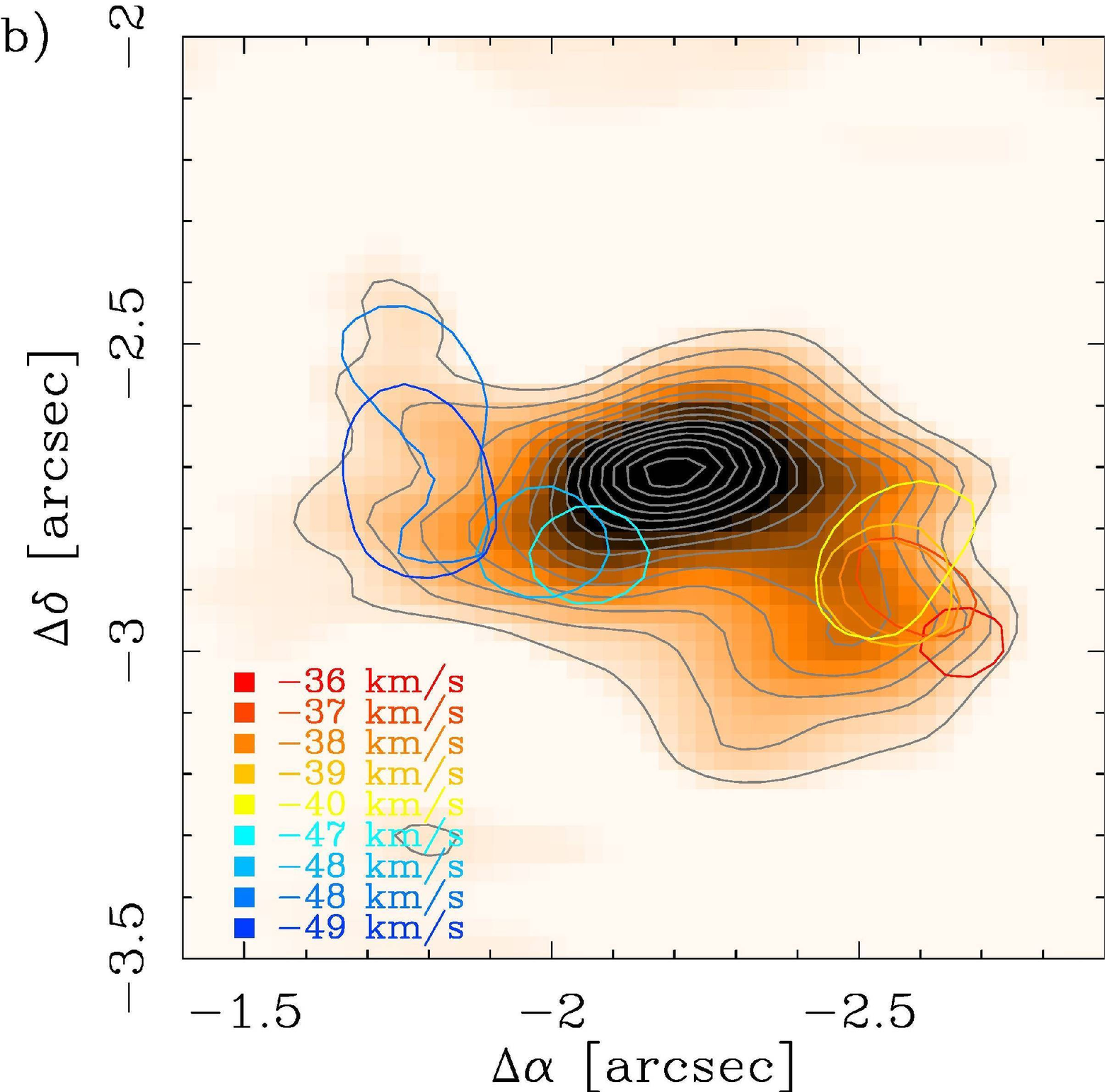}
      \caption{\textsl{a)} The color scale is the same as in Fig.\,\ref{fig:dust}{\sl c} showing the thermal dust emission with the larger scale core emission removed in order to enhance the structure of the inner envelope. The color lines show the 50\% contour of the peak emission of $\varv_{\rm t}=1$ \methanol\ line at different velocities starting from $-50.5$\,\kms\ (blue) to $-38$\,\kms\  (red). The corresponding velocity of every second contour is shown in the figure legend. \textsl{b)} Same as the left panel only showing the 90\% contours of most extreme velocity components in red and blue. The velocities corresponding to each contour are shown in the figure legend.}
              \label{fig:renzo}%
    \end{figure}

In particular, the torsionally excited, $\varv_{\rm t}=1$, line with the upper energy level of 315\,K, 
typically requires an infrared radiation field at 20--50\,$\mu$m in order to populate {its upper} state. Despite the high average volume densities, assumed to be above ${n}>10^7$\,cm$^{-3}$,
radiative excitation could be necessary to
populate rotational levels in the $\varv_t=1$ state. 
Heated dust in the vicinity of the protostar would naturally provide the infrared photons, 
therefore, it is  very surprising that this line does not peak on the dust continuum at the position of the protostar,
instead it peaks on the inner envelope (Fig\,\ref{fig:methanol}).
This particular pattern could be explained by a decrease of \methanol\ abundance towards the protostar, 
and in fact our results {in Sect.\,\ref{sec:rot_dia}} suggest somewhat larger \methanol\ column densities offset from the dust peak.   
Considering that the  H$_2$ column density increases towards the {continuum peak}, {and since in Sect.\,\ref{sec:methanol2} we rule out optical depth effects of the discussed transitions,} this suggests a decreasing \methanol\ abundance towards the position of the protostar.  
{The observed high \methanol\ column densities and their emission peak could also
pinpoint
local heating from shocks, which would naturally lead to an 
increase both in the temperature and in the molecular abundance, especially for} \methanol.
{This is because \methanol} has been found to show an increase in abundance by orders of magnitude in various shock conditions as the molecule gets released from the grain surfaces via sputtering \citep{Flower2010, Flower2012}.

Shocks are produced in a discontinuity in the motion of the gas, and in a collapsing envelope they can emerge in various conditions. 
One possibility is an origin associated with the outflowing gas hitting the ambient medium of the envelope. Such an increase in the \methanol\ abundance has been observed in both low-  \citep{Bachiller1995} and high-mass star forming regions \citep{Liechti1997, Palau2017}. We therefore compare in more detail the kinematics of the \methanol\ emission in Fig.\,\ref{fig:pv_comp} with that of outflow tracers (see Sect.\,\ref{sec:so2-hc3n}) to exclude its association with outflow shocks. 
While the CO (3--2) emission is not useful for velocities below $v_{\rm lsr}\pm6$\,\kms\ due to strong
self absorption, the HC$_3$N $\varv_7=0$ line shows a velocity gradient starting at a position close to the protostar and shows increasing velocities at larger distances.
The $\pm$4.5\,\kms\ peaks of the CH$_3$OH $\varv_{\rm t}=1$ line are offset from the HC$_3$N $\varv_7=0$ transition, showing no evidence for the \methanol\ emission to follow the high-velocity emission from the outflowing gas.
    \begin{figure}[!ht]
   \centering
   \includegraphics[width=0.45\linewidth]{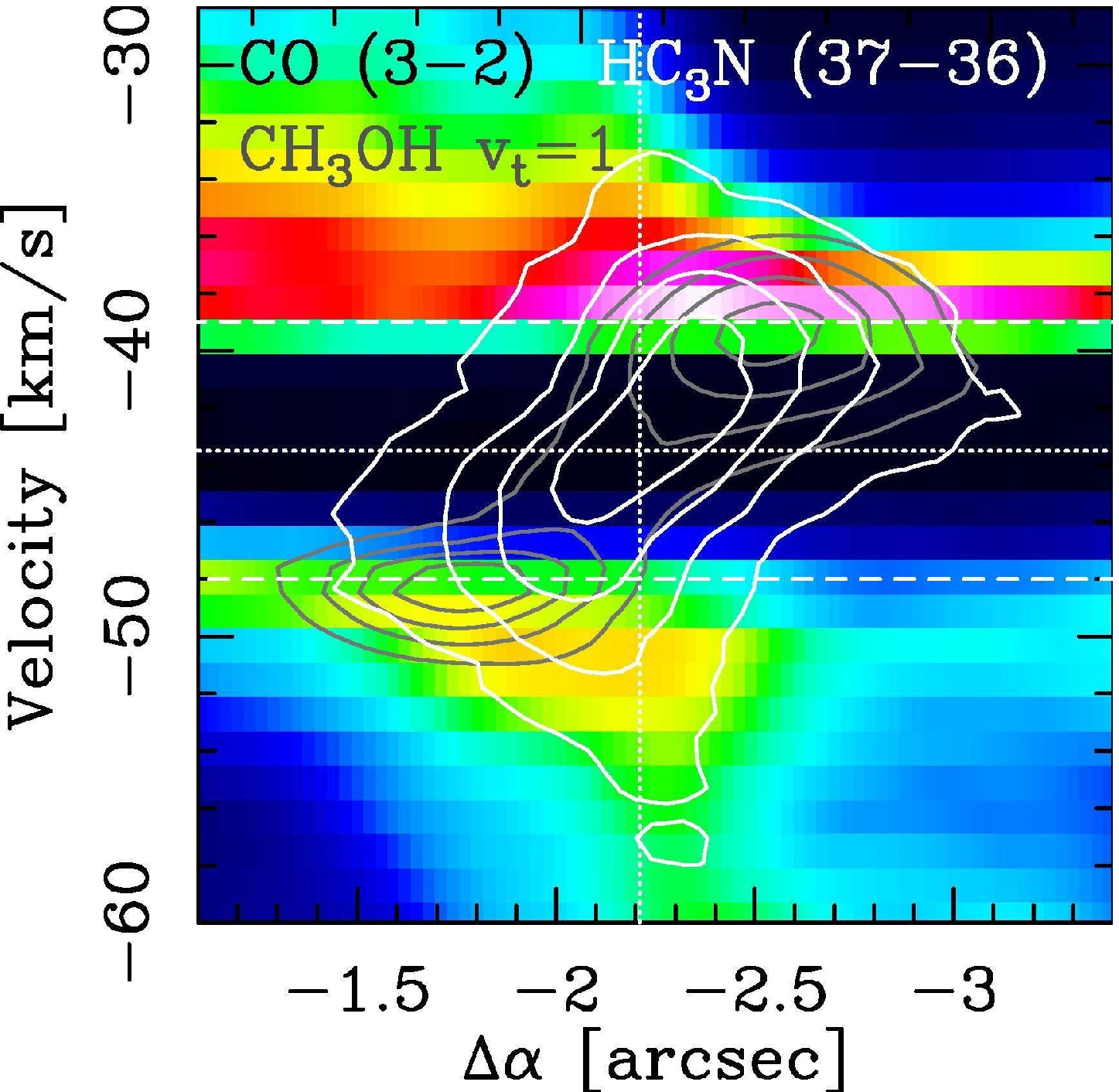}
   \includegraphics[width=0.45\linewidth]{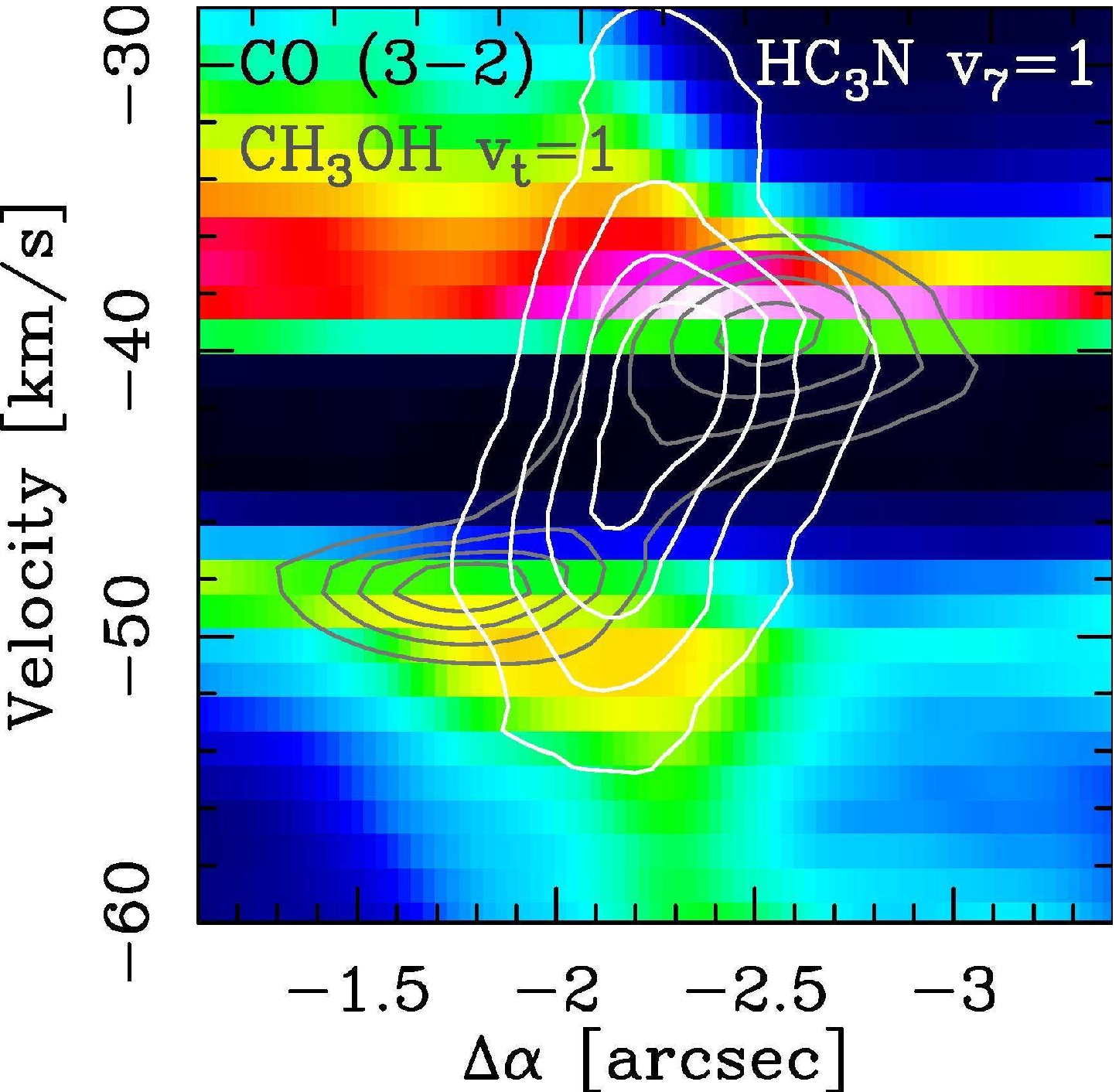}
      \caption{Comparison of the position velocity map along the $\Delta\alpha$ axis and averaged over a region of $\sim$2.5\arcsec. The color scale shows the CO (3--2) emission which is heavily affected by missing spacings and self absorption at the source $v_{\rm lsr}$. The dark gray contours show the CH$_3$OH $\varv_{\rm t}=1$, white contours the HC$_3$N (37--37) (left), and the HC$_3$N $\varv_{\rm 7}=1$ (right) {transitions}. }
              \label{fig:pv_comp}%
    \end{figure}

Interestingly, a comparison with the HC$_3$N $\varv_7=1$ line (Fig.\,\ref{fig:pv_comp}, right panel), shows that the higher excitation gas is more compact, and is well 
confined between the spatial and velocity axes outlined by \methanol. The velocity gradient across the line is visible, its peak is, however, offset from the \methanol\ peaks. This supports a picture where the \methanol\ and HC$_3$N trace different components, the former one tracing more the inner envelope, while the HC$_3$N $\varv_7=1$ line {probes} the regions closer to the protostar.

Another possibility to explain the methanol peaks would be the launching of the outflow that would also lead to shocks liberating \methanol\ into the gas phase. The launching mechanism of the outflowing gas is highly unexplored territory in high-mass star formation, the location of the outflow launching site is therefore not constrained. Towards low-mass protostars, methanol has been observed to be associated with the launching of the jet/disk wind in the close vicinity, within $<135$ au distance from the protostar \citep{Leurini2016}, which is a considerably smaller scale than probed by our observations. There is some indication that towards low-mass protostars outflow can be launched at the outer regions of the Keplerian accretion disk, a  recent study of a low-mass Class\,I type protostar presents an example where the outflow is launched at a distance beyond the disk edge, from the inner envelope \citep{Alves2017}.  
With our current angular resolution we do not probe such small scales, which would be unresolved, and peaking on the protostar. However, we cannot exclude that the observed \methanol\ spots may have contribution from the surface of the flattened envelope. 

Given the extent of the torsionally excited \methanol\ emission, the best explanation 
for our observations is that a significant amount of \methanol\ is liberated into the gas phase by shocks associated with the inner envelope itself.  Around the low-mass Class 0 protostar, L1157, \methanol\ has been detected tracing shocks within the infalling gas \citep{Goldsmith1999,Velusamy2002}. In both of the latter two scenarios explaining the \methanol\ emission, the observed maximum velocity offset corresponds to the line-of-sight rotational velocity of the gas at the innermost regions of the envelope.

\subsection{Indirect evidence for a Keplerian disk}\label{sec:d3}
The brightest spots of the torsionally excited methanol emission 
can be interpreted as tracing shocks emerging in the innermost regions of the envelope, hence associated with 
the centrifugal barrier.
This happens when the inflowing material from the envelope hits material with a smaller radial velocity component that corresponds to an accretion disk surrounding the central protostar. Such a transition between the envelope and the disk material at the centrifugal barrier has been directly observed in nearby low-mass protostars \citep{Sakai2014,Oya2016,Alves2017}. {Towards these objects} both the gas kinematics and the gas chemistry  change at the inner envelope; some studies interpret {the extent of this region} as a sharp (c.f.\,\citealp{Alves2017}), while others as a more gradual transition region \citep{Oya2016}. Here we observe it as a more extended {emission}, which also appears to be asymmetric. 

Direct evidence for the presence of the disk is hindered by the angular and spectral resolution of our dataset; however, putting all pieces of evidence together, we find several indications supporting the scenario of accretion shocks at the centrifugal barrier.
In Sect.\,\ref{sec:cont} we already noted a marginally resolved, compact dust component that could correspond to the disk with a resolved major axis of 250\,au seen in projection.
The elongation of this residual is resolved, and is perpendicular to the outflow 
axis within $\sim$10$^\circ$. Both its orientation and extent 
are consistent with the location of the two \methanol\ peaks at a projected distance of 300 to 800 au considering the uncertainty {resulting from} our angular resolution of $\sim${400}\,au.  

In addition, we find that the highest excitation molecular emission observed by us in the 
vibrationally excited HC$_3$N $\varv_7=1$ line, is very compact, peaking on the compact dust component. This transition has an upper level energy of 645\,K, {and also requires 
an infrared radiation field at 20--50\,$\mu$m in order to populate its vibrationally excited states. }
Therefore, it
more likely traces a region considerably closer to the protostar than the torsionally excited methanol line. 
Although this molecule is also present in the outflow, as suggested by the vibrationally ground state transition, 
we propose that the high excitation vibrationally excited HC$_3$N $\varv_7$=1 emission  
is a good candidate for tracing emission from the accretion disk. 

Because the accretion disk is expected to be in Keplerian rotation, we 
can compare the observed $pv$-diagrams with a simple toy model adapted from 
\citet{Ohashi1997} to describe an axisymmetric rotating thin disk for the HC$_3$N $\varv_7$=1 line, and a ring of gas at the centrifugal barrier for the \methanol\ $\varv_{\rm t}=1$ line (Fig.\,\ref{fig:models}). We show here two models: the CH$_3$OH lines trace the rotating envelope with $v_{\rm rot}(r)\sim r^{-1}$, and the HC$_3$N $\varv_7=1$ line traces the gas in Keplerian motion. 

We use an x,y grid size of $2\times2500$\,au, and a velocity axis with 0.7\,km\,s$^{-1}$ resolution, a power-law density distribution with $n\propto r^{-1.5}$, constant molecular abundance with respect to H$_2$, and a central protostellar mass of 15\,$M_{\odot}$ (see Sect\,\ref{sec:d1}). For the geometry of the CH$_3$OH emission we use a ring between 300 and 900\,au corresponding to the extent of the marginally resolved compact component, and $v_{\rm rot}$=4.5\,km s$^{-1}$ at the inner radius of 300\,au. For the HC$_3$N $\varv_7=1$ line, we use a Keplerian model with a disk size up to 600\,au. 

The models are shown in Fig.\,\ref{fig:models} for the ring (top row) and the disk (bottom row) component. The observed spots of \methanol\ corresponding to the highest column densities are reasonably well reproduced with these models, the observed asymmetries in the distribution of the emission and projection effects are, however, not included in our models. We break the degeneracy between the  location of the ring of \methanol\ emitting gas and the mass of the central object by measuring the position of the methanol shocks. Since we do not include a correction for the inclination angle between the source and our line of sight, the determined parameters are uncertain within a factor of a few. 

These models demonstrate 
that the \methanol\ emission can be well explained by a ring of emitting gas from the infalling envelope that is more extended than the observed HC$_3$N $\varv_7=1$ line, and the compact dust continuum source. This scenario is similar to {what} has been observed towards the high-mass object, AFGL2591, by \citet{Jimenez-Serra2012}, who find a ring of \methanol\ emission at a velocity that is consistent with the Keplerian velocity of the estimated source mass of 40\,\msol. 
Due to the observed asymmetry of the HC$_3$N $\varv_7=1$ line, and our poor velocity resolution, our models only show that the extent of the observed velocity range of the HC$_3$N $\varv_7=1$ line 
could be consistent with a disk in Keplerian rotation based on the physical constraints of the model.

Finally, in Fig.\,\ref{fig:keplerian} we show analytic estimates for the Keplerian velocity at a range of central mass between 10 and 20\,$M_{\odot}$, and a range of centrifugal barriers around 300-800\,au, which corresponds to the parameter range that could still be consistent with the observations. To obtain the rotational velocity at this radius, we correct the observed  4.5\,km s$^{-1}$  for an inclination angle, $i$, of  56$^\circ$  based on the 
axis ratio of the measured envelope size in Sect.\,\ref{sec:cont}. We see that the observed velocity offset of the \methanol\ peaks fits well the Keplerian velocity for the plausible mass range and the range of centrifugal barrier within a resolution element. 

   \begin{figure}[!]
   \centering
   \includegraphics[width=0.43\linewidth,angle=0]{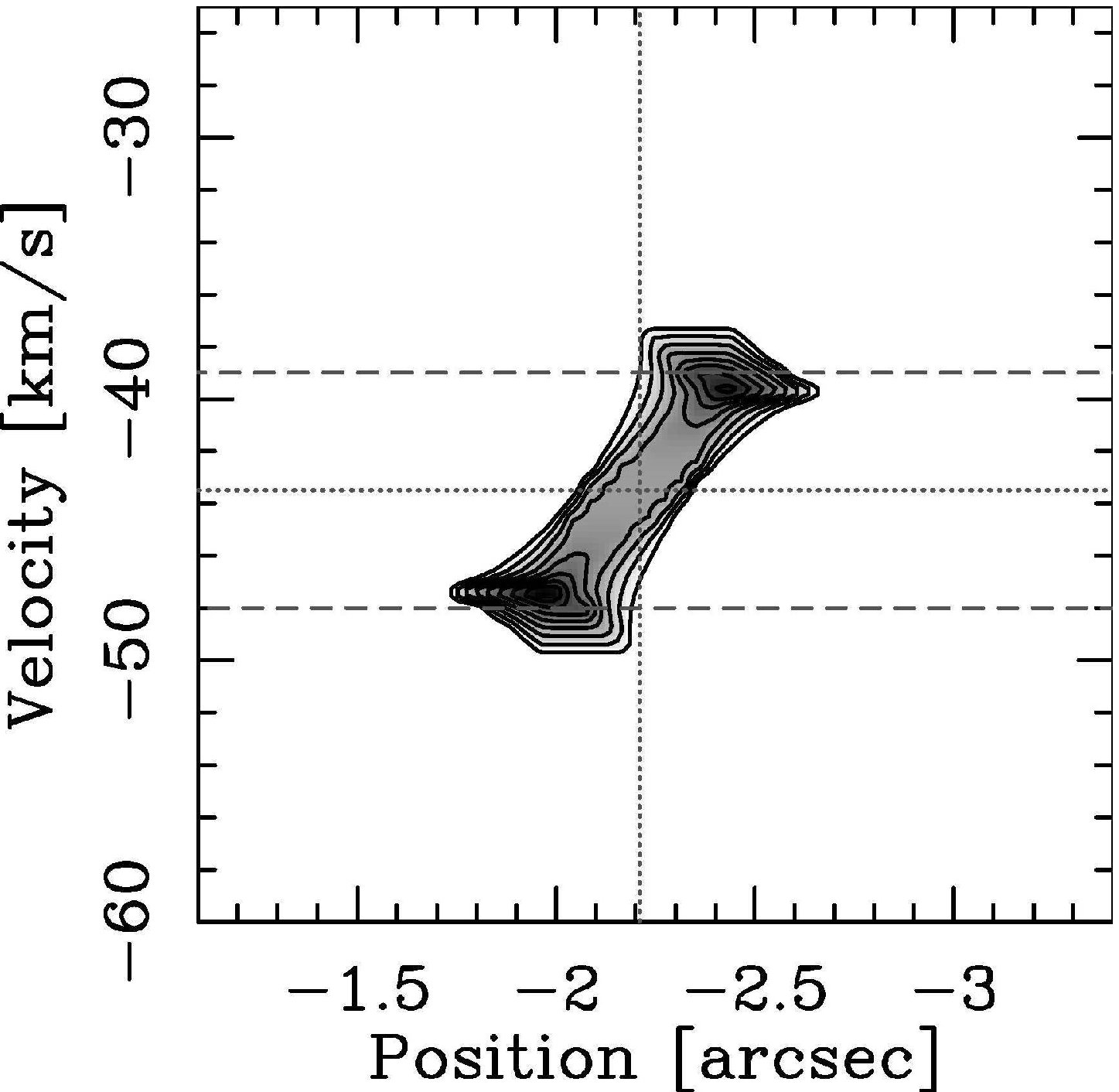}
    \includegraphics[width=0.43\linewidth,angle=0]{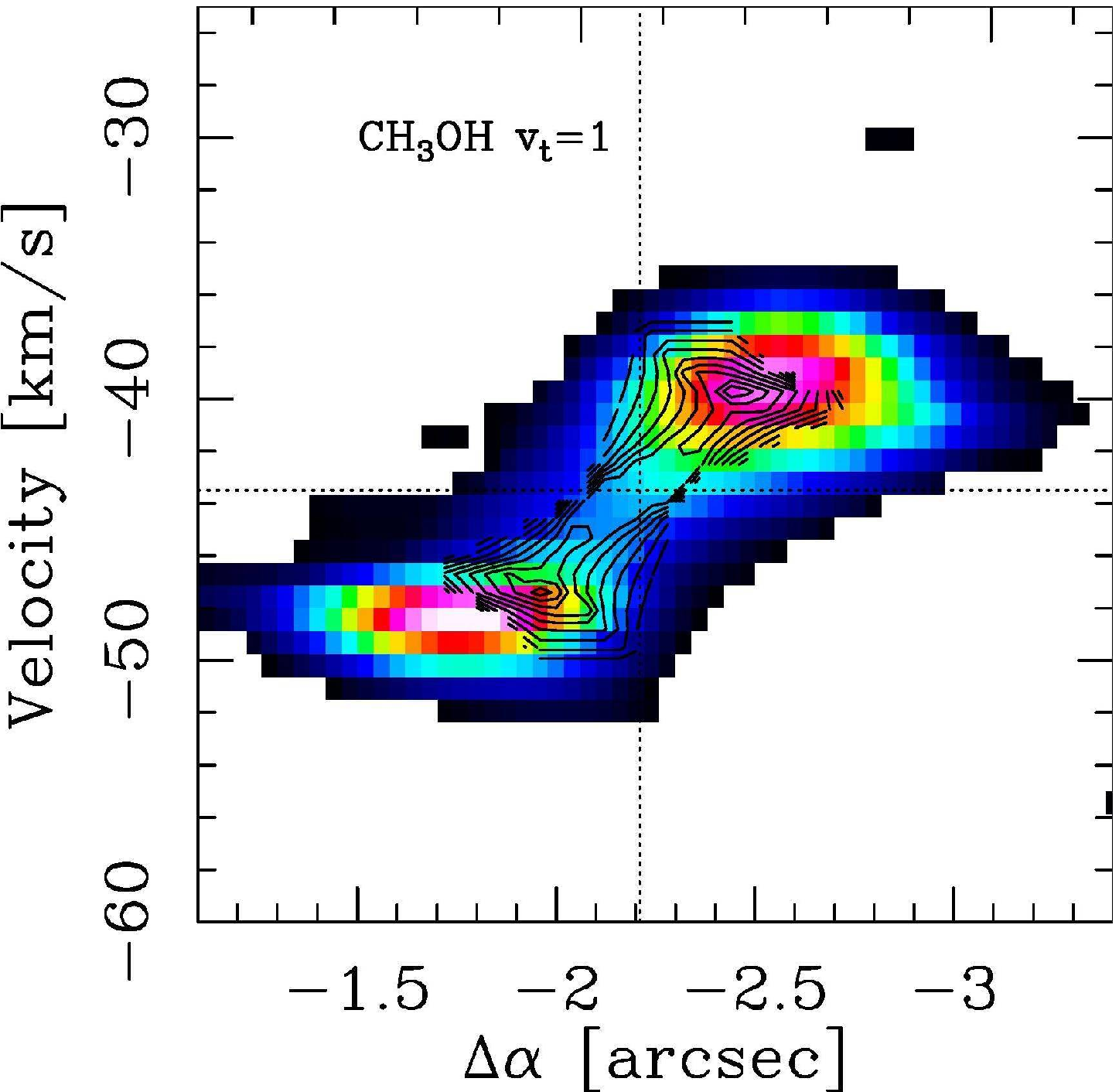}
  \includegraphics[width=0.43\linewidth,angle=0]{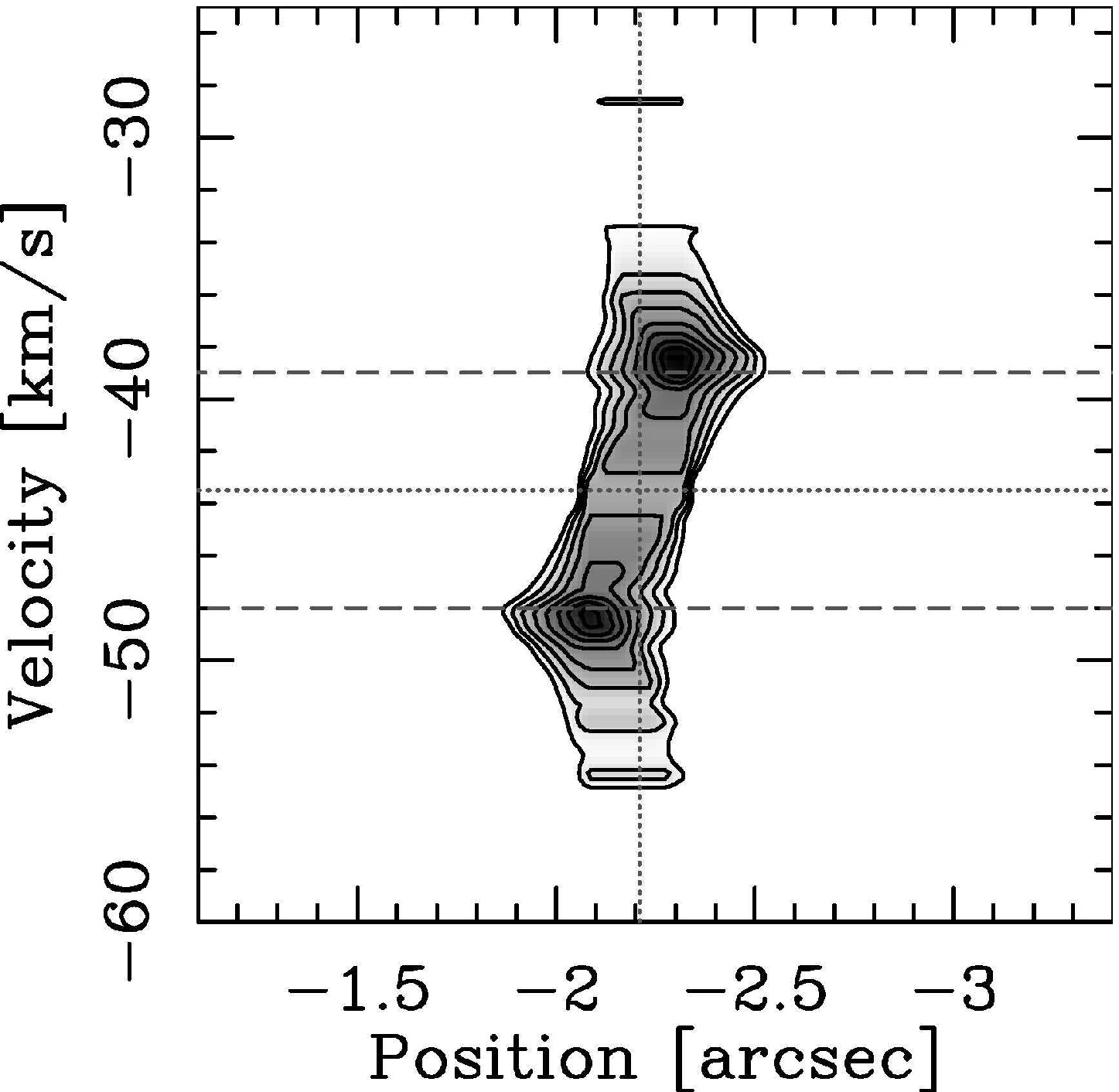}
   \includegraphics[width=0.43\linewidth,angle=0]{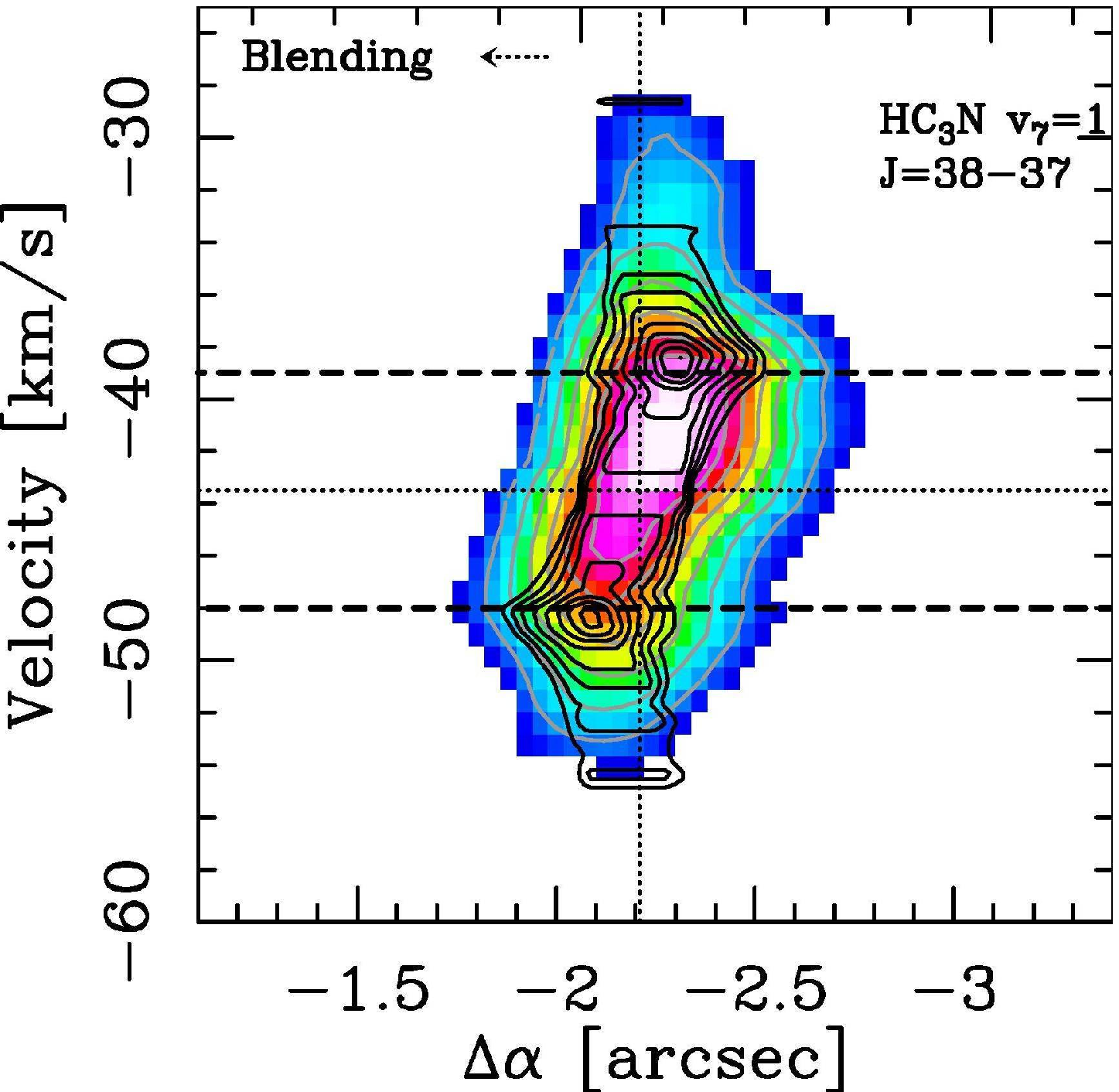}
      \caption{\textsl{Left column:}  $pv$-diagram of models with a rotating ring with $v_{\rm r}\sim r$ (top), and with a disk in Keplerian rotation (bottom). \textsl{Right column:} The color scale shows the 0th moment maps for the 334.42\,GHz CH$_3$OH $\varv_t=1$ line (top), and the 346.456\, GHz HC$_3$N $\varv_7=1$ line (bottom). Contours are the same as in the left column and show the model prediction.}
                   \label{fig:models}%
    \end{figure}
     
   \begin{figure}[!]
   \centering
    \includegraphics[width=0.9\linewidth,angle=0]{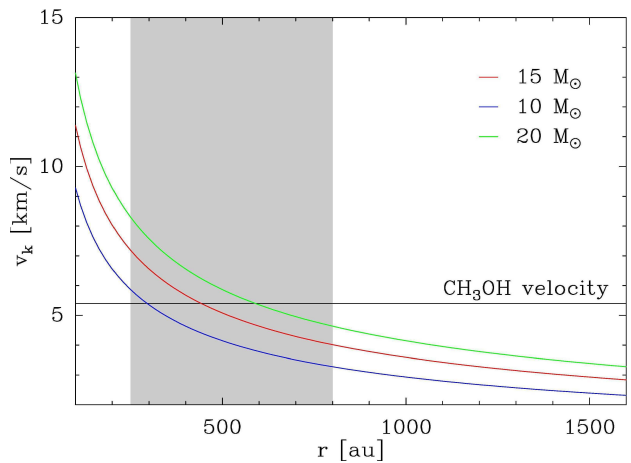}
       \caption{Keplerian velocity as a function of radius, $r$, from the central object for 10, 15, and 20\,$M_{\odot}$. The projection corrected velocity determined from the 334.42\,GHz CH$_3$OH $\varv_t=1$ line is marked as a gray horizontal line. The grey shaded area corresponds to the projected distance range for the position of the CH$_3$OH $\varv_t=1$ spots.}
              \label{fig:keplerian}%
    \end{figure}

\subsection{Physical properties of disks around high-mass protostars}\label{sec:d4}

Although 
massive rotating toroids towards precursors of OB-type stars have been frequently detected 
\citep{Beltran2004, Cesaroni2014, Beltran2014, SM2013, Cesaroni2017},
clear signatures of accretion disks around high-mass protostars are still challenging to identify, in particular towards the precursors of the most massive, O-type stars \citep{Beltran2016}. 
Our results suggest the presence of an
accretion disk around a still deeply embedded young high-mass protostar that is likely to be a precursor of an O4-O5 type star. 
Our findings suggest
that a disk may have formed already at this early stage,
providing observational support to 
numerical simulations which predict that  despite 
 the strong radiation pressure exerted by high-mass protostars, accretion through flattened structures, and disks enable the formation of the highest mass stars  
\citep{Krumholz2009, Kuiper2010, Kuiper2011}. 

Together with other examples of envelope-outflow-disk systems (e.g.\,\citealp{Johnston2015, Beltran2016}), this suggests a physical picture of high-mass star formation on the core scale that is qualitatively very similar to that of 
low-mass objects. As observed towards
L1527 \citep{Sakai2014},  TMC-1 \citep{Aso2015}, and VLA1623A \citep{Murillo2013},
we also see evidence for shocks induced by the infall from the envelope 
to the disk imposing a
change in the chemical composition of the infalling gas at the centrifugal barrier (see also \citealp{Oya2017}).
The accretion disk around the protostar of L1527 has been confirmed since then with direct imaging \citep{Sakai2017}.

   \begin{figure}[!ht]
   \centering
   \includegraphics[width=0.65\linewidth,angle=90]{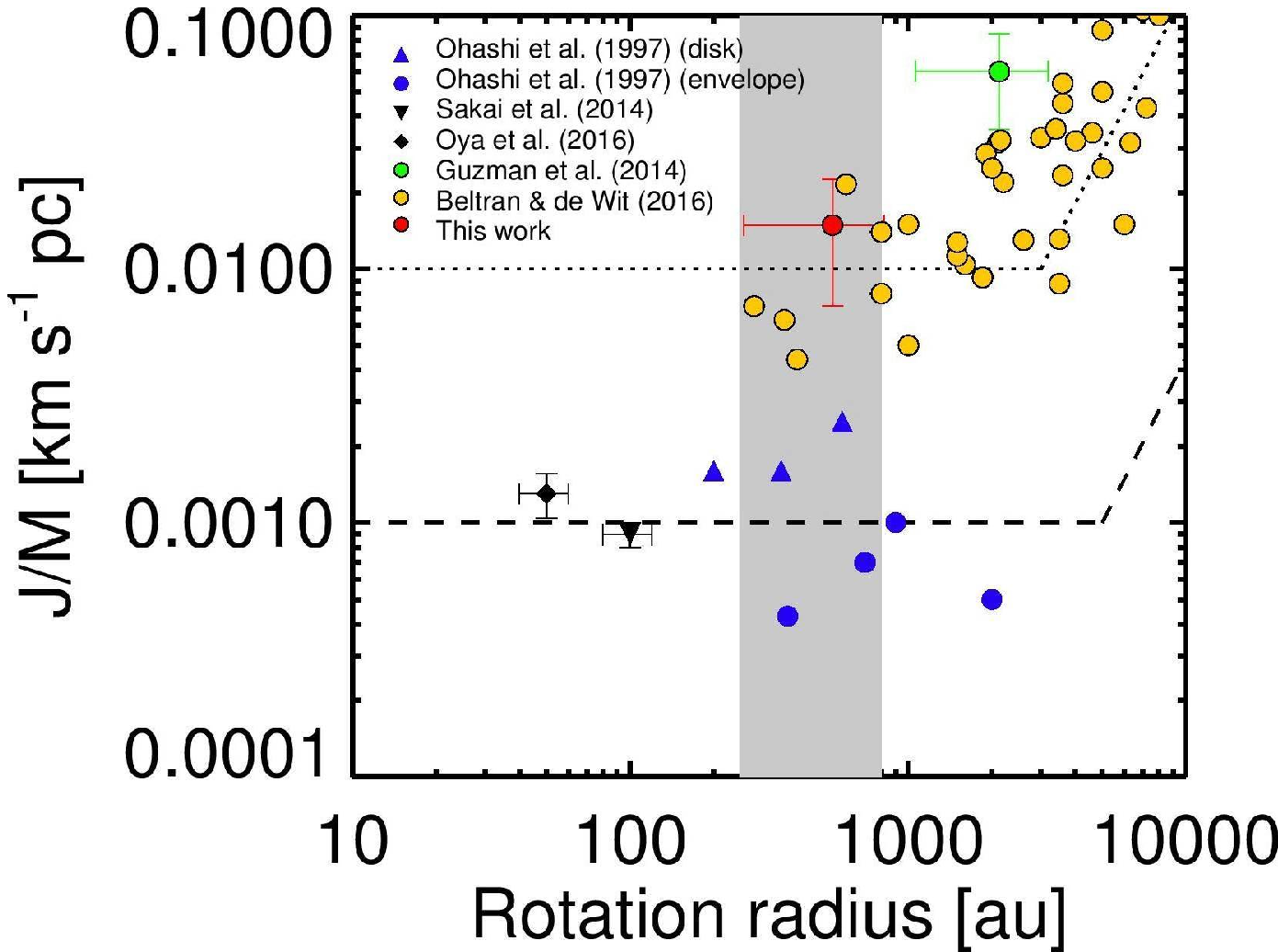}
      \caption{Local specific angular momentum ( $j/m= R \times v_{\rm rot}$) as a function of radius for a sample of low- and high-mass protostars and YSOs.  The high-mass sample \citep{Guzman2014, Beltran2016} corresponds to high-mass YSOs with disk candidates, the low-mass objects \citep{Ohashi1997, Sakai2014, Oya2016} correspond to envelopes and disks identified around low-mass Class 0 protostars. Where no explicit information was available, we adopt a 50\% uncertainty for the rotation radius, and a linearly propagated 50\% uncertainty on the estimated specific angular momentum. For our target, we estimate a 52\% uncertainty on these values based on the uncertainty in the disk radius estimate. On this plot we made no attempt to correct for the source's inclination angle. Dashed line shows the j=0.001\,km\,s$^{-1}$ line \citep{Belloche2013}, and an angular velocity, $\Omega$=1.8\,km\,s$^{-1}$\,pc$^{-1}$. For the high-mass sample we show the j=0.01\,km\,s$^{-1}$\,pc$^{-1}$ line, and $\Omega$=50\,km\,s$^{-1}$\,pc$^{-1}$. The grey shaded area shows our range of disk size estimate. }
              \label{fig:jm}%
    \end{figure}

Our findings suggest, however, considerably
different physical conditions and chemical environment for high-mass protostars. 
The observed prominent bright peak of \methanol,  that we explain by shocks at the centrifugal barrier, 
blends with the more diffuse emission from the envelope. At our spatial resolution, these shocks do not outline a sharp boundary, and leave us with a relatively large range of disk radii that is still consistent with the observations.
Based on the size of the compact dust continuum source, we measure a  minimum projected radius of $\sim$250\,au for the disk major axis, while a maximum outer radius is constrained by the peak of the \methanol\ emission located somewhere between 300 and 800\,au. 
We correct these values for an orientation angle of $\phi\sim$12$^\circ$ based on the fitted position angle of the dust residual emission, giving   
a disk radius between 255 and 817\,au, that is
a factor of few larger than {recent ALMA observations suggest} for disks around some
low-mass Class 0 protostars (e.g.\,\citealp{Oya2016, Sakai2017}). For the rotational velocity at this outer radius we take 
the velocity offset of the \methanol\ peaks corrected for the inclination angle (Sect.\,\ref{sec:d2}). Taking an average disk size between the minimum and maximum expected 
values implies that the local specific angular momentum, $j/m=R \times v_{\rm rot}=4.5\times10^{21}$\,cm$^{2}$\,s$^{-1}$, 
where $R$ is the disk radius, and $v_{\rm rot}$, 
is the rotational velocity of the disk at the given radius. 

We compare our measurement in 
Fig.\,\ref{fig:jm} to values from the literature following  \citet{Ohashi1997}, and \citet{Belloche2002}, and complement it with high-mass disks and toroids from \citet{Beltran2016}.
We recognise that the local specific angular momentum is considerably higher towards the high-mass case compared to the low-mass {case}, although the so far observed disk candidates are still at larger physical scales, and typically towards objects that are likely in a more evolved stage than the protostar of \mysou. 
This suggests that the kinetic energy may be larger  at the onset of the collapse in the case of high-mass star formation. The larger kinetic energy could be explained if the collapse sets in at the clump, thus at $>$0.3\,pc scales compared to the core-scale collapse that is typical for the formation of low-mass protostars. The large specific angular momentum towards high-mass protostars is therefore in agreement with the scenario of global collapse or models based on cloud-cloud collisions at the origin of high-mass stars.

A dust based mass estimate for the disk is very uncertain, not only due to the uncertainty of the temperature, but also because of the unknown dust opacity. Assuming the same dust parameters as for the envelope, and taking $T_{\rm dust}=150$\,K, we obtain sub-solar mass estimations around $M_{\rm disk}<0.25$\,\msol. Such an elevated temperature is expected for the disk in the close vicinity of the protostellar embryo, it is, however, still consistent with the SED because of the large optical depth of the cooler dust. Since our disk mass estimate is sub-solar, it is likely that the disk mass is below 10\% of the mass of the central object and thus gravitationally stable. The stability of the disk itself is an important question, unstable massive disks could either lead to episodic accretion bursts, or undergo fragmentation \citep{Vorobyov2010}. Both phenomena are observed towards low-mass protostars; in particular multiplicity within low-mass cores has been recently explained by disk fragmentation \citep{Tobin2016}. On the other hand, the high-mass disk candidate AFGL 4176 \citep{Johnston2015} is more extended, but also more massive. Fragmentation of massive or unstable disks and toroids around high-mass stars could explain why short period binaries have the highest frequency among O-type stars \citep{Sana2017}.

\subsection{Implications for high-mass star formation}
The fact that we observe at the same time a massive core which is not fragmented down to our resolution limit of $\sim$400\,au, and find strong evidence for a centrifugal barrier at a large radius of 300-800\,au, 
may appear contradictory and needs to be discussed. The observed indication for the Keplerian disk together with the large angular momentum suggest that 
magnetic braking has not been efficient to evacuate and redistribute the angular momentum.
Numerical simulations predict that, in particular, at the early phase of the collapse, the angular momentum from the accretion disk can be efficiently removed due to magnetic braking, and thereby suppress the formation of large disks (e.g.\,\citealp{Seifried2011, Myers2013, Hennebelle2016}). Our observations therefore point to a relatively weak magnetic field. 
On the other hand, despite its large mass which is two orders of magnitude larger than the thermal Jeans mass, the core did not fragment and appears to be collapsing monolithically, which is consistent with the Turbulent Core model \citep{MT03}. This would require additional support to complement the thermal pressure which can either be magnetic or turbulent. Neither the line widths, nor the large angular momentum are consistent with strong enough turbulence or magnetic fields, therefore the properties of the core embedded in \mysou\ appear difficult to explain. 

However, if turbulence or strong ordered motions are present, the misalignment between the magnetic field lines and the angular momentum vectors can limit the effect of magnetic braking leading to a less efficient removal of the angular momentum \citep{Myers2013}. Alternatively, the present day properties of the collapsing core and the physical state of the pre-stellar core prior to collapse may have been significantly different at onset of the collapse. The small scale properties of high-mass protostars, and the physical properties of their accretion disk, such as in \mysou\ may thus challenge high-mass star formation theories. Clearly, more observational examples of accretion disks around high-mass protostars are needed to put further constrains on the formation and properties of disks, and the collapse scenario. 

\section{Summary and conclusions}
We presented a case study of one of the targets from the SPARKS project, which uses high angular-resolution observations from ALMA to study the sample of the most massive mid-infrared quiet massive clumps selected from the ATLASGAL survey.  Our observations reveal a single {massive protostellar envelope} associated with the massive clump, \mysou. Based on protostellar evolutionary tracks, we estimate the current protostellar mass to be between 11 and 16\,\msol, surrounded by a massive core of $\sim$120\,\msol. The estimated envelope mass is an order of magnitude larger than the currently estimated protostellar mass, making this object an excellent example of a high-mass protostar in its main accretion phase, similar to the low-mass Class\,0 phase. 

We discovered torsionally excited \methanol\ spots offset from the protostar with a velocity offset of $\pm$4.5\,\kms\ compared to the source $v_{\rm lsr}$. These peaks are best explained by shocks from the infalling envelope onto the centrifugal barrier. Based on the observed unblended methanol transitions, we estimate the physical conditions on these spots,
and find $T_{\rm kin}=160-170$\,K, and $N$(CH$_3$OH)={$1.2-2\times10^{19}$}\,cm$^{-2}$, suggesting large \methanol\ column densities.

Our analysis of the dust emission reveals {azimuthal} elongations associated with the dust continuum peak, 
and a compact component with a marginally resolved beam deconvolved $R_{90\%}$ radius of $\sim$250\,au measured along is major axis.
This component is consistent with an accretion disk within the centrifugal barrier outlined by the \methanol\ shock spots at a distance between $\sim$300 and 800\,au offset from the protostar. Furthermore, we propose the vibrationally excited HC$_3$N {$\varv_{\rm 7}=1e$ $J$=38--37} line as a potential new tracer for the emission from the accretion disk. 

Our results allow for the first time to dissect a clearly massive protostellar envelope potentially forming an O4-O5 type star. The physical picture is qualitatively very similar to that of the low-mass star formation process, however, quantitatively both the physical and the chemical conditions show considerable differences. Our estimate of the specific angular momentum carried by the inner envelope at its transition to an accretion disk is an order of magnitude larger than that observed around low-mass stars. This is consistent with the scenario of global collapse, where the larger collapse scale would naturally lead to a larger angular momentum compared to the core collapse models.

\begin{acknowledgements}
{We thank the referee for the careful reading of the manuscript.}
This paper makes use of the ALMA data: ADS/JAO.ALMA 2013.1.00960.S. ALMA is a partnership of ESO (representing its member states), NSF (USA), and NINS (Japan), together with NRC (Canada), NSC and ASIAA (Taiwan), and KASI (Republic of Korea), in cooperation with the Republic of Chile. The Joint ALMA Observatory is operated by ESO, AUI/NRAO, and NAOJ. T.Cs. acknowledges support from the \emph{Deut\-sche For\-schungs\-ge\-mein\-schaft, DFG\/}  via the SPP (priority programme) 1573 'Physics of the ISM'. 
HB acknowledges support from the European Research Council under the Horizon 2020 Framework Program via the ERC Consolidator Grant CSF-648505.
LB acknowledges support by CONICYT Project PFB06.  A.P. acknowledges financial support from UNAM and CONACyT, M\'exico.
 \end{acknowledgements}
   \bibliographystyle{aa} 
   \bibliography{scibib} 

\begin{appendix} 
\section{Dust spectral energy distribution and modelling}\label{app:sed}
We construct the spectral energy distribution (SED) of the protostar embedded in the clump \mysou, from the mid-infrared wavelengths up to
 the radio regime (Fig.\,\ref{fig:sed}) in order to estimate its bolometric luminosity ($L_{\rm bol}$), and constrain a representative dust temperature ($T_{\rm d}$) for the bulk of the mass using a model of greybody emission. 

In Fig.\,\ref{fig:sed} we show the flux densities from the GLIMPSE catalog at the shortest indicated wavelengths \citep{Benjamin2003}, as well as the WISE band 4 photometry
at 22\,$\mu$m \citep{Cutri2012}. {To illustrate the complexity of the region, we show the  emission from the far-infrared and millimetre wavelength range in Fig.\,\ref{fig:herschel}
using Herschel/Hi-GAL data \citep{Molinari2010}, and the ATLASGAL-Planck combined data at 870\,$\mu$m \citep{Csengeri2016a}. This shows that the emission is largely dominated by extended structures at all these wavelengths.}

As a comparison we show the corresponding sources identified in
the Herschel/Hi-GAL point source catalog \citep{Molinari2016}, and the ATLASGAL Gaussclumps source catalog \citep{Csengeri2014}. The 870\,$\mu$m flux density measurement by ALMA reveals the MDC
at a size-scale of 0.06\,pc (see also \citealp{Csengeri2017b}) and therefore puts constraints 
on the extent of the embedded source. To derive therefore the properties of the gas
 representative of the embedded protostar, our aim is to extract and scale the flux densities corresponding to a source at $\sim$8\,\arcsec\ (the geometric mean of the PACS~70\,$\mu$m beam, \citealp{Molinari2016}) with a density profile of $n\sim r^{-2}$, {and neglecting any temperature gradient}. To do this, we perform aperture photometry on the PACS~70\,$\mu$m, and 160\,$\mu$m maps by first measuring the peak intensity  at the position of our target within a single beam (taking 5.8\arcsec$\times$12.1\arcsec, and 11.4\arcsec$\times$13.4\arcsec, respectively), and then measure the background emission in 3 different annuli at increasing distance from 15\arcsec\ to 35\arcsec\ with respect to the source. For the SPIRE~250, 350, and 500\,$\mu$m data, we use the values from the  
 Herschel/Hi-GAL point source catalog, and scale them adopting the geometric mean of the measured major and minor axes 
 \citep{Molinari2016}.  

To obtain the bolometric luminosity of the protostar, we add up all
emission from the near/mid-infrared to 870\,$\mu$m, 
and obtain $1.3\times10^{4}$\,$L_{\rm bol}$.  
As a comparison, we also calculated the protostar's internal luminosity using the empirical relation between $L_{\rm bol}$ and the flux density measured at 70\,$\mu$m \citep{Dunham2008}, and obtain $1.2\times10^{4}$\,$L_{\rm bol}$. The significant confusion due to extended emission from the mid-infrared to the submillimeter wavelengths adds, however, some uncertainty to our estimate, the measured values likely correspond to an upper limit to the luminosity.

To obtain an estimate of the dust temperature, $T_{\rm d}$, we perform a greybody fit to the far-infrared points of the SED between 70\,$\mu$m and 870\,$\mu$m. Since at 70\,$\mu$m the emission is mostly dominated by the heated dust in the vicinity of the protostar, we use a 
two component greybody to model the SED, which reveals the temperature corresponding to the cold gas component dominating the bulk of the emission, and puts strong constrain on the warm gas temperature, as well as the fraction of the heated gas mass.
We use $\kappa_{\rm{345\,GHz}}=0.0185$\,g\,cm$^{-2}$ and an emissivity index of $\beta=2$, where $\kappa_\nu=\frac{\nu}{\rm{345\,GHz}}^{-\beta}$. We obtain a cold gas component at 22\,K, and a warm gas component at 48\,K that contains $<$5\% of the total gas mass. The result of the SED fit is shown in Fig.\,\ref{fig:sed}.

   \begin{figure}
   \centering
   \includegraphics[width=5cm,angle=90]{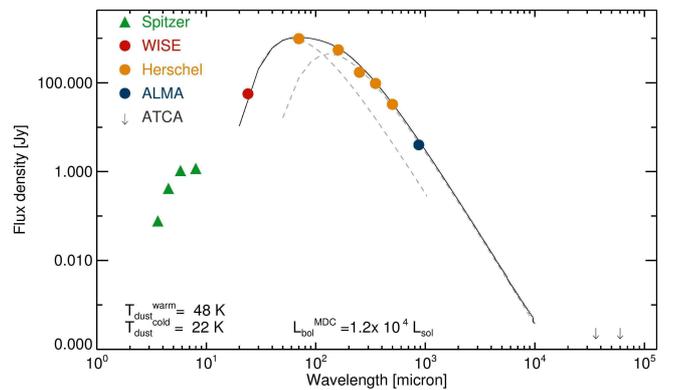}
      \caption{{SED of the embedded protostar within the ATLASGAL clump, {\mysou}. The origin of the shown flux densities are labeled in the figure legend, and are described in the text. Solid line shows the result of a two (warm and cold) component grebody fit with 48\,K and 22\,K. The individual components are shown in a dashed gray line. }}
              \label{fig:sed}%
    \end{figure}
   \begin{figure*}
   \centering
   \includegraphics[width=0.9\linewidth]{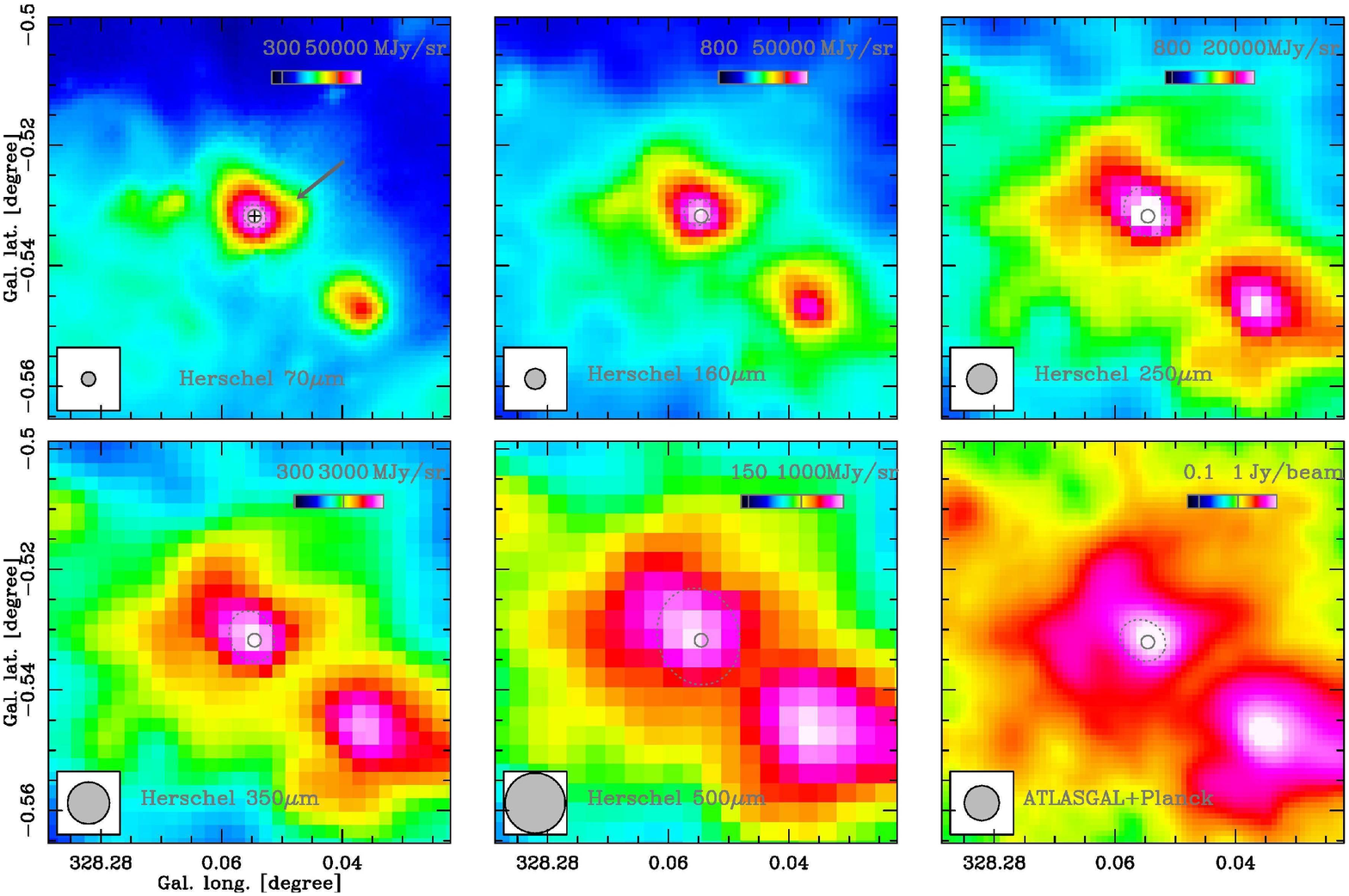}
      \caption{Far-infrared multi-wavelength view of the ATLASGAL clump, {\mysou} using the {\sl Herschel} Hi-GAL images \citep{Molinari2016}, and the APEX/LABOCA and Planck combined maps \citep{Csengeri2016a}. The position of the identified source is marked with a black cross, dotted gray ellipses show the corresponding sources from the Hi-GAL, and ATLASGAL catalogues (\citealp{Molinari2016, Csengeri2014}, respectively). The solid circle shows half the 8\arcsec\ $FWHM$ corresponding to the scale of the MDC. The FWHM beam widths are shown in the lower left corner of each panel. The color scale goes on a logarithmic scale, {the black line shows the labeled flux density value.}}
              \label{fig:herschel}%
    \end{figure*}

\end{appendix}

\end{document}